\documentclass[english]{article}
\usepackage[a4paper]{geometry}

\usepackage[T1]{fontenc}
\usepackage[latin1]{inputenc}
\usepackage{float}
\usepackage{mathrsfs}
\usepackage{amsmath}
\usepackage{amssymb}
\usepackage{graphicx}
\usepackage{hyperref}
\usepackage[capitalize,noabbrev]{cleveref}
\crefformat{equation}{(#2#1#3)}

\hypersetup{
    colorlinks=true,
    linkcolor=blue,
    filecolor=magenta,      
    urlcolor=cyan,
    citecolor = red,
}

\floatstyle{ruled}
\newfloat{algorithm}{tbp}{loa}
\providecommand{\algorithmname}{Algorithm}
\floatname{algorithm}{\protect\algorithmname}

\usepackage{mathrsfs}
\usepackage{amsfonts}
\usepackage{bbm}
\usepackage{epsfig}
\usepackage{bm}
\usepackage{mathrsfs}
\usepackage{enumerate}
 \usepackage{color}

\usepackage{subfig}\usepackage[all]{xy}

\newcommand{\bbR}{\mathbb R}

\newtheorem{rem}{Remark}[section]
\newtheorem{example}{Example}[section]

\newcommand{\bbE}{\mathbb{E}}
\newcommand{\bbC}{\mathbb{C}}
\newcommand{\bbP}{\mathbb{P}}

\newcommand{\cN}{\mathcal{N}}
\newcommand{\cL}{\mathcal{L}}
\newcommand{\cO}{\mathcal{O}}

\newcommand{\note}[1]{{{#1}}}

\usepackage{xspace}
\usepackage{tabu}
\usepackage{booktabs}

\def\bbR{\mathbb{R}}
\def\bbP{\mathbb{P}}
\def\bbE{\mathbb{E}}

\def\cD{\mathcal{D}}

\title{Sparse online variational Bayesian regression}
\author{Kody {J.H. Law}\thanks{Department of Mathematics, University of Manchester, UK.}
\and Vitaly Zankin\footnotemark[1]}

\begin{document}
\maketitle

\begin{abstract}
This work considers {variational Bayesian inference} 
as an inexpensive and scalable alternative to a fully 
Bayesian approach in the context of sparsity-promoting priors.
In particular, the priors considered arise from scale mixtures of 
Normal distributions with a generalized inverse Gaussian mixing distribution. 
This includes the variational Bayesian LASSO as an inexpensive and scalable alternative 
to the Bayesian LASSO introduced in \cite{park2008bayesian}.
It also includes a family of priors which more strongly promote sparsity. 
For linear models the method requires only the iterative solution of 
deterministic least squares problems. 
 \note{Furthermore, for $p$ unknown covariates 
 the method can be implemented exactly online with a cost of $\cO(p^3)$ 
 in computation and $\cO(p^2)$ in memory per iteration -- in other words, 
 the cost per iteration is independent of $n$, 
 and in principle infinite data can be considered.}
For large $p$ an approximation is able to achieve promising results for a cost of 
 $\cO(p)$ per iteration, in both computation and memory.
 Strategies for hyper-parameter tuning are also considered. 
 The method is implemented for real and simulated data.
  It is shown that the performance in terms of variable selection and 
  uncertainty quantification of the variational Bayesian LASSO 
  can be comparable to the Bayesian LASSO for problems which are tractable with that method, and for a fraction of the cost. 
 The present method comfortably handles 
 \note{$n= 65536,\ p = 131073$ on a laptop in less than $30$ minutes, 
 and $n=10^5,\ p=2.1\times10^6$ overnight.}
\end{abstract}

\section{Introduction}
\label{sec:intro}

Regression is a quintessential and ubiquitous task of machine learning.
The simplest method one can use to solve regression tasks is a linear model 
with Gaussian noise and prior \cite{box2011bayesian}.
The most attractive feature of linear Gaussian models is analytical tractability, 
from both frequentist and Bayesian viewpoints.
However, once one employs basis function expansions, they also become quite flexible.
There are numerous methods in which linear models can be embedded, such as 
total least squares \cite{zhu2011sparsity} and mixtures of regressions \cite{khalili2007variable}, for example.
Sparsity promoting priors have proven to be very useful for identifying useful features
and avoiding overfitting, perhaps most notably the LASSO \cite{tibshirani1996regression} and its incarnation
as total variation (TV) regularization in imaging \cite{vogel1996iterative,strong2003edge}.
However, as soon as a non-Gaussian prior is introduced then analytical tractability
is lost and, in particular, the Bayesian solution becomes very expensive \cite{park2008bayesian}, 
requiring Markov chain Monte Carlo (MCMC) methods \cite{robert, green2015bayesian}.
Furthermore, sparsity promoting priors are not differentiable, which prevents
the use of simple Gaussian approximations such as Laplace approximation \cite{bishop}.

\subsection{Linear models}
\label{sec:linear}

Let $\cD_n = \{(x_i,y_i)\}_{i=1}^n$,with $x_i\in \bbR^p$ and $y_i\in \bbR$ (for simplicity),
and let $X_n = [x_1, \dots, x_n]^T$ and $Y_n=(y_1,\dots,y_n)^T$.
\note{Consider the following statistical model, in a Bayesian linear regression context
\begin{equation}\label{eq:basic}
y_i = x_i^T \beta + \epsilon_i \, , \quad \beta \perp \epsilon_i \sim N(0, \gamma^2) ~{\rm i.i.d.} \, ~ {\rm for}~ i=1,\dots, n
\, ,
\end{equation} 
where $N(m, C)$ denotes a multivariate Gaussian distribution 
with mean $m\in \bbR^l$, $l\geq 1$, and covariance $C\in \bbR^{l\times l}$.
The notation $N(z; m, C)$ will be used to denote the corresponding density
with argument $z\in \bbR^l$.
The Bayesian formulation of this problem is to identify the posterior distribution
on $\beta$
\begin{equation}\label{eq:posterior}
\bbP(\beta | \cD_n) =
\frac{\bbP(Y | X, \beta) \bbP(\beta)}{\bbP(Y | X)} \, .
\end{equation}
If $\bbP(\beta) = N(\beta ; m_{0}, C_{0})$, 
then the posterior distribution $\beta | \cD_n \sim N(m_n,C_n)$ 
is given in closed form. Otherwise it is not.}

Linear models of the form $x^T \beta$ are quite flexible, 
once one considers basis function expansions. 
In other words,  
$x = (1,\psi_{0}(s), \psi_1(s), \dots, \psi_p(s))$ for data $s\in \bbR^d$ and some functions $\{\psi_i\}$, 
which can be a subset of polynomials \cite{uqhandbook, chkifa2015discrete, guo2020constructing}, 
wavelets and other ``$x$-lets'' \cite{chen2001atomic}, 
radial basis functions \cite{broomhead1988radial},
random feature models \cite{rahimi2007random}, or any number of other choices. 
The book \cite{bishop} provides a concise and easy to read summary for regression applications.
In fact, there are complete bases for many function-spaces. 
For example, if $\Omega = [0,1]^d$ then the Fourier series forms a complete basis for 
$L^2(\Omega) = \{f: \Omega \rightarrow \bbR ; \int_\Omega f(s)^2 ds < \infty \}$ 
 \cite{georgi1976fourier} and a subset of such features can therefore be used to construct 
 a convergent approximation. 
 In fact, since shift-invariant kernel operators 
 (those defined only in terms of differences, $K(x,y)=k(x-y)$)
 are diagonalized by the Fourier basis,
 then the expectation of the product of two such features with an
 appropriately distributed  
 random frequency is equal to the kernel evaluation. 
Monte Carlo approximation of such expectations is the basis of 
 random feature models \cite{rahimi2007random}, 
 which are another popular class of linear models in the machine learning literature. 

An issue is how many terms to include, and perhaps more importantly, 
how to select a subset of the important terms from a sufficiently rich set, 
without incurring a loss in accuracy due to overfitting,
as can occur with too much flexibility. 
The latter issue is often referred to as ``variance-bias tradeoff'':
a model which is too flexible (negligible bias) 
may be too strongly influenced by a particular data set, hence incurring a large
variance over a range of different data sets \cite{geman1992neural}.
This well-known issue can be dealt with by beginning with a sufficiently rich
class of approximating functions (e.g. a large enough basis) and then 
introducing prior assumptions, or regularization, in order to let the model
select the most parsimonious representation 
\cite{tikhonov1963solution, hoerl1970ridge, tibshirani1996regression, 
neal1996bayesian, muller2004nonparametric, stuart2010inverse}.

\subsection{Sparsity priors}
\label{sec:sparse}

In the context of prior selection, often the Gaussian assumption is considered
too restrictive. In particular, it has become very popular at the end of the last and 
the beginning of this millenium to utilize a sparsity promoting prior.
\note{Motivated by sparsity penalties which have been successful in frequentist 
bridge regression 
\cite{tibshirani1996regression, chen2001atomic, donoho2006compressed}
early sparsity priors simply replace the quadratic density associated to a Gaussian prior 
(ridge regression)
with another density of the form $\exp(-R(\beta))$, 
where $R(\beta) = |L\beta|_r^r$, for $r \in (0,1]$, $L : \bbR^{\tilde{p}} \rightarrow \bbR^p$, 
and $|\beta|_r^r = \sum_{i=1|}^p |\beta_i| ^r$
\cite{polson2014bayesian,park2008bayesian,armagan2009variational}.
This can be extended to the case $r=0$, which corresponds to counting measure 
on the non-zero elements $|\beta|_0 = \sum_{i=1}^p {\bf 1}_{\{\beta_i \neq 0\}}$.} 
Note that if there is a $W: \bbR^p \rightarrow \bbR^{\tilde{p}}$ such that $WL=I_{\tilde{p}}$, then 
one can always redefine $X_n \rightarrow X_nW$ and $\beta \rightarrow L\beta$.
Therefore, we assume without loss of too much generality that $L=I_p$.
This will be discussed further in the examples.

\note{General sparsity-promoting priors of the type $R(\beta)=f(|\beta|)$ are also possible,
and this type of prior will be the focus of the present work, to be introduced 
in the following subsection.
Collectively, this family of priors 
have come to be known as ``shrinkage'' priors because the resulting 
maximum a posteriori (MAP) estimator 
(or frequentist penalized maximum likelihood estimator (MLE)) 
tends to ``shrink'' all the coefficients towards zero. In particular, 
sufficiently small coefficients are exactly zero, 
which is the main impetus underlying their use,
while the excess shrinkage leads to a non-desirable bias.
In recent years ``spike and slab'' priors \cite{mitchell1988bayesian} 
have become very popular, as they soften this non-desirable property. 
Such priors are hierarchical, 
consisting of a mixture distribution with a Dirac mass at 0 (a spike) 
and a continuous distribution such as the ones considered here (a slab). 
See \cite{bai2020spike} for a recent review, 
focused on the spike and slab LASSO
(SSLASSO).

There is a computational burden to performing inference with these more
exotic priors, even in the case when we 
abandon uncertainty quantification (UQ)
and settle for a MAP estimator. In the best case of $R(\beta) = |\beta|_1$
we have a convex optimization problem, which can be 
solved efficiently by a number of modern methods, such as
iterative soft thresholding \cite{daubechies2004iterative, bredies2008linear}
and alternating direction method of multipliers \cite{gabay1976dual, boyd2011distributed}.
These methods are able to achieve a comparable cost 
to the solution of a least squares problem, i.e. $\cO(np)$ at best,
and there is scope for per-iteration parallelization.
However, there are limitations and drawbacks to this choice,
and it is often desirable to promote sparsity more strongly, 
e.g. using non-convex 
$R(\beta)$, where there are no such default recipes \cite{donoho2006compressed}. 
The recently introduced SSLASSO
\cite{rovckova2018spike} is a notable algorithm 
that delivers a point estimate which 
promotes sparsity more strongly than $R(\beta) = |\beta|_1$
and at a comparable cost to the methods above.}

Considering the full Bayesian posterior, the situation is even more daunting.
Indeed once one adopts such a sparsity prior then the posterior is no longer characterized 
in closed form with finitely many parameters, as in the Gaussian case (in the finite/discrete/parametric case).
Laplace approximation \cite{bishop} requires derivative and Hessian of the log-posterior,
which may not exist.
Computationally-intensive methods such as Markov chain Monte Carlo (MCMC) \cite{robert}
are required for consistent estimation, as proposed for the Bayesian LASSO 
(BL) in \cite{park2008bayesian}.
There has been a lot of activity in this direction in the past 10 years -- see, e.g.
\cite{lucka, roininen, calvetti, marzouk}
for some examples from the applied mathematics and inverse problems communities
\note{and \cite{pereyra1, pereyra2, jacob} for some examples from machine learning and statistics.}
Here we propose to employ a variational approach to recover the best Gaussian
approximation to the sparsity-promoting posterior, in a sense to be defined precisely below.
\note{This approach provides approximate UQ for a 
substantially smaller cost than fully Bayesian approaches. 
Indeed the cost is only slightly larger than point estimation methods.}

\subsection{Contribution}
\label{ssec:introcont}

\begin{figure}[!htbp]
	\centering\includegraphics[width=0.49\textwidth]{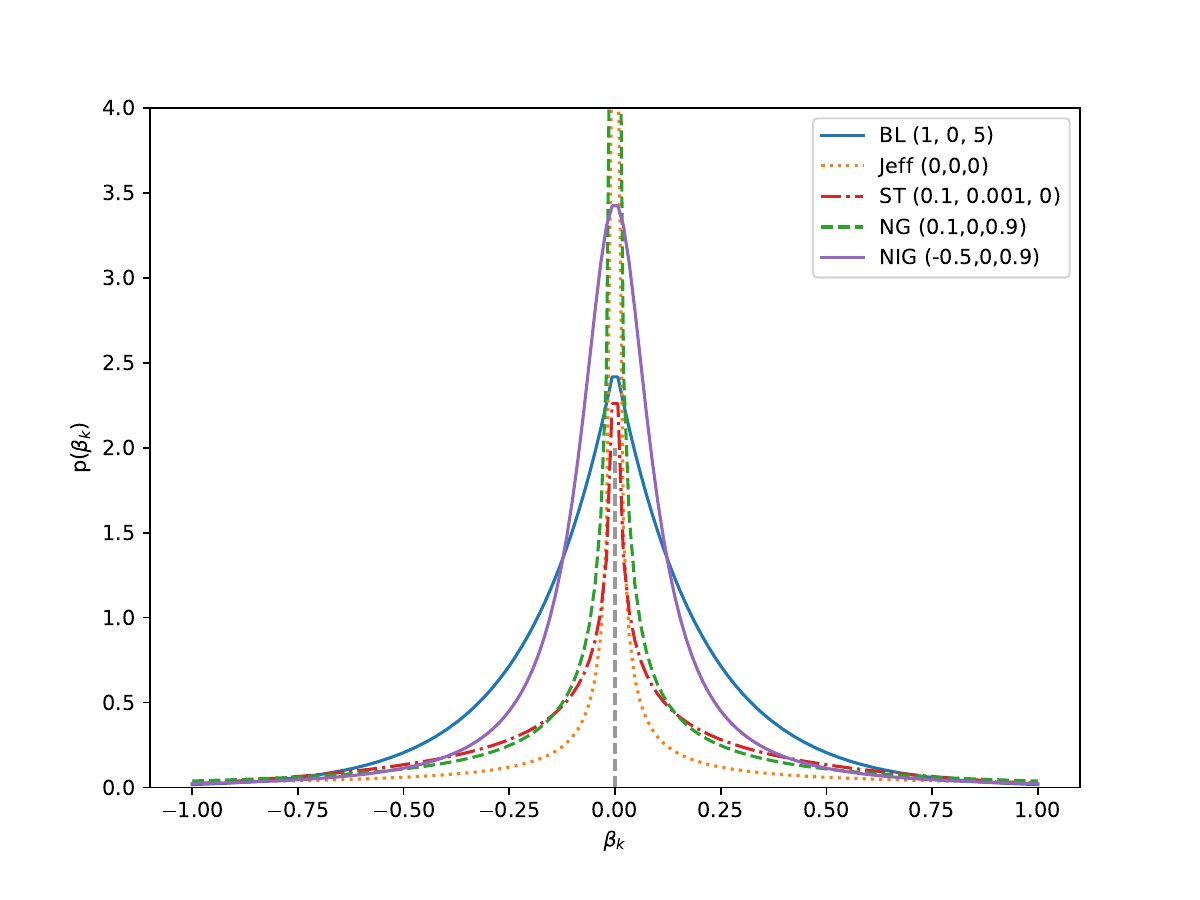}
	\includegraphics[width=0.49\textwidth]{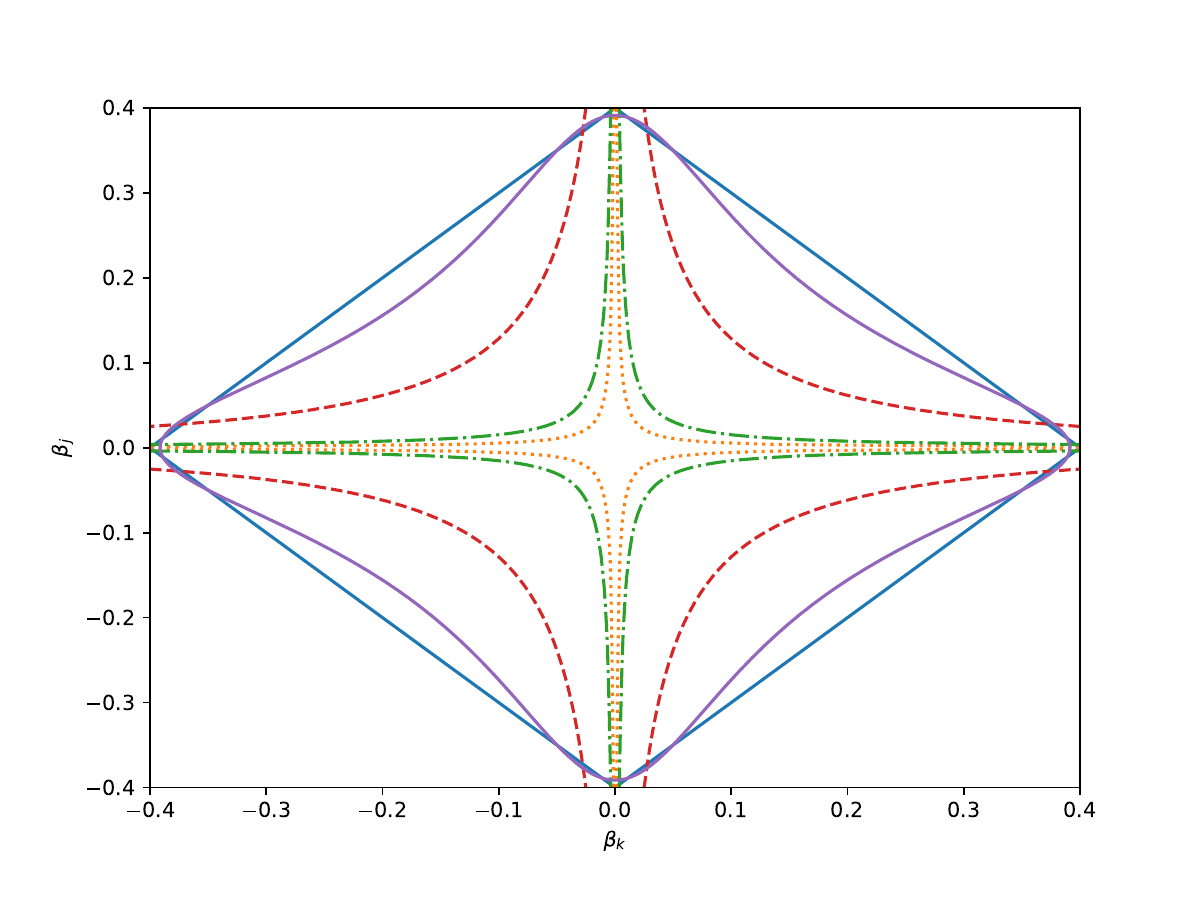}
	\caption{\note{
	The left panel shows 1d density profiles for N-GIG priors
	on $\beta_j$ considered in the present work.
	The right panel shows a contour for each example 
	on a pair of covariates $(\beta_j,\beta_k)$. The range of permissible values of $(\nu,\delta,\lambda)$, as well as some properties of the examples, are given in \cref{tab:examples}. The particular values of $(\nu,\delta,\lambda)$ in the plot are given in the legend of the left panel.}}
	\label{fig:examples}
\end{figure}

\note{The present work is focused on the case of 
Normal scale-mixtures of generalized inverse Gaussians (GIG),
which will be referred to as Normal-GIG (N-GIG) priors. In particular, 
the prior of interest is the $\beta$ marginal of the following hierarchical model
\begin{equation}\label{eq:statmod}
\bbP(\beta | \theta) = \prod_{j=1}^p N(\beta_j; 0, \theta_j) \, , \qquad 
\bbP(\theta) = \prod_{j=1}^p
\mathcal{GIG}(\theta_j ; \nu, \delta, \lambda) \, ,
\end{equation}
where the distribution on $\theta_j>0$ is given by
\begin{equation}\label{eq:gig}
\mathcal{GIG}(\theta_j ; \nu, \delta, \lambda) \propto
\theta_j^{\nu-1}\exp\left [ -\frac12 (\delta^2/\theta_j + \lambda^2 \theta_j) \right] \, .
\end{equation}}
\note{The $\mathcal{GIG}$ distribution 
is generalized above by defining it only as proportional to the right-hand side. 
One requires $\nu>0$ if $\delta=0$ and $\nu<0$ if $\lambda=0$
for a proper prior. Otherwise there are no constraints on the parameters.
This family of priors was considered before in \cite{andrieu}.
Some relevant examples which will be considered in the present work are:
BL \cite{andrews1974scale, roy2017selection}, 
Jeffrey's (Jeff) \cite{figueiredo2003adaptive}, 
Student-t (ST) \cite{tipping1999relevance},
Normal-Gamma (NG) and Normal inverse Gaussian (NIG) 
\cite{caron2008sparse, griffin2011bayesian, caron2012sparsity}.
See \cref{fig:examples} for an illustration 
and \cref{tab:examples} for a summary of some relevant properties.}

\begin{table}[h]
\centering
\note{\begin{tabular}{ |c|c|c|c|c|c|c| } 
 \hline
& BL & Jeff & ST & NG & NIG \\ 
  \hline
$(\nu,\delta,\lambda)$ 
& $(1,0,\lambda)$  & $(0,0,0)$ & $(\nu<1/2,\delta,0)$
& $(\nu,0,\lambda)$ 
& $(-1/2,\delta,\lambda)$ \\
\hline
 Singular at $\beta_j=0$ & No   & Yes & No & Yes & No \\ 
 \hline
Tail behavior & exponential & algebraic & algebraic & exponential & exponential \\ 
 \hline
\end{tabular}}
\caption{\note{N-GIG priors considered here and some relevant properties.}}
\label{tab:examples}
\end{table}

A variational Bayesian expectation maximization (VBEM) method 
\cite{ghahramani2003variational, attias2000variational} 
will be employed for approximation of the resulting posterior,
which requires only the solution of (unconstrained) linear systems, and provides approximations of the mean and covariance of the target.
Additionally, in parallel we will perform classical expectation maximization (EM) \cite{dempster1977maximum}
to obtain the maximum a posteriori (MAP) estimator. 
This approach will deliver the {\em variational Bayesian LASSO} (VBL),  
a principled Gaussian approximation to the BL,
\note{as well as variational approximation associated to the 
other N-GIG priors listed in \cref{tab:examples}.}

\note{In the Bayesian community, 
recent theoretical results have revealed that the
BL is suboptimal in both parameter estimation and variable selection \cite{castillo2015bayesian}.
However \cite{song2017nearly} have shown that many
shrinkage priors are consistent.
The necessary condition is that the tail should decay 
algebraically and not exponentially, which includes ST and Jeff 
above.
Also, it is well-known that the mean of sparsity priors, 
such as the total variation (TV) prior,
may not promote sparsity \cite{lassas2004can}.
In fact, in this work it is shown that in the limit of $p \rightarrow \infty$,
for different choices of $\lambda$ depending on $p$, 
the posterior associated to a TV prior is either a Gaussian 
or has a diverging mean.
Indeed even the BL 
point estimator \cite{park2008bayesian}, 
which returns the median instead of the mean, is not as sparse
as the standard LASSO, i.e. its MAP estimator.
Despite these shortcomings, VBL is still considered here as a prominent 
example, due to the persistent ubiquity of (B)LASSO as a model of choice
across science and engineering applications.}

\note{
VBEM is a particular instance of a more general methodology known as
{\em mean-field variational Bayes}.
See \cite{blei} for a recent review. 
The ST example
is referred to as automatic relevance detection
\cite{neal1996bayesian,mackay1996bayesian},
and variational inference 
for this model has been introduced, independently, in 
\cite{bishop2000variational} and \cite{drugowitsch2013variational}.
The work \cite{armagan2009variational} considers variational inference in the 
context of the stable mixing distribution on $\theta$ in
\cref{eq:statmod}.
The distribution cannot be explicitly represented in general, but yields
a nice family of marginal priors of the form 
$\propto \exp(-\lambda |\beta|^r_r)$ for $r\in (0,2)$,
which is referred to as Bayesian bridge regression, and includes BL.
It is important to note that such variational mean-field models 
are only an approximation of the original full Bayesian model, 
and as such the UQ delivered is not precise
-- 
see e.g. \cite{neville2014mean} for a careful study of some 
related models in this context.}

The approach presented here provides fast UQ 
in the context of sparse regression and linear inverse problems.
The key points are:
\begin{itemize}
\item Fast and inexpensive variational Bayesian solution with
{\bf sparsity} promoting priors.
The Bayesian formulation provides {\bf UQ}
in the estimate, 
which is becoming increasingly important in science and engineering applications 
\cite{smith2013uncertainty, uqhandbook}.
\item {\bf Online} variational Bayesian solution 
means arbitrarily large $n$ can be handled
with a $\cO(p^2)$ memory cost
and $\cO(p^3)$ compute cost per update, 
where $p$ is the width of the design matrix (see below). 
The method provides {\bf exact} MAP estimate (for convex prior)
and {\em variational approximation to the posterior}, 
with little more than a pair of parallel Kalman filters \cite{kalman1960new}. 
We note that the results presented are constrained to the case of regression, 
but will be extended to classification and non-trivial dynamics in future work.
\item {\bf Adaptive} online learning of hyper-parameters via EM approach 
provides improved accuracy at a marginal additional cost. 
\item A further ``low rank $+$ diagonal'' {\bf approximation} 
provides a reduction in cost to $\cO(p)$ in case $p$ is prohibitively large, 
in a similar spirit to the ensemble Kalman filter (EnKF) \cite{evensen1994sequential}.
\end{itemize}

\note{
The rest of the paper is organized as follows. 
The basic model is introduced in \cref{sec:basic}:
after introducing the basic model
in \cref{sec:bayes},
\cref{ssec:em} describes MAP estimation with the EM algorithm,
and \cref{ssec:vbem}
describes posterior approximation using VBEM,
culminating in our Gaussian approximation of sparsity-promoting 
posteriors and VBL in \cref{ssec:sparse}.
Further enhancements to the basic model are introduced in \cref{sec:enhance}. In particular, the online version 
is introduced in \cref{ssec:sparseonline}, and hyper-parameter
tuning is considered in \cref{sec:hyper}.
Numerical results are presented in \cref{sec:numerics},
including a small basic  dataset  relating to diabetes in \cref{ssec:vbld},
a comparison-study with state-of-the-art competitors
on a prototype problem used before to evaluate 
variable selection capability in \cref{ssec:varsel},
and more computationally intensive (generalized) total variation (TV) 
denoising/deblurring examples in \cref{ssec:vbltv}.}

\section{The basic model and method}\label{sec:basic} 
\note{
\subsection{The model: a Bayesian formulation of sparsity priors}
\label{sec:bayes}

The basic model considered here is given by \cref{eq:basic},
\cref{eq:statmod}, and \cref{eq:gig}, iterated compactly here for clarity
\begin{align}
Y_n | X_n, \beta &\sim N(X_n \beta, \gamma^2 I_n) \, ,  \label{eq:statmodfull}\\
\beta | \theta &\sim N(0, D(\theta)) \, , \quad 
D(\theta) := {\rm diag}(\theta_1, \dots, \theta_p) \, , \notag\\
\theta_j &\sim \mathcal{GIG}(\theta_j ; \nu, \delta, \lambda) \, , ~{\rm i.i.d.} \, ~ {\rm for}~ j=1,\dots, p \, . \notag
\end{align}
The conditionals of the joint are known exactly 
\begin{align}
\beta | \theta, X_n,Y_n &\sim \cN(\beta; m_n^\theta, C_n^\theta) \, , \label{eq:betacond}\\
\theta_j | \beta ,X_n,Y_n &\sim \mathcal{GIG}(\theta_j; \nu -1/2, \sqrt{\delta^2+\beta_j^2},\lambda) \, ,
\label{eq:thetacond}
\end{align}
where $m_n^\theta, C_n^\theta$ are given by
\cite{box2011bayesian}
\begin{align}
m_n^\theta &= \left(X_n^TX_n + \gamma^2 D(1/\theta) 
\right)^{-1} 
X_n^TY_n
\label{eq:monolith} \\
& =  
D(\theta)X_n^T \left(\gamma^2I_n + X_nD(\theta)X_n^T\right)^{-1}
Y_n
\, , \notag
\\
C_n^\theta &= \left(\frac1{\gamma^2}X_n^TX_n + D(1/\theta)
\right)^{-1} \label{eq:monolithC}\\
&= 
\left (I_p - D(\theta) X_n^T \left(\gamma^2I_n + X_nD(\theta)X_n^T\right)^{-1} X_n 
\right ) D(\theta) \, . \notag
\end{align}
By choosing the appropriate version above,
the computation cost is $\cO(pn\min\{p,n\})$, 
and the memory cost is $\cO(pn)$. 
In case $n>p$, and in particular if $n \rightarrow \infty$, 
then the Kalman filter \cite{kalman1960new} provides exact solution
online with a memory and (per iteration) computation cost of $\cO(p^2)$.
See \cref{app:enkf}.

Since the conditionals are known, 
Gibbs sampling can be used to sample exactly from the %
posterior 
\cite{andrieu, park2008bayesian}. 
Sequential Monte Carlo methods \cite{doucetsequential, moral2004feynman} 
have also been designed to sample from the 
full posterior sequentially in \cite{andrieu, caron2012sparsity}, 
and a sequential expectation maximization (EM)  \cite{dempster1977maximum} 
method has been used 
to approximate MAP estimates of $\beta | (X_n,Y_n)$ 
in \cite{caron2008sparse, caron2012sparsity}.

\begin{example}
Assume that the prior consists of $p$ independent random variables $\beta_j$,
each with Laplace distribution $\cL(\beta_j; \lambda)$, the BL model.
The MAP estimator associated to this model corresponds to L1 regularized
regression, or LASSO \cite{tibshirani1996regression}.
It is well-known that the Laplace distribution can be expressed 
as a scale mixture of GIG with parameters $(1,0,\lambda)$:
\begin{equation}\label{eq:lap}
\cL(\beta_j; \lambda) = \int_{\bbR_+} \cN(\beta_j; 0, \theta_j) \mathcal{GIG}(\theta_j ; 1,0, \lambda) d\theta_j \, .
\end{equation}
\end{example}

\begin{rem}
If \cref{eq:statmodfull} is modified as 
$\bbP(\beta | \theta) = \cN(\beta ; 0, (C_0^{-1} + D(1/\theta))^{-1} )$
and one replaces $D(1/\theta) \leftarrow C_0^{-1} + D(1/\theta)$ in 
\cref{eq:monolith} and \cref{eq:monolithC}, then 
marginally one has an elastic net prior for $\delta=0$ and $\nu=1$:
$\bbP(\beta) \propto \cN(\beta ; 0, C_0) \cL(\beta ; \lambda)$ \cite{roy2017selection}.
\end{rem}}

\subsection{MAP estimation by Expectation maximization}
\label{ssec:em}

For the next sections we suppress $X$ and $n$ in the notation where convenient.
Suppose we want to maximize
\begin{equation}\label{eq:jointlb}
\log \bbP(Y,\beta) = \log \int \bbP(Y,\beta,\theta) d\theta 
\geq \int \log \left ( \frac{\bbP(Y,\beta,\theta)}{q(\theta)} \right ) q(\theta) d\theta \, ,
\end{equation}
where the inequality arises (for {\em any} probability density $q(\theta)>0$) 
from an application of Jensen's inequality, 
and $\bbP$ here denotes a probability density.
The expectation maximization (EM) algorithm \cite{dempster1977maximum} 
proceeds as follows. 
Define $q^t(\theta) = \bbP(\theta| \beta^t, Y)$, 
\begin{equation}\label{eq:qb}
{Q}(\beta|\beta^t) = 
\int \log \left ( \frac{\bbP(Y,\beta,\theta)}{\bbP(\theta| \beta^t, Y)} \right ) 
\bbP(\theta| \beta^t, Y) d\theta \, ,
\end{equation}
and let $\beta^{t+1} = {\rm argmax}_\beta Q(\beta|\beta^t)$.

In our context this entails iteratively computing 
\begin{equation}%
Q(\beta | \beta^t) = \frac12 \beta^T D(1/\theta^{t+1}) \beta  + 
\frac1{2\gamma^2} |Y_n - X_n \beta|_2^2 + 
\kappa(\beta^t, X_n,Y_n) 
\label{eq:e}
\end{equation}
where we recall that $D(1/\theta^{t+1})$ is the diagonal matrix with $1/\theta^{t+1}_j$ on the diagonal, \note{and $\kappa(\beta^t, X_n,Y_n)$ 
is a constant depending on $\beta^t, X_n,Y_n$
but not $\beta$}.
For example, in the case of $\delta\geq 0$ and $\nu=1$,
$\theta^{t+1}$ is defined element-wise as 
\begin{equation}\label{eq:emtheta}
1/\theta^{t+1}_j :=  \bbE\left [{1}/{\theta_j}  | \beta^t, X_n,Y_n \right ] 
= {\lambda }((\beta_j^t)^2+\delta^2)^{-1/2} \, .
\end{equation}
The calculation of \cref{eq:e} is given in \cref{ap:em}
along with a slightly lengthier explanation of EM.
Note we have assumed $\nu=1$ but allowed $\delta \neq 0$ in the hyperprior \cref{eq:gig}, 
which relaxes the marginal Laplace identity \cref{eq:lap}.
The general form of \cref{eq:emtheta} is given in equation \cref{eq:ethetai},
and the case $\nu=0$ is given in \cref{eq:ethetainu1}.
For $\nu=1$, one then has the iteration
\begin{equation}\label{eq:m}
\mu^{t+1}_n = \left ( X_n^TX_n + \gamma^2 {\lambda }D\left(\left((\mu_n^t)^2+\delta^2\right)^{-1/2}\right) \right)^{-1} X_n^TY_n\, .
\end{equation}
These analytical calculations have 
been shown and used before in several works, 
including \cite{figueiredo2003adaptive, caron2008sparse, griffin2011bayesian}. 
From this, one obtains the MAP estimator at convergence 
$\mu_n^t \rightarrow \mu_n$.

\begin{rem}\label{rem:irls}
An iteratively reweighted least squares (IRLS) algorithm 
\cite{holland1977robust} can be 
derived in order to approximate regularization with $r \in (0,1]$ by a sequence of problems
with $r=2$, based on the following observation 
$$
|\beta|_r^r = \sum_{i=1}^p |\beta_i|^r = \sum_{i=1}^r |\beta_i|^{r-2} \beta_i^2 \, .
$$
The resulting iteration for $r=1$ is exactly as in \cref{eq:m}, where $\delta^2>0$
is interpreted as a regularization parameter.
It can be shown under appropriate assumptions 
that $\mu_n^t \rightarrow \mu_n$ as $t \rightarrow \infty$, 
where $\mu_n$ is sparse if such solution exists, and convergence
is linear (exponentially fast) for $\mu_n^t$ sufficiently close to $\mu_n$ \cite{daubechies2010iteratively}.
\end{rem}

\subsection{Posterior approximation}

\subsubsection{Variational Bayesian Expectation maximization}
\label{ssec:vbem}

Here we propose to use the variational Bayesian 
expectation maximization (VBEM) algorithm, introduced in the context of graphical models in 
\cite{attias2000variational, ghahramani2003variational}.
We show how it works
elegantly in our context to provide a Gaussian approximation to problems with sparsity 
priors, which is optimal in a certain sense. 
Suppose we return to \cref{eq:jointlb}, and this time multiply/divide by some density 
$q(\beta)>0$ and integrate over $\beta$ as well.
Then we have the evidence lower bound
\begin{equation}\label{eq:elbo}
\log \bbP(Y) = \log \int \bbP(Y,\beta,\theta) d\theta d\beta
\geq \int \log \left ( \frac{\bbP(Y,\beta,\theta)}{q(\theta)q(\beta)} \right ) q(\theta)q(\beta) d\theta d\beta\, .
\end{equation}
Coincidentally, maximizing this with respect to the {\em densities} $q(\theta)q(\beta)$
coincides with minimizing the KL divergence between 
this variational approximation and the joint posterior, i.e. 
\begin{align}\nonumber
\log \bbP(Y) - \int \log \left ( \frac{\bbP(Y,\beta,\theta)}{q(\theta)q(\beta)} \right ) q(\theta)q(\beta) d\theta d\beta
&= - \int \log \left ( \frac{\bbP(\beta,\theta|Y)}{q(\theta)q(\beta)} \right ) q(\theta)q(\beta) d\theta d\beta \notag \\
&=: KL\left [q(\theta)q(\beta) || \bbP(\beta,\theta|Y)\right] \, .\label{eq:kld}
\end{align}
The objective functions for each of $q(\theta)$ and $q(\beta)$ given the other are convex
and can be minimized exactly,
as observed in \cite{attias2000variational, ghahramani2003variational}, 
leading to the iterative algorithm
\begin{align}
q^{t+1}(\theta) &\propto \exp\left (\int \log {\bbP(Y,\beta,\theta)} q^{t}(\beta) d\beta\right) \, , \notag \\
q^{t+1}(\beta) &\propto \exp\left (\int \log {\bbP(Y,\beta,\theta)} q^{t+1}(\theta) d\theta\right) \, . \label{eq:coordinatedescent}
\end{align}
Furthermore, following from convexity of the intermediate targets 
this gives a descent direction for \cref{eq:kld}
$KL\left [q^{t+1}(\theta)q^{t+1}(\beta) || \bbP(\beta,\theta|Y)\right] \leq 
KL\left [q^{t}(\theta)q^{t}(\beta) || \bbP(\beta,\theta|Y)\right]$.
Observe that constraining to $q^{t+1}(\beta) = \delta_{\beta^{t+1}}(\beta)$, 
where $\delta(\cdot)$ is the Dirac delta function and $\beta^{t+1}$ is the point of maximum probability above, yields the original EM algorithm. 
Also, observe that \cref{eq:coordinatedescent} may itself be intractable
in general, although it is shown in \cite{ghahramani2003variational}
that it is simplified somewhat for conjugate exponential models and may be analytically soluble.
Fortunately, the present situation is the best case, where it is analytically soluble.
Notice that the objective function \cref{eq:kld} corresponds to an independence assumption
between $\theta$ and $\beta$, however from \cref{eq:coordinatedescent}
it is clear that the solution solves a coupled system, 
and in fact {\em probabilistic dependence is replaced with a 
deterministic dependence on each others' summary statistics}, 
as noted in \cite{ghahramani2003variational}.

\begin{figure}[!htbp]
	\centering\includegraphics[width=0.49\textwidth]{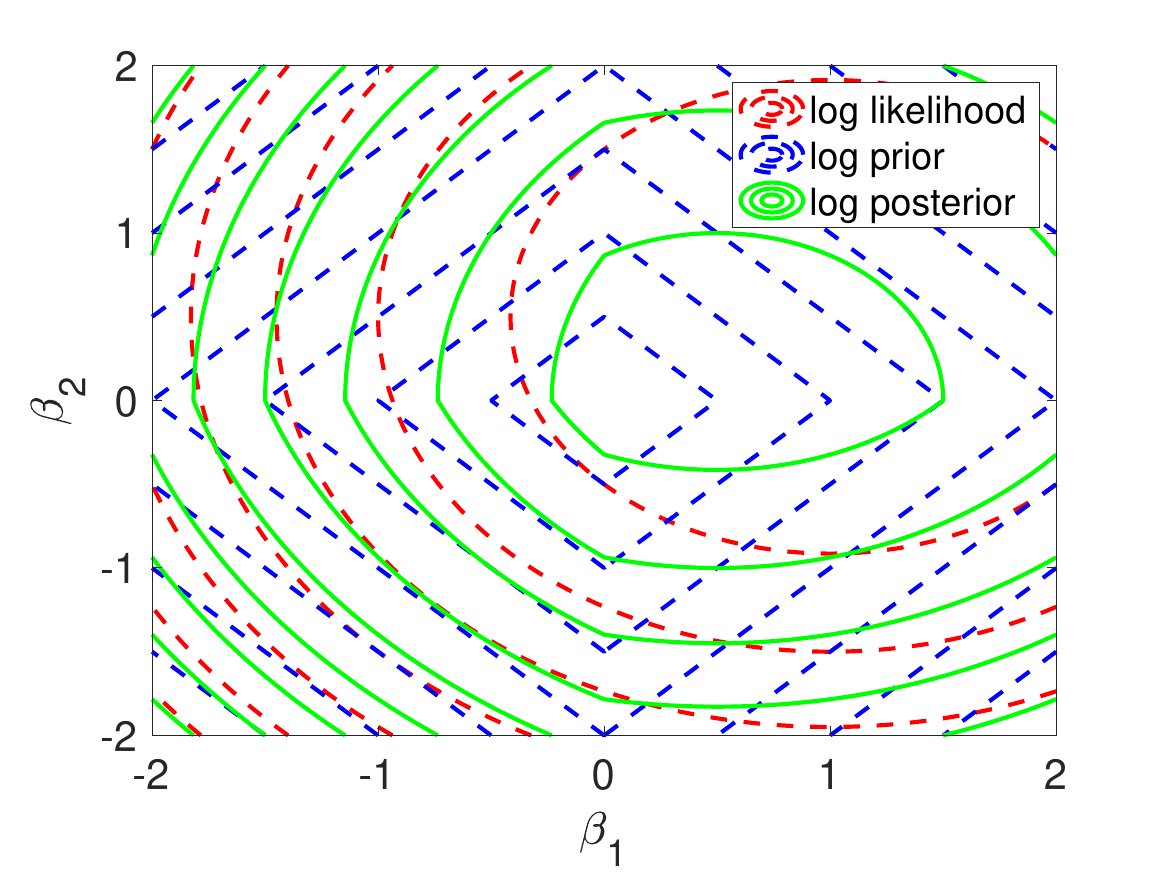}
	\includegraphics[width=0.49\textwidth]{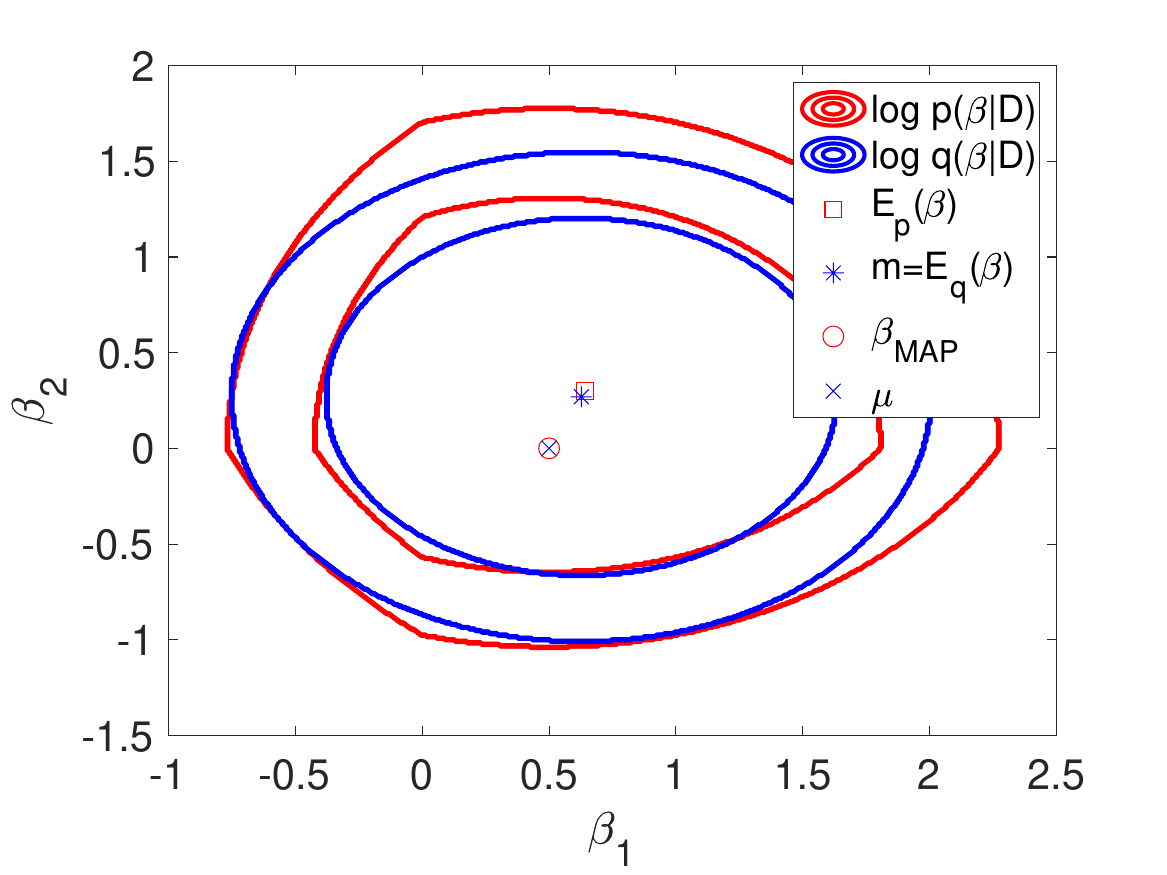}
	\centering\includegraphics[width=0.49\textwidth]{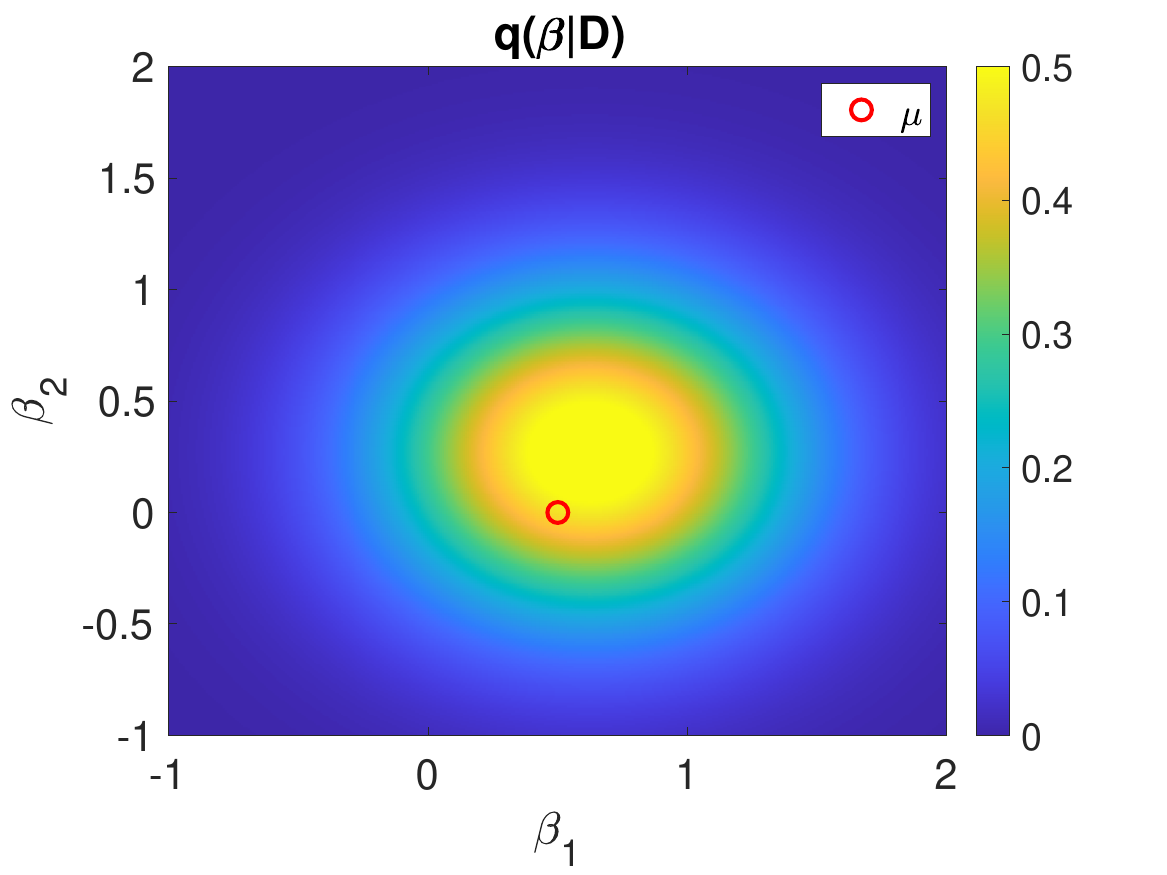}
	\includegraphics[width=0.49\textwidth]{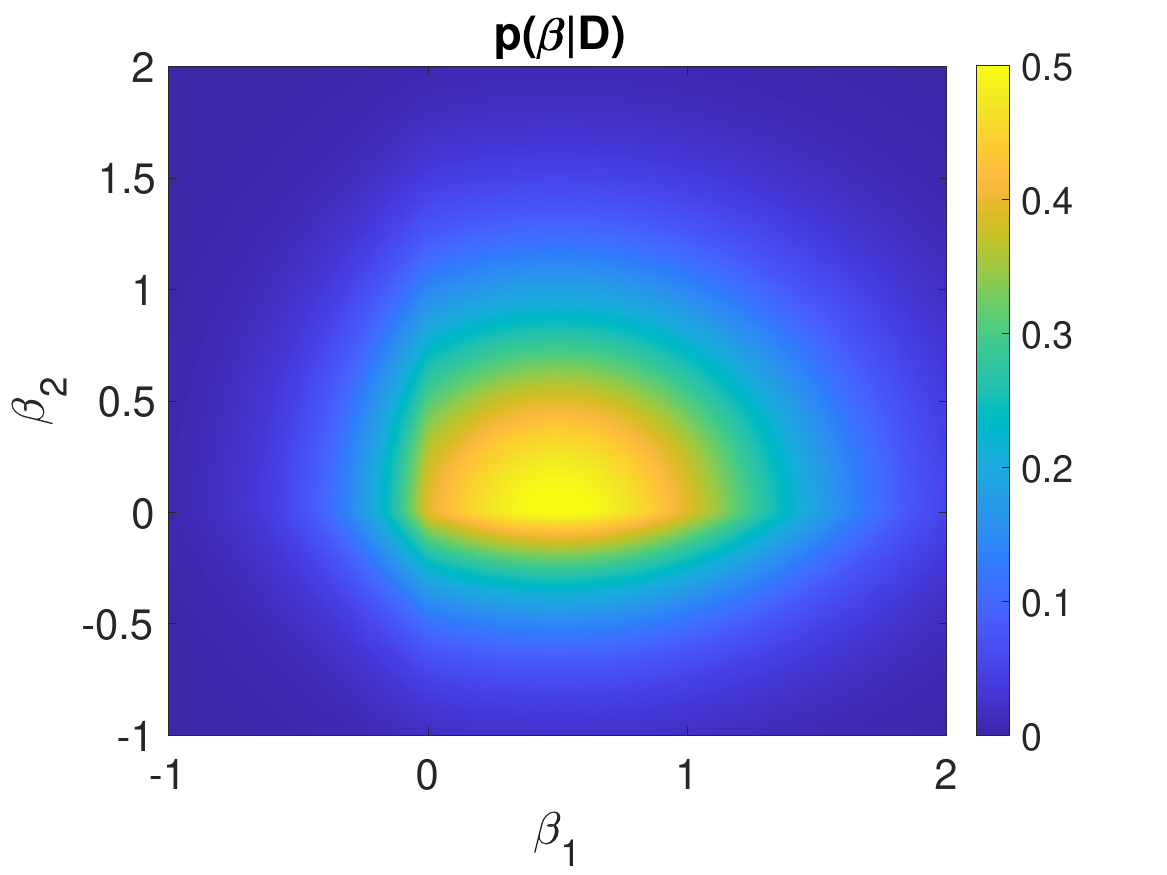}
	\caption{Illustration of the variational Bayesian LASSO (VBL).
	The top left figure shows the contours of the prior, the likelihood, and the posterior.
	The top right figure shows the $\alpha=0.8$ and $\alpha=0.95$ HPD credible 
	contours of the posterior $\bbP(\beta|\cD)$
	and the variational approximation $q(\beta| \cD)$, 
	along with means and MAPs. %
	For this example, 
	the posterior median is very close to the mean, but slightly towards the MAP. 
	The bottom row shows the density of VBL (left, with MAP estimate $\mu$) 
	and the posterior (right).}
	\label{fig:vbem}
\end{figure}

\subsubsection{Gaussian approximation to a sparsity promoting posterior} %
\label{ssec:sparse}

We approximate the model in \cref{eq:statmodfull} using
the variational Bayesian approach of \cref{ssec:vbem}.
Equations \cref{eq:coordinatedescent} are given by
\begin{align}
q^{t+1}(\theta) &\propto \exp\left (-\frac12 \sum_{j=1}^p \bbE_{t}[\beta_j^2]/\theta_j \right) 
\bbP(\theta|\lambda) \, , \label{eq:vbem} \\
q^{t+1}(\beta) &\propto \exp\left (- \frac12 \sum_{j=1}^p \beta_j^2 \bbE_{t+1}[1/\theta_j] -
 \frac1{2\gamma^2} |Y_n - X_n \beta|_2^2 \right)  \, , \notag
\nonumber
\end{align}
where $\bbE_t$ is used to (degenerately) denote expectation with respect to
the iteration $t$ intermediate variational distribution, 
with respect to its argument, $\beta$ or $\theta$.
This is referred to as %
{\em coordinate ascent variational inference} \cite{blei}. 
The first equation looks similar to the EM algorithm, 
however with the important difference
\begin{equation}\label{eq:vbthetaprob}
q^{t+1}(\theta_j) = \mathcal{GIG}\left(\theta_j; \nu -1/2, 
\sqrt{\delta^2+C_{n,jj}^t + (m_{n,j}^t)^2},\lambda\right) \, ,
\end{equation}
where $(m_n^t, C_n^t)$ are the mean and covariance of 
$q^t(\beta) = N(\beta; m_n^t, C_n^t)$
(note the appearance of the variance $C_{n,jj}^t$ instead of just $(m_{n,j}^t)^2$).

This means that for the case $\nu=1$ we have
\begin{equation}\label{eq:vbtheta}
1/\theta^{t+1}_{n,j} := \bbE_{t+1} \left [ {1}/{\theta_j} \right ] 
= {\lambda}(C_{n,jj}^t+ (m_{n,j}^t)^2+\delta^2)^{-1/2} \, .
\end{equation}
The update equations are given
by
\begin{align}
m_n^{t+1} &= %
C_n^{t+1} \left (\frac1{\gamma^2}X_n^TY_n \right) \, , \label{eq:mtheta}\\
C_n^{t+1} &= %
\left ( \frac1{\gamma^2} X_n^TX_n + 
D(1/\theta^{t+1}_n) \right)^{-1} \, . \label{eq:ctheta}
\end{align}
Here we can explicitly observe the deterministic dependence between the marginally optimal
$\theta$ and $\beta$ distributions via each others' summary statistics.
The general form of the update \cref{eq:vbtheta} is given in \cref{eq:ethetai},
and the case $\nu=0$ is given in \cref{eq:ethetainu1}.

Note that this algorithm runs for approximately the same cost as 
the former, and provides a Gaussian approximation $(m_n^*,C_n^*)$ of the 
posterior. 
The former provides an approximation of the MAP estimator $\mu_n^*$. 
In the case $\nu=1$ we refer to this triple $(\mu_n^*,m_n^*,C_n^*)$ 
as the variational Bayesian LASSO (VBL). %
This is summarized in \cref{alg:monovbl}. 
In the context of UQ, %
one may consider the sparse solution to be reasonable if, 
for any index $j$ such that $\mu_{n,j}=0$, 
one has $0 \in (m_{n,j} - 2\sqrt{C_{n,jj}}, m_{n,j} + 2\sqrt{C_{n,jj}})$, i.e.
the origin is within the credible interval of the variational
Bayesian marginal for those coordinates which are predicted to vanish. 
In general, one may flag as unusual any circumstances where 
$\mu_{n,j} \notin (m_{n,j} - 2\sqrt{C_{n,jj}}, m_{n,j} + 2\sqrt{C_{n,jj}})$.
See \cref{fig:vbem} for an illustration of the VBL 
applied to a simple example with $p=n=2$ 
(``exact'' values are calculated with numerical quadrature on the domain 
$[-4,4]^2$ with $1000$ grid points in each direction).
\note{The $\alpha$ highest posterior density (HPD) credible contour
of a density $\bbP$ is defined by $\{\beta ; \bbP(\beta) = c(\alpha) \}$,
where $c(\alpha)$ solves %
$$\int_{\{\beta;\bbP(\beta) \geq c(\alpha) \}} \bbP(\beta)d\beta  = \alpha \, .$$
The top right panel shows the $\alpha$ HPD credible contours of $\bbP(\beta|\cD_n)$
and $q(\beta|\cD_n)$ associated to $\alpha=0.8$ and $\alpha=0.95$.}

\begin{algorithm}
\textbf{Input}: Design matrix $X$, labels $Y$, 
parameters $\gamma, \lambda, \delta, \nu$, 
initial guess $(\mu^0, m^0, C^0) \in \bbR^{2p+p^2}$, 
convergence criteria $\epsilon, T>0$ and 
distance function $d: \bbR^{(2p+p^2) \times (2p+p^2)} \rightarrow \bbR_+$.
\begin{enumerate}
\item Specify functional forms $\theta^{t+1}_{\rm VBEM}(\beta)$ and 
$\theta^{t+1}_{\rm EM}(\beta)$ based on $\delta, \nu$, as given in \cref{eq:ethetai}.
\item Set $t=0$, and $(\mu^{-1}, m^{-1}, C^{-1}) = {\bf 0}_{2p+p^2}$ (all zeros).
\item \textbf{While} $t \leq T$ \textbf{and} 
$d((\mu^t, m^t, C^t),(\mu^{t-1}, m^{t-1}, C^{t-1}))>\epsilon$;
\begin{enumerate} 
\item Compute $\theta^{t+1}_{\rm VBEM}$%
and $\theta^{t+1}_{\rm EM}$ %
(arguments suppressed);
\item Compute 
\begin{eqnarray}\label{eq:vblgains}
G^{t+1}_{\rm VBEM} &=& 
D(\theta^{t+1}_{\rm VBEM})X^T(X D(\theta^{t+1}_{\rm VBEM}) X^T + \gamma^2 I_n)^{-1} \, , \\
G^{t+1}_{\rm EM} &=& 
D(\theta^{t+1}_{\rm EM})X^T(X D(\theta^{t+1}_{\rm EM}) X^T + \gamma^2 I_n)^{-1} \, . 
\end{eqnarray}
\item Compute 
\begin{eqnarray}\label{eq:vblupdates}
m^{t+1} &=& G^{t+1}_{\rm VBEM} Y \, ,\\
C^{t+1} &=& (I-G^{t+1}_{\rm VBEM}X)D(\theta^{t+1}_{\rm VBEM}) \, , \\
\mu^{t+1} &=& G^{t+1}_{\rm EM} Y \, .
\end{eqnarray}
\item t=t+1.
\end{enumerate}
\end{enumerate}
\textbf{Output}: $(\mu^*, m^*, C^*) \in \bbR^{2p+p^2}$.
\caption{Variational Bayesian N-GIG.} 
\label{alg:monovbl}
\end{algorithm}

\begin{rem} Step (3) of \cref{alg:monovbl}
requires a stopping criterion. 
\note{A good, if somewhat cumbersome, option 
is $d((\mu^t, m^t, C^t),(\mu^{t-1}, m^{t-1}, C^{t-1})) = 
|{\sf ELBO}_{t}-{\sf ELBO}_{t-1}|/|{\sf ELBO}_{t}|$,
where ${\sf ELBO}_t$ is the
lower bound appearing on the right-hand side of \cref{eq:elbo}, 
which can be computed in closed form.
Another simpler option is %
$d((\mu^t, m^t, C^t),(\mu^{t-1}, m^{t-1}, C^{t-1})) = \|X m^t - Y\|/\gamma$.}
We will see that the algorithm returns a good estimator very quickly, 
but may take a long time to converge, 
and may even return a worse estimator at convergence 
Therefore, a maximum number of iterations is often
also employed as an alternative in practice.
\end{rem}

\section{Enhanced model}
\label{sec:enhance}

This section focuses on enhancing the model by enabling 
sequential/online inference and hyper-parameter optimization.
\note{The static version, with fixed $n$, 
given in \cref{eq:monolith} and \cref{eq:monolithC} 
will henceforth be referred to as {\em monolithic}
so that the distinction is clear.}

\subsection{Online Gaussian approximation to
a sparsity
posterior}
\label{ssec:sparseonline}

In the following we focus the description on $(m_n,C_n)$ of the VBEM formulation 
\cref{eq:mtheta}, \cref{eq:ctheta}, but note that analogous equations hold for 
$\mu_n$ \cref{eq:m}. There are two distinct scenarios to consider here. 
First we will consider 
the case of moderate $p$ and $n \gg p$,
where the online method reproduces EM and VBEM exactly at a cost of $\cO(p^3)$
per iteration.
The second case we will consider is that of very large $p$, and possibly $n \leq p$, 
where it is necessary to impose an approximation in order to control the cost by $\cO(p)$.
Both approaches are amenable to online implementation, i.e. $n \rightarrow \infty$.

\subsubsection{Small/moderate $p$ exact method}

Suppose we are assimilating batches of size $M$, and denote 
$$
\bar{X}_n=X_{nM}\, , \quad \bar{Y}_n=Y_{nM} \, , \quad 
\tilde{X}_n = [x_{(n-1)M+1},\dots,x_{nM}]^T \, , \quad
\tilde{Y}_n = [y_{(n-1)M+1},\dots,y_{nM}]^T \, ,
$$
so that e.g. $\tilde{X}_n^T\tilde{X}_n = \sum_{i=(n-1)M+1}^{nM} x_ix_i^T$.
Sequential batches are presented but this may not always be a sensible choice 
and some permutation of the indices may make sense. See \cref{rem:batch}.
We can compute batch updates of the required matrices in \cref{eq:monolith}
exactly with a total of at most $\cO(p^2 M)$ operations:
$$
\bar{X}_n^T\bar{X}_n = \bar{X}_{n-1}^T\bar{X}_{n-1} + \tilde{X}_n^T\tilde{X}_n \, , \quad
\bar{X}_n^T \bar{Y}_n = \bar{X}_{n-1}^T\bar{Y}_{n-1} + \tilde{X}_n^T \tilde{Y}_n \, , \quad %
\bar{Y}_n^T \bar{Y}_n = \bar{Y}_{n-1}^T\bar{Y}_{n-1} + \tilde{Y}_n^T \tilde{Y}_n \, .$$ 
The cost may be smaller if the intermediate quantities are sparse (many zeros) or low-rank.
We have the following equation for the precision
$(C_n^t)^{-1} = \frac1{\gamma^2} \bar{X}_n^T\bar{X}_n + D(1/\theta_n^t)$
and can proceed directly with iterating \cref{eq:mtheta} and \cref{eq:ctheta}.
In the worst case scenario, the inversion required to compute \cref{eq:mtheta}
will incur a cost of $\cO(p^3)$, so one would aim to take $M=p$. 
We iterate that the focus here is the case $n \gg p$.
For large $p$, which will be discussed now, 
$\bar{X}_n^T\bar{X}_n$ must be sparse or low-rank. 
In this case, inversion can be done approximately with a cost of %
as little as $\cO(p)$.
In case $\cO(p^2)$ is prohibitive for computation or memory, 
then online computation and storage of the component matrices 
must be controlled as well. This is discussed further in the following section. 

\subsubsection{Large $p$ approximate method}

In the case of large $p$ and/or $n\leq p$, the problem is different.
It is preferable to directly confront the monolithic problem \cref{eq:monolith} if it is possible,
for example in case that $n<\infty$ is fixed and $X_n$ is sufficiently sparse and/or low-rank to allow the direct 
use of an iterative Krylov-type solver \cite{hestenes1952methods, saad1981krylov}.
On the other hand, if this cannot be handled directly, then a sequential/online
strategy can be adopted as follows.

\note{It is first instructive to observe the following recursive formulation
of \cref{eq:monolith} 
\begin{align}
m_n =&  \left(\frac1{\gamma^2}%
\bar{X}_{n}^T\bar{X}_{n} + D(1/\theta) \right)^{-1}
\left(\frac1{\gamma^2}(\tilde{X}_n^T\tilde{Y}_n + \bar{X}_{n-1}^T \bar{X}_{n-1} m_{n-1}) + D(1/\theta) m_{n-1}\right) \notag \\
&= m_{n-1} + D(\theta) \bar{X}_n^T \left(\gamma^2I_n + \bar{X}_n D(\theta)\bar{X}_n^T \right)^{-1}
\left(\widehat{Y}_n - \bar{X}_n m_{n-1}\right) \,  , \label{eq:KFR}
\end{align}
where $\widehat{Y}_n := ((\bar{X}_{n-1} m_{n-1})^T, \tilde{Y}_n)^T$.
This observation is obviously not useful by itself, 
as it incurs a cost of $\cO(\min\{p,n\}^3)$ per iteration, 
whereas the Kalman filter delivers $\cO(p^2)$ 
updates in primal/covariance form (see \eqref{eq:KF}).
However, the precision $X_{n}^TX_{n} + D(1/\theta)$ 
is required for our VBEM method 
in order to update $\theta$.
The
representation above facilitates 
a recursive ``rank $M$ + diagonal'' approximation
in similar spirit to the ensemble Kalman filter \cite{burgers1998analysis},
which allows us to 
effectively pass information forward in an online fashion.}

In particular, suppose we have 
$\widehat{X}_{n-1} \in \bbR^{M\times p}$ s.t. 
$\bar X_{n-1}^T\bar X_{n-1} \approx \widehat{X}_{n-1}^T\widehat{X}_{n-1}$,
so that 
$$
(C_{n-1}^*)^{-1} \approx \frac1{\gamma^2}\widehat{X}_{n-1}^T\widehat{X}_{n-1} + D(1/\theta^*_{n-1})\, .
$$ 
Now, recall equation \cref{eq:KFR} and define 
$$
\widehat{Y}_n := \begin{pmatrix} 
\widehat{X}_{n-1} m_{n-1} \\
\tilde{Y}_n 
\end{pmatrix}
\in \bbR^{2M} \, , \quad
\mathsf{X}_n := 
\begin{pmatrix}
\widehat{X}_{n-1} \\
\tilde{X}_n 
\end{pmatrix}
\in \bbR^{2M \times p} \, .
$$
The $\cO(2Mp)$ update which replaces \cref{eq:mtheta} and \cref{eq:ctheta}
is given by 
\begin{eqnarray}\label{eq:onlineapp}
m_n^{t+1} &=& m_{n-1} + D(\theta^{t+1}_{n}) \mathsf{X}_n^T \left(\gamma^2I_{2M} + 
\mathsf{X}_n D(\theta^{t+1}_{n}) \mathsf{X}_n^T\right)^{-1}\left(\widehat{Y}_n - \mathsf{X}_n m_{n-1}\right) \,  , \\
C_n^{t+1} &=& \left (I_p - D(\theta^{t+1}_{n}) \mathsf{X}_n^T \left(\gamma^2I_{2M} + 
\mathsf{X}_n D(\theta^{t+1}_{n}) \mathsf{X}_n^T\right)^{-1}\mathsf{X}_n \right)D(\theta^{t+1}_{n}) \, .
 \,
 \label{eq:onlineappC}
\end{eqnarray}
Finally we need $\widehat{X}_{n} \in \bbR^{M\times p}$ s.t. 
$\widehat{X}_{n}^T\widehat{X}_{n} \approx \mathsf{X}_n^T\mathsf{X}_n$
in order to proceed to the next iteration.
This is achieved by (i) computing a reduced rank-M eigendecomposition
$\mathsf{X}_n \mathsf{X}_n^T \approx U \Sigma^2 U^T$, 
with $\Sigma \in \bbR^{M\times M}$ diagonal and $U \in \bbR^{2M \times M}$
orthogonal, and (ii) defining $\widehat{X}_{n} := U^T \mathsf{X}_n$. 
This approximation in principle costs $\cO(4M^2(p+2M))$ but with 
a memory cost of only $\cO(Mp)$. All the steps above are summarized in \cref{alg:onlinevbl}. 
It is clear that in terms of cost one wants to choose $M$ small,
but in terms of accuracy one wants to choose $M$ large, 
so these considerations should be balanced.

\begin{rem}[EnKF]
We note the similarity between the above procedure and a (square-root)
EnKF
\cite{law2015data} for solution of \eqref{eq:KF} and \eqref{eq:KFC},
which proceeds with a low-rank (or low-rank plus diagonal) approximation of the covariance $C_n$.
In our scenario, the above framework is more natural and is expected to provide a better 
approximation. 
It may also be useful 
for quadratic/Tikhonov regularization where $\theta$ is held constant,
as an ``EnKF for regression'', since it delivers a natural 
non-degenerate covariance approximation.
This is described further in \cref{app:enkf}.
\end{rem}

\begin{rem}[Batching strategy] \label{rem:batch}
In the offline scenario where the data size $n>0$ is fixed
and the sequential method is employed then choice of batches is important.
If the inputs are i.i.d. random samples then sequential batching is sensible, 
i.e. $(1,2,\dots,M)$, $(M+1,\dots, 2M)$, etc.
Otherwise if there is structure in the inputs (e.g. in the context of inverse problems)
then it makes more sense to use
random sampling without replacement or 
evenly spread out batches $(1, 1+b, 1+2b, \dots, n)$, $(2, 2+b, 2+2b, \dots, n)$, etc.,
where $b=n/M$.
\end{rem}

\begin{algorithm}
\textbf{Input}: Design matrix $X$, labels $Y$ 
(possibly infinite and arriving online), 
parameters $\gamma, \lambda, \delta, \nu$, 
initial guess $(\mu^0, m^0, C^0) \in \bbR^{2p+p^2}$, 
inner convergence criteria $\epsilon, T>0$, 
distance function $d: \bbR^{(2p+p^2) \times (2p+p^2)} \rightarrow \bbR_+$, 
batch size $M$ and rule for batching (see \cref{rem:batch}).
\begin{enumerate}
\item Set $n=1$, $\widehat{X}_1=X_1$. 
{\bf Do} \cref{alg:monovbl}, and output 
$(\mu^*_{1}, m^*_{1}, C^*_{1})$ and functional forms
$\theta^{t+1}_{\rm VBEM}(\beta)$ and 
$\theta^{t+1}_{\rm EM}(\beta)$, as given in \cref{eq:ethetai}.
\item For $n=2,\dots$
\begin{enumerate}
\item Set $t=0$, 
$(\mu^0_n, m^0_n, C^0_n) = (\mu^*_{n-1}, m^*_{n-1}, C^*_{n-1})$,  
$(\mu^{-1}_n, m^{-1}_n, C^{-1}_n) = {\bf 0}_{2p+p^2}$,
$$
\widehat{Y}_n^{\rm VBEM} = 
\begin{pmatrix} 
\widehat{X}_{n-1} m_{n-1}^* \\
\tilde{Y}_n 
\end{pmatrix}
\, , \quad
\widehat{Y}_n^{\rm EM} = 
\begin{pmatrix} 
\widehat{X}_{n-1} \mu_{n-1}^* \\
\tilde{Y}_n 
\end{pmatrix} \, , \quad
\mathsf{X}_n =
\begin{pmatrix} 
\widehat{X}_{n-1} \\
\tilde{X}_n
\end{pmatrix} \, .
$$
\item \textbf{While} $t \leq T$ \textbf{and} 
$d((\mu^t_n, m^t_n, C^t_n),(\mu^{t-1}_n, m^{t-1}_n, C^{t-1}_n))>\epsilon$;
\begin{enumerate} 
\item Compute $\theta^{t+1}_{\rm VBEM}$ %
and $\theta^{t+1}_{\rm EM}$ %
(arguments suppressed);
\item Compute 
\begin{eqnarray}\label{eq:onvblgains}
\mathsf{G}^{t+1}_{\rm VBEM} &=& 
D(\theta^{t+1}_{\rm VBEM})\mathsf{X}_n^T(\mathsf{X}_n 
D(\theta^{t+1}_{\rm VBEM}) \mathsf{X}_n^T + 
\gamma^2 I_{2M})^{-1} \, , \\
\mathsf{G}^{t+1}_{\rm EM} &=& 
D(\theta^{t+1}_{\rm EM})\mathsf{X}_n^T(\mathsf{X}_n 
D(\theta^{t+1}_{\rm EM}) \mathsf{X}_n^T + 
\gamma^2  I_{2M})^{-1} \, . 
\end{eqnarray}
\item Compute 
\begin{eqnarray}\label{eq:onvblupdates}
m^{t+1}_n &=& m_{n-1}^* + \mathsf{G}^{t+1}_{\rm VBEM} 
(\widehat{Y}_n^{\rm VBEM} - \mathsf{X}_n m_{n-1}^*) \, ,\\
C^{t+1}_n &=& (I_{p}-\mathsf{G}^{t+1}_{\rm VBEM}\mathsf{X}_n)D(\theta^{t+1}_{\rm VBEM}) \, , \\
\mu^{t+1} &=& \mu_{n-1}^* + \mathsf{G}^{t+1}_{\rm EM} 
(\widehat{Y}_n^{\rm EM} - \mathsf{X}_n \mu_{n-1}^*) \, .
\end{eqnarray}
\item t=t+1.
\end{enumerate}
\item Compute rank $M$ approximation 
$U\Sigma^{2}U^T \approx\mathsf{X}_n \mathsf{X}_n^T$,
and set $\widehat{X}_n := U^T\mathsf{X}_n$.
\end{enumerate}
\end{enumerate}
\textbf{Output}: $(\mu^*_n, m^*_n, C^*_n) \in \bbR^{2p+p^2}$, at any time $n$ (or rank $M$ version of the latter).
\caption{Online Approximate Variational Bayesian N-GIG (large $p$).} 
\label{alg:onlinevbl}
\end{algorithm}

\subsubsection{Further discussion}

Note that in practice one would hope that after some iterations 
$(\mu_n,m_n,C_n)$ will not be changing very much with the iterative re-weighting,
and few inner updates will be required, if any.  If the model is stationary, then 
one may also not need to allow $n \rightarrow \infty$.
There are some other modifications which could be made along the way to 
further improve efficiency,
such as thresholding, i.e. $m_n \rightarrow {\bf 1}_{\{ \epsilon < |m_n|\}} m_n$, 
for small $\epsilon>0$, where ${\bf 1}_{A}$ is the indicator function on the set $A$, 
and it acts elementwise on the entries of $m_n$ (and similar for $\mu_n)$. 
Suppose that $m_n$ has essentially converged, and $p' \ll p$
parameters are non-zero. We can then discard the $0$-valued parameters, thereby
either vastly speeding up the algorithm or making way for inclusion of $p-p'$ new parameters.
Similar things have been done before. 
See e.g. \cite{yen2014sparse} \note{and \cite{armagan2009variational}.
In the latter article it is noted that this also mitigates a problem with singularity,
which is not an issue here because we use the dual formulation of the problem --
see \eqref{eq:vblgains}, \eqref{eq:vblupdates}.
}

All of the present technology is well-suited to an online scenario,
where one assumes a fixed static problem but data arrives sequentially in time, 
and may continue indefinitely (i.e. $n \rightarrow \infty$). 
If this model is meant to emulate a computer simulation,
for example which is called by another computer simulation for a particular value of inputs,
as in the common case of coupled multi-physics systems and whole device modelling,
then one can decide whether the emulator is suitable for a given query input, 
for example by evaluating the uncertainty under the current approximation
$x(s)^T C_n x(s)$. If this is below a certain level then the model returns 
$m_n^Tx(s)$ (and possibly also $x(s)^T C_n x(s)$ if the requesting program is capable of handling uncertainty),
otherwise the emulator requests a label from the computer simulation and is updated accordingly,
as above.
It may also be of interest in an offline scenario to build up a database of labeled data 
and revise the emulator as this is done. 
Such a task is called experimental design, 
and greedy or myopic sequential experimental design
can be posed elegantly within the sequential framework above.

\subsection{Learning hyperparameters}
\label{sec:hyper}

\note{Here we define the vector of parameters 
$\phi=(\gamma^2,\nu,\delta^2,\lambda^2)$ to be optimized.
Some of these may be fixed, but this provides a general framework.
In particular, when some parameters are fixed the objective function
may be convex or even analytically soluble. Nonetheless,
we will consider global optimization for a 4 parameter objective function
a solved problem and present the general method.}

For parameter estimation, 
we introduce an {\em iterated nested VBEM algorithm}, 
which works as follows.
For each $\tau=1,\dots$ the inner VBEM algorithm is as in 
\cref{eq:vbem} and \cref{alg:monovbl}, and yields 
$$
q^{t,\tau}(\beta| \phi^\tau) \rightarrow 
q^{*,\tau}(\beta| \phi^\tau) = 
N(m_{n}^{*,\tau},C_n^{*,\tau}) \, .$$ 
This is then %
used in an outer standard EM on $\phi$ to find
$$
\phi^{\tau+1} = 
{\rm argmax}_{\phi} \int \log q^{*,\tau}(\beta, Y | \phi)  
q^{*,\tau}(\beta | Y , \phi^\tau )d\beta \, .
$$
The details of how this is done 
will be described in detail in \cref{ssec:hyper1st} below.
The procedure is iterated until convergence. 
A more computationally efficient variant on this is given by executing %
single steps of the outer and inner algorithm iteratively 
(hence only one index is needed):
\begin{equation}\label{eq:third}
\phi^{t+1} = 
{\rm argmax}_{\phi} \int \log q^{t}(\beta, Y | \phi)  
q^{t}(\beta | Y , \phi^t )d\beta \, .
\end{equation}

\note{One can alternatively augment the variational distribution with an additional factor 
$q(\phi)$, 
which is learned in an additional step after \cref{eq:vbem}.
This approach seems somewhat more elegant but it turns out to be messy
for our model, and for that reason is not considered further.}

Note that all these approaches discussed above can be easily incorporated into \cref{alg:monovbl} and \cref{alg:onlinevbl} as optional steps 
 that change values of $\phi$ %
 in between the consecutive iterations of the main algorithm. 

\subsubsection{Some preliminaries}
\label{ssec:hyperprelim}

Before describing the method, it will be useful to recall some basic results relating to 
$$\bbP(\theta| \beta, \phi) = 
\mathcal{GIG}(\theta_j; \nu -1/2, \sqrt{\delta^2+\beta_j^2},\lambda) \, ,$$ 
which can be found for instance in \cite{abramowitz1948handbook}.

The general calculation for $\mathcal{GIG}$ distributions 
(with $\delta^2,\lambda^2 \neq 0$) which we need is
\begin{equation}
\bbE(\theta_j^{-1} | \beta_j, \phi) = \frac{\lambda}{\sqrt{\delta^2 + \beta_j^2}} 
\frac{K_{\nu + \frac12}\left (\lambda\sqrt{\delta^2 + \beta_j^2}\right)}
{K_{\nu - \frac12}\left (\lambda\sqrt{\delta^2 + \beta_j^2}\right)}
+ \frac{1-2\nu}{\delta^2 + \beta_j^2} \, ,
\label{eq:ethetai}
\end{equation}
where $K_\alpha(z)$ denotes the modified Bessel function of the second kind.
Note $\theta_n^{t+1} := 1/\bbE_t(\theta_{n,j}^{-1})$ as in \cref{eq:vbtheta} and \cref{eq:emtheta}.	
Important special cases with analytically tractable expressions include $\nu=1$ and 
$\nu=0$, in which case 
\begin{equation}\label{eq:bessrat}
\frac{K_{\frac32}(z)}{K_{\frac12}(z)} =  \frac{z+1}{z} \, , \qquad  
\frac{K_{\frac12}(z)}{K_{-\frac12}(z)} =  1 \, .
\end{equation}
Evaluating the expressions above at $z=\lambda\sqrt{\delta^2+\beta_j^2}$ 
gives 
\begin{equation}
\bbE(\theta_j^{-1}|\beta_j, \nu=1) = \frac{\lambda}{\sqrt{\delta^2 + \beta_j^2}} \, , \qquad
\bbE(\theta_j^{-1} |\beta_j, \nu=0) = \frac{\lambda}{\sqrt{\delta^2 + \beta_j^2}} 
+ \frac{1}{\delta^2 + \beta_j^2} \, \, .
\label{eq:ethetainu1}
\end{equation}

\note{If $\lambda=0$, as in the ST and Jeff cases, 
then the ratio in the first term of \cref{eq:ethetai} and \cref{eq:ethetainu1} is not defined. 
The resulting calculation shows that the first term vanishes,
and so we require $\nu < 1/2$.} 

\subsubsection{Detailed approach}
\label{ssec:hyper1st}

As before, $X$ and $n$ will be suppressed where not needed.
Assume we run the algorithm of \cref{ssec:sparse} for a fixed value of the hyper-parameters
$\phi^\tau$, resulting in a joint variational distribution 
$$q^{*,\tau}(\beta,Y|\phi) = 
\bbP(Y|\beta,\gamma^2)  N(\beta ; 0, D (1/\theta^{*,\tau}(\phi))\, ,$$
\note{where $q^{*,\tau}(\beta| Y, \phi) = N(\beta ; m_n^{*,\tau}, C_n^{*,\tau})$
is the variational posterior associated to this joint,
and all relevant information about $\phi^\tau$ is now 
encoded by $\theta^{*,\tau}(\phi)$, which 
appears in $m_n^{*,\tau}, C_n^{*,\tau}$ via 
\cref{eq:mtheta}, \cref{eq:ctheta}.} %
In particular, $(\theta^\tau(\phi))^{-1}$ is given in general by \cref{eq:ethetai}, 
 or for the particular cases of $\nu=0$ and $\nu=1$ by 
\cref{eq:ethetainu1}. %
An EM step for the MLE of $\phi$ is 
\begin{align}
Q(\phi| \phi^\tau ) &= 
\int \log(q^{*,\tau}(\beta,Y|\phi)) q^{*,\tau}(\beta| Y, \phi^\tau) d\beta \, , \notag \\
\phi^{\tau+1} &= 
{\rm argmax}_{\phi} Q(\phi | \phi^\tau ) \, . \label{eq:outerem}
\end{align}
The objective function for $\gamma^2$ is given by
\begin{equation}\label{eq:gamma}
f(\gamma^2) := - n \log \gamma + \frac1{2\gamma^2} \left(s_n - 2 v_n^T m_n^{*,\tau} + 
{\rm tr}[A_n(C_n^{*,\tau}+ m_n^{*,\tau}(m_n^{*,\tau})^T)]\right) \, , %
\end{equation}
where the following can be computed recursively in an online scenario
\begin{equation}\label{eq:an}
A_n := %
X_n^TX_n \, , \quad v_n := X_n^TY_n \, , \quad s_n := Y_n^T Y_n \, .
\end{equation}
This can be optimized independently of the remaining variables,
giving 
$$
(\gamma_n^{t+1})^2 = \frac1n \bbE_\tau|Y_n - X_n \beta |^2 = 
\frac1n \left(s_n - 2 v_n^T m_n^{*,\tau} + {\rm tr}[A_n(C_n^{*,\tau}+ m_n^{*,\tau}(m_n^{*,\tau})^T)]\right) \, ,
$$
where the expectation is with respect to $q^{*,\tau}(\beta| Y, \phi^\tau)$.
Note that in cases where $n<p$, one 
can use the identity 
$${\rm tr}[A_n(C_n^{*,\tau}+ m_n^{*,\tau}(m_n^{*,\tau})^T)] = 
{\rm tr}[X_n C_n^{*,\tau} X_n^T] + |X_n m_n^{*,\tau}|^2 \, .$$ 
These computations are easily adapted to the online case described in \cref{ssec:sparseonline}.

The objective function for $\phi$ is also easily computed as
\begin{equation}
 \sum_{j=1}^p \left ( \bbE_\tau(\beta_j^2) /\theta^\tau_j(\phi) + \log (\theta^\tau_j(\phi)) \right  ) \, ,
\end{equation}
where we recall again that $(\theta^\tau(\phi))^{-1}$ 
is given in general by \cref{eq:ethetai}, 
 or for the particular cases of $\nu=0$ and $\nu=1$ by 
\cref{eq:ethetainu1}.
\note{We consider global optimization 
for $3$ (or fewer) parameters of cheap-to-evaluate functions 
to be essentially a solved problem
\cite{floudas2013state}, e.g. via combination of 
basic local optimizers \cite{nocedal} initialized with multiple dispersed initial conditions.  
The derivative and Hessian are available in closed form, which is useful.

Some particular cases are convex and/or
even analytically soluble. For example, 
in the BL} case of $\nu=1$ and fixed $\delta\geq 0$, 
one has
\begin{equation}\label{eq:hypernu1}
(\lambda^{\tau+1})^{-1} %
= \frac1p \sum_{j=1}^p (C_{n,jj}^{*,\tau} + (m_{n,j}^{*,\tau})^2)  /\sqrt{ \delta^2 +  
C_{n,jj}^{*,\tau} + (m_{n,j}^{*,\tau})^2} \, .
\end{equation}
While in the case of $\nu=0$ and fixed $\delta\geq 0$, one has
\begin{equation}\label{eq:hypernu0}
(\lambda^{\tau+1})^{-1} = 
\frac1{p-\sum_{i=1}^p \frac{C_{n,jj}^{*,\tau} + (m_{n,j}^{*,\tau})^2}
{\left(\delta^2 +  C_{n,jj}^{*,\tau} + (m_{n,j}^{*,\tau})^2\right)}}
\sum_{j=1}^p \frac{C_{n,jj}^{*,\tau} + (m_{n,j}^{*,\tau})^2}{\sqrt{ \delta^2 +  C_{n,jj}^{*,\tau} + (m_{n,j}^{*,\tau})^2}} \, .
\end{equation}
\note{Finally, the ST case of $\lambda=0$ and fixed $\delta \geq 0$
leads to 
\begin{equation}\label{eq:hypernum1}
(1-2\nu^{\tau+1})^{-1}%
= \frac1p \sum_{j=1}^p (C_{n,jj}^{*,\tau} + (m_{n,j}^{*,\tau})^2)  /(\delta^2 +  
C_{n,jj}^{*,\tau} + (m_{n,j}^{*,\tau})^2) \, .
\end{equation}
It is reassuring to note that if $\delta=0$ then the scale factor
$1-2\nu^{\tau+1} \equiv 1$, i.e. $\nu^{\tau+1} \equiv 0$,
just as in the case of the scale-invariant Jeffrey's prior.
In other words, if we generalize Jeffrey's to allow any $\nu<1/2$
then we would find that standard Jeffrey's is optimal.}

Despite less attractive theoretical properties in comparison to the full VBEM,
this is a clean and simple approach for 
optimizing the hyperparameters. 
The objective function \cref{eq:outerem} is convex and analytically soluble (for the cases above). 
An obvious issue is the nested EM algorithms, which is undesirable.
Hence, the second option may be preferred, 
which is to simply iterate between a single iteration of 
\cref{eq:outerem} and a single iteration of VBEM, as described in \cref{eq:third}.

\section{Numerical Results}
\label{sec:numerics}

In this section we will explore the approach presented on some 
simulated and real data. Code which implements the methods is available at GitHub repository.\footnote{\href{https://github.com/zankin/SOVBR}{\bf \texttt{https://github.com/zankin/SOVBR}}}

\subsection{Diabetes data set}
\label{ssec:vbld}

\begin{figure}[!htbp]
	\centering\includegraphics[width=1\textwidth]{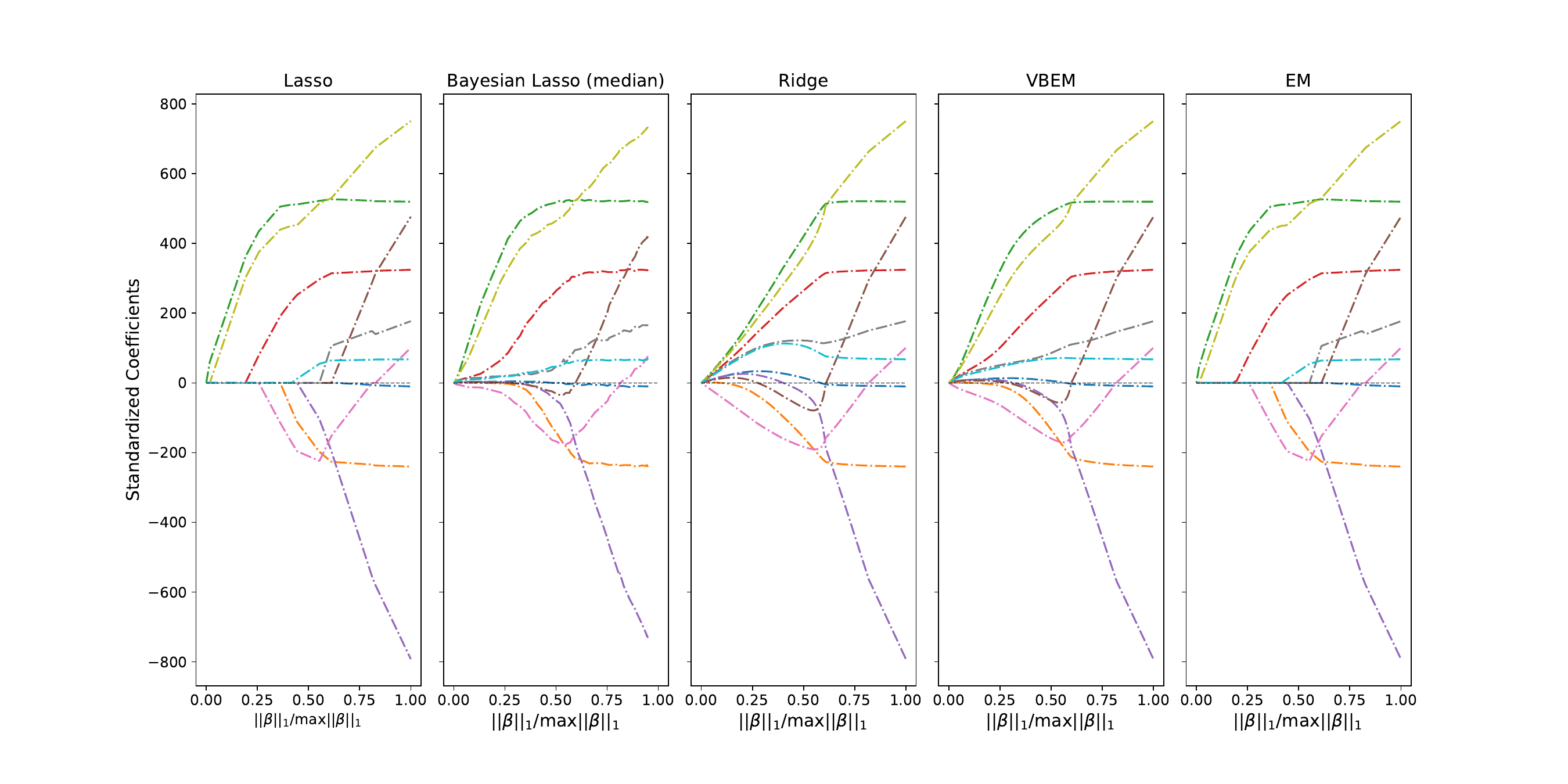}
	\includegraphics[width=0.7\textwidth]{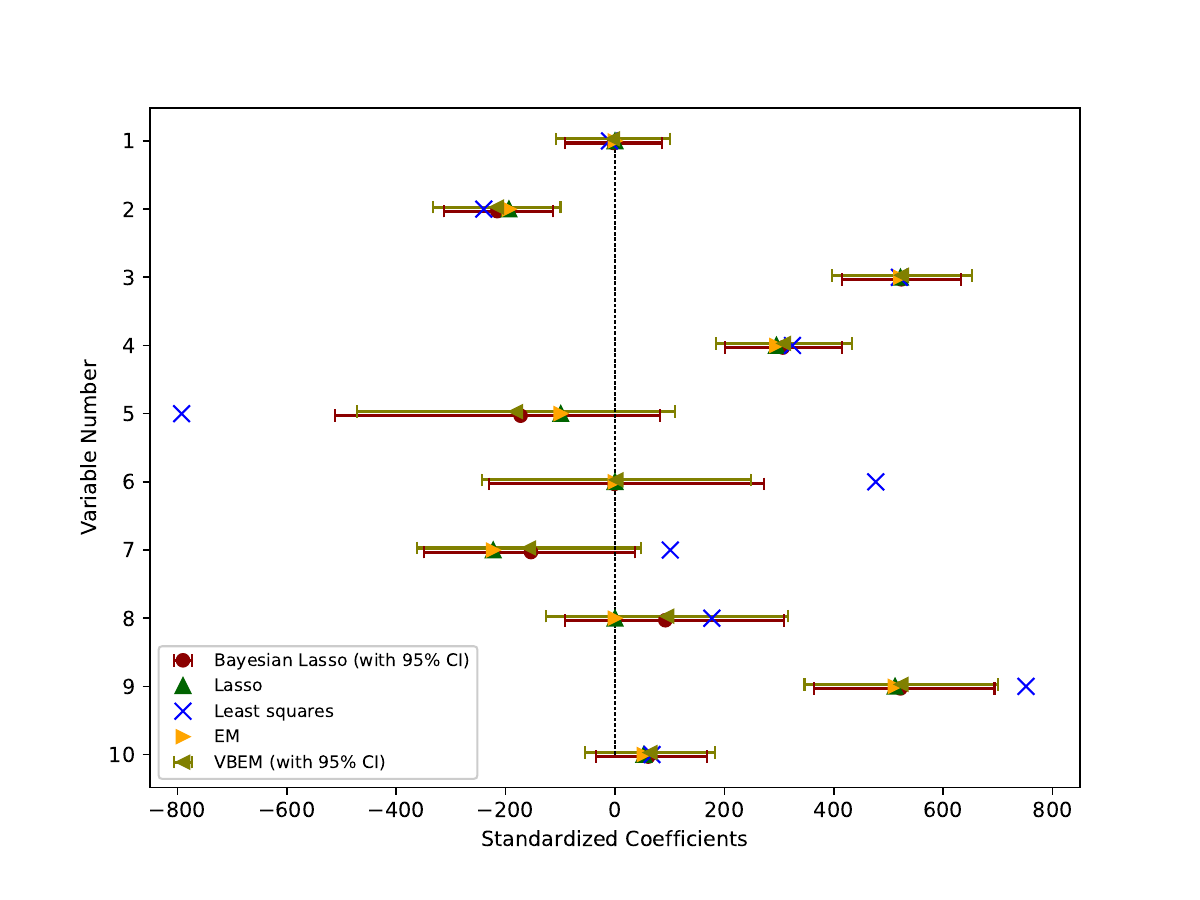}
	\caption{Illustration of the variational Bayesian LASSO (VBL) on the diabetes dataset.
	The top figure is analogous to Figure 1 of \cite{park2008bayesian}, except it adds VBL in the last 2 columns.
	The bottom one is analogous to Figure 2 of \cite{park2008bayesian}, 
	except with VBL (and 95\% credible interval) added.}
	\label{fig:VBLd}
\end{figure}

Here we present the VBL model and compare to the Bayesian LASSO (BL) of \cite{park2008bayesian}.
We use the simple diabetes data set from \cite{efron2004least}, with $n=484$ and $p=10$, 
which was used in \cite{park2008bayesian}.
The fully Bayesian methodology is quite expensive, and yet tractable for this very small problem,
which allows us to compare our very cheap variational approach.
In turn, the VBL is applicable for problems with several orders of magnitude larger values for $n$ and $p$,
where even the mightiest supercomputers will struggle to achieve the full Bayesian solution.
 The results are shown in \cref{fig:VBLd}.
     
The estimates of hyperparameter $\lambda$ for VBEM and EM models are obtained by respectively using the second approach from \cref{sec:hyper} with $\nu=1$ \cref{eq:hypernu1}.
 The resulted hyperparameters $(\gamma, \lambda)$ are given by %
 $\left(\widehat\gamma_{\text{VBEM}}, \widehat\lambda_{\text{VBEM}}\right) = (53.62, 0.0041)$. 
 For the BL %
 we take the optimal value $\widehat{\lambda}_{\text{BL}}=0.237$ for the BL selected by maximum marginal likelihood as in \cite{park2008bayesian} for the model there, 
 which has the following relationship to our model 
 $\widehat{\lambda}_{\text{BL}}/\gamma = \lambda$, 
 i.e. they scale the parameter in their model $\widehat{\lambda}$ by $\gamma$. 
 See \cite{moran2019variance} for discussion of the 
 benefits and drawbacks of the different formulations.
 
Another aspect of comparison between the BL and VBL (more precisely, the VBEM part) involves computational costs. \Cref{tab:cost} displays the inference time\footnote{This comparison was made on the laptop with Intel I7-7700HQ processor} (averaged over 30 runs) and the root-mean-squared error (obtained by 5-fold cross-validation) for VBEM and BL that were run with fixed hyperparameters $\widehat\lambda_{\text{VBEM}}$ and $\widehat\lambda_{\text{BL}}$. The number of consecutive iterations of the BL Gibbs sampler is 10000 (after 1000 burn-in) as in \cite{park2008bayesian}, and the maximum number of iterations of the VBEM is limited by 10. It is worth highlighting that while the RMSEs are very similar (and correspond well with the noise estimation $\widehat\gamma_{\text{VBEM}}$ or $\widehat\gamma_{\text{EM}}$) the difference in inference time reaches an impressive 1000 times speed up. 

\begin{table}[h]
\centering
\begin{tabular}{ |c|c|c| } 
\cline{2-3}
\multicolumn{1}{c|}{} & \bf{Time, ms} & \bf{RMSE} \\ 
 \hline
 VBEM, $\gamma\cdot\lambda = 0.220$   & 7.6 $\pm$ 0.7   & 54.611 \\ 
 \hline
 BL, $\lambda = 0.237$        & 7560 $\pm$ 602 & 54.612 \\ 
 \hline
\end{tabular}
\caption{VBEM and BL inference time and cross-validation errors.}
\label{tab:cost}
\end{table}

\note{
\subsection{Variable-selection: comparison of various examples with competitors}
\label{ssec:varsel}

Next, we consider the toy model of \cite{bai2020spike}, which is meant to 
assess skills in variable selection. As in \cite{bai2020spike}, we set $n = 100$ and $p = 1000$ in \cref{eq:basic}. Then, we sample the design matrix $X$ from zero-mean multivariate Gaussian distribution $N(0_p, \mathbf{\Sigma})$. The covariance matrix $\mathbf{\Sigma}$ is a block-diagonal one $\mathbf{\Sigma} = \text{bdiag}\left(\widetilde{\Sigma},\ldots,\widetilde{\Sigma}\right)$ with blocks $\widetilde{\Sigma} = \left\{\widetilde{\gamma}\right\}^{50}_{i,j=1}$, where $\widetilde{\gamma}_{ij} = 1$ if $i=j$ and $\widetilde{\gamma}_{ij} = 0.9$ otherwise. The true sparse vector of coefficients 
$\beta^{\sf true}$ is defined as
$\beta^{\sf true}_i = \left\{-3.5,-2.5,-1.5,1.5,2.5,3.5\right\}$ for $i\in\left\{1,51,101,151,201,251\right\}$ and zeros in all remaining ($p-6$) positions. The response vector $y$ is generated according to \cref{eq:basic}, where we set $\gamma^2 = 3$.

We compare the 5 choices of N-GIG priors
given in \cref{tab:examples} against the SSLASSO (mixture of LASSOs) \cite{bai2020spike}, 
Sparse VB (mean-field variational spike and slab) \cite{ray2021variational},
Horseshoe \cite{carvalho2010horseshoe}, 
variational Bayesian Bridge (for $r \approx 0$) 
\cite{armagan2009variational}, and automatic differentiation variational Bayes (ADVI) for ST (inverse Gamma mixing distribution, but without a priori specified factorization) \cite{kucukelbir2015automatic}. The competing methods were applied using the corresponding R packages: \texttt{SSLASSO}, \texttt{sparsevb}, \texttt{horseshoe}, \texttt{BayesBridge}, and \texttt{rstan} (R interface to probabilistic programming language Stan for ADVI method). 
The experiment was repeated 100 times, where each time we generated a new design matrix $X$ and the corresponding vector of responses $y$. During each experiment, we tracked the following quantities. 
\begin{itemize}
\item Mean squared error (MSE) and mean prediction error (MPE), defined as
$$
\text{MSE} = \frac{1}{p}||\widehat{\beta} - \beta^{\sf true}||_2^2 \quad \text{and} \quad \text{MPE} = \frac{1}{n}||X(\widehat{\beta} - \beta^{\sf true})||_2^2 \, ,
$$
where $\widehat{\beta}$ is corresponding point estimator of $\beta|\cD$,
and $\beta^{\sf true}$ is the frequentist truth defined above, which is 
used to simulate the data.
\item The false discovery rate (FDR) and the false negative rate (FNR)
$$
\text{FDR} = \frac{\text{FP}}{\text{FP}+\text{TP}} \quad \text{and} \quad \text{FNR} =\frac{\text{FN}}{\text{FN}+\text{TP}},
$$
where TP, TN, FP, and FN denote the number of true positives, true negatives, false positives, and false negatives, respectively.
Positive means a discovery that the null hypothesis is false with 95\% probability, where the null hypothesis is that 
$\beta_j=0$, i.e. a positive refers to the discovery of a covariate or selection of the variable $\beta_j \neq 0$.
\item  Empirical coverage (EC) for the individual coefficients:
$$
\text{EC} = \frac{1}{p}\sum_{j=1}^{p}\mathbbm{1}_{A_j}(\beta_j^{\sf true}),
$$
where $A_j$ is the $95\%$ posterior credible interval, 
$\mathbbm{1}$ is an indicator function,
and $\beta_j^{\sf true}$ is the frequentist truth used to simulate the data.
For the examples which deliver Gaussian approximations $N(m,C)$, 
this is given by 
$$
A_j = (m_j - 2\sqrt{C_{jj}}, m_j + 2\sqrt{C_{jj}}) \, .
$$
For the MCMC examples, it is calculated based on the order statistics of the MCMC simulations
$\beta^{(i)}$, $i=1,\dots,N$, 
as $A_j = (a_{\rm min}^j, a^j_{\rm max})$, where 
$$\frac1N \sum_{i=1}^{N} \mathbbm{1}_{\{\beta^{(i)}_j < a^j_{\rm min}\}} =0.025 \, , ~~ {\rm and} \quad
\frac1N \sum_{i=1}^{N} \mathbbm{1}_{\{\beta^{(i)}_j > a^j_{\rm max}\}} =0.025 \, .$$
\end{itemize}

The results are presented in \cref{tab:variable_selection}. 
Our method is faster than other VB methods, with better coverage
and comparable accuracy. 
It is notable that the methods which achieve the best FDR and FNR
have, respectively, worse FNR and FDR than our method.
It, therefore, provides a nice balance of speed, UQ, and accuracy.
}

\begin{table}[H]
\centering
\scriptsize
\note{
\begin{tabular}{l|lllllll}
\hline & \text { MSE } & \text { MPE } & \text { FDR } & \text { FNR } & \text { Runtime, s} & \text { EC, \% }\\
\hline 
\hline \text {BL (MAP)} & 0.009 (0.002) & 0.701 (0.332) & 0.006 (0.004) & 0.183 (0.132) & 2.44 (0.2) & -\\
\hline \text {BL (Mean)} & 0.012 (0.004) & 0.832 (0.422) & 0.009 (0.005) & 0.145 (0.072) & 2.44 (0.2) & 99.42 (0.18)\\
\hline \text {Jeff (MAP)} & 0.006 (0.003) & 0.743 (0.455) & 0.007 (0.006) & 0.220 (0.122) & 2.72 (0.5) & -\\
\hline \text {Jeff (Mean)} & 0.011 (0.005) & 0.923 (0.431) & 0.011 (0.004) & 0.110 (0.093) & 2.72 (0.5) & 99.49 (0.21) \\
\hline \text {ST (MAP)} & \textbf{0.004 (0.002)} & 0.623 (0.317) & 0.004 (0.004) & 0.183 (0.128) & 2.57 (0.3) & -\\
\hline \text {ST (Mean)} & 0.007 (0.004) & 0.651 (0.422) & 0.006 (0.005) & 0.110 (0.093) & 2.57 (0.3) & \textbf{100.0 (0.00)}\\
\hline \text{NG (MAP)} & 0.011 (0.007) & 0.770 (0.413) & 0.009 (0.006) & 0.145 (0.092) & 3.01 (0.5) & -\\
\hline \text{NG (Mean)} & 0.016 (0.008) & 0.801 (0.519) & 0.012 (0.009) & 0.122 (0.103) & 3.01 (0.5) & 99.30 (0.26)\\
\hline \text{NIG (MAP)} & 0.018 (0.006) & 0.864 (0.317) & 0.007 (0.006) & 0.103 (0.092) & 3.17 (0.4) & -\\
\hline \text{NIG (Mean)} & 0.022 (0.005) & 0.983 (0.422) & 0.013 (0.004) & 0.081 (0.078) & 3.17 (0.4) & 99.40 (0.00)\\
\hline
\hline \text{SSLASSO}   & 0.006 (0.006) & 0.696 (0.552) & \textbf{0.001(0.001)} & 0.171 (0.148) & \textbf{0.28(0.1)} & -\\
\hline \text{Sparse VB} & 0.016 (0.011) & 1.497 (0.852) & 0.007 (0.007) & 0.335 (0.176) & 4.52 (1.6) & 99.38 (0.28)\\
\hline \text{Horseshoe} & 0.004 (0.005) & \textbf{0.619(0.446)} & 0.047 (0.076) & 0.030 (0.064) & 47.3 (6.1) & 99.88 (0.09)\\
\hline \text{BB (MAP)} & 0.030 (0.007) & 2.732 (0.405) & 0.046 (0.003) & 0.166 (0.144) & 4.72 (1.6) & -\\
\hline \text{BB (Mean)} & 0.016 (0.007) & 4.669 (2.572) & 0.318 (0.040) & 0.033 (0.083) & 206.6 (11.3) & 99.67 (0.15)\\
\hline \text{ADVI (Stan)} & 0.026 (0.011) & 2.993 (0.064) & 0.166 (0.006) & \textbf{0.015(0.007)} & 632.1 (25.7) & 99.79 (0.11)\\
\hline
\end{tabular}}
\caption{\note{MSE, MPE, FDR, FNR, Runtime, and Empirical Coverage (where relevant). For each quantity, we report average values across 100 experiments and corresponding standard deviations (in parentheses).}}\label{tab:variable_selection}
\end{table}

\subsection{Total variation (TV) deblurring} 
\label{ssec:vbltv}

Now we will consider the problem of image deconvolution. 
Consider that we would like to recover an image $\tilde \beta \in \bbR^{\tilde{p}}$ 
comprised of $\tilde p=p_0^2$ pixels, from observations $Y = [y_1,\dots, y_n]^T \in \bbR^n$ 
with $n \leq \tilde p$. 
The design matrix is defined as follows, for $l=1,\dots, n$,
\begin{equation}\label{eq:fftX}
\tilde x_l^T \tilde \beta := z_{i_l,j_l} \, , \qquad  
z = F^{-1}\left( D(\exp(-\omega |k|^2)) \right) F \tilde\beta \, .
\end{equation}
where $\{(i_l,j_l)\}_{l=1}^n$ is a subset of pairs of indices associated to spatial observations
of the degraded image/signal $\tilde\beta$, denoted $z$ 
(both represented as 2 dimensional $p_0\times p_0$ arrays here), 
\note{$F$ denotes the discrete Fourier transform} 
(which will be computed with fast Fourier transform (FFT) \cite{fft} at a cost of 
$\cO(p_0^2 \log p_0)$),
and $k = (k_1,k_2) \in \{-p_0/2,\dots, p_0/2-1\}^2$ is the multi-index of wave-numbers 
associated to the transformed signal. 
This simply corresponds to convolution in physical space with a Gaussian kernel with \note{kernel width}
proportional to $\omega$. 
The observations are then defined as usual
$$
y =  \tilde X \tilde \beta  + \epsilon \, , \qquad   \epsilon \sim N(0,\gamma^2 I_n) \, .
$$
\note{Note that periodic boundary conditions are implicitly assumed once 
Fourier transform is used, however that constraint can be removed by 
padding with $p_0$ additional zeros in each direction, sometimes referred to 
as {\em circulant embedding} \cite{dietrich1997fast}.}

Recall the discussion in \cref{sec:sparse}. 
We now are interested not in sparse signals per se, but rather in edge-preservation,
or in other words sparse gradient. 
For this purpose a popular choice is the (non-isotropic) total variation prior given by 
$\prod_{j=1}^{p_0^2-1} \cL((D_1\beta)_j; \lambda) \cL((D_2\beta)_j; \lambda)$,
\note{where $D_i$ is some discrete approximation of the derivative with respect to 
coordinate $j$, for $j=1,2$.
Often the finite difference is used, but here the natural choice is a Fourier approximation,
given by $D_j = F^{-1} (-i) k_j F$, $i=\sqrt{-1}$, } 
and the missing degree of freedom corresponds to the constant
wavenumber $k=(0,0)$.
This can be constructed as a marginal just like \cref{eq:lap},
using a pair of 
Normals $\cN((D_1\beta)_j; 0, \theta_j)$ and
$\cN((D_2\beta)_j; 0, \theta_j)$ for each $j=1,\dots, \tilde p-1$,
This change of variables proves to be messy within the VBEM
(although it works just fine for EM/TV, even with a standard finite difference 
approximation in the spatial domain).

We adopt an alternative approach as follows, 
which we have found cleaner and more computationally expedient.
Note that the TV prior can be alternatively written as a standard LASSO prior 
on $\bar \beta := {\bf D} \tilde \beta \in \bbR^{p-1}$, where $p = 2 \tilde p +1$
and ${\bf D} = (D_1^T, D_2^T )^T \in \bbR^{p -1 \times \tilde p}$.
This matrix also has the vector ${\bf 1}_{\tilde p} \in \bbR^{\tilde p}$ of ones in its kernel.
Denote the coefficient of ${\bf 1}_{\tilde p}$ as $\beta_0$.
We can redefine the forward model on ${\beta} := (\bar \beta^T, \beta_0)^T$
as ${X} := \tilde X ({\bf D}^\dagger, {\bf 1}_{\tilde p})$, 
where ${\bf D}^\dagger = ({\bf D}^T {\bf D})^{-1} {\bf D}^T$ is the left pseudo-inverse 
of ${\bf D}$.
Our data for the transformed model is 
\begin{equation}\label{eq:tvdata}
Y = X \beta + \epsilon \, .
\end{equation}

Typically $p_0$ will be large, for example $p_0=256,512$ or even larger, 
which precludes $\cO(p^2)$ calculations.
In terms of computation, when $n$ is small, then 
$X^T$ can be computed explicitly with $n$ FFTs, which allows explicit computation of 
$G := C_0 X^T (X C_0 X^T + \gamma^2 I_n)^{-1}$ for small enough $n$.
This allows computation of \cref{eq:monolith}.
In order to compute the diagonal of $C_n$ in \cref{eq:monolithC} we observe that
the second term can be written as $(G \circ (C_0 X^T)) {\bf 1}_n$, where $\circ$ denotes the
element-wise product of two matrices. 

\note{
\subsubsection{1D signal deblurring}

Here we consider a simple 1D version, and compare VBL as well as 
the other models from \cref{tab:examples} 
on the model data given by \cref{eq:tvdata}.
The setting is exactly the same as described above, 
however $D=1$, so there is a single index, 1D FFT, a single derivative, $p=\tilde{p}+1$, 
and a discretization of $\tilde{p}= p_0 = 200$ nodes are used between $[-4,24]$.
The signal is the 1D Bernholdt function $f(s)$ for $[-4,10]$ \cite{bernholdt}, and
padded with zeros on $[10,24]$. %
Note that the domain has been doubled in order to accommodate non-periodic boundary conditions.}

\begin{table}[h]
\centering
\scriptsize
\note{
\begin{tabular}{l|lllllll}
\hline & \text { BL (MAP) } & \text { Jeff (MAP) } & \text { ST (MAP) } & \text { BL (Mean) } & \text { Jeff (Mean) } & \text { ST (Mean) }\\
\hline 
\hline \text {MSE, $\times 1e^{-3}$} & 1.097 & 1.005  & 1.070 & 1.136 & 1.113 & 1.149\\
\hline \text {MPE, $\times 1e^{-6}$} & 3.903 & 3.461  & 3.506 & 4.020 & 3.623 & 3.689\\
\hline
\end{tabular}}
\caption{\note{MSE and MPE errors for BL, Jeffreys, and ST priors.}}\label{tab:1d_TV}
\end{table}

\begin{figure}[!htbp]
	\centering\includegraphics[width=0.49\textwidth]{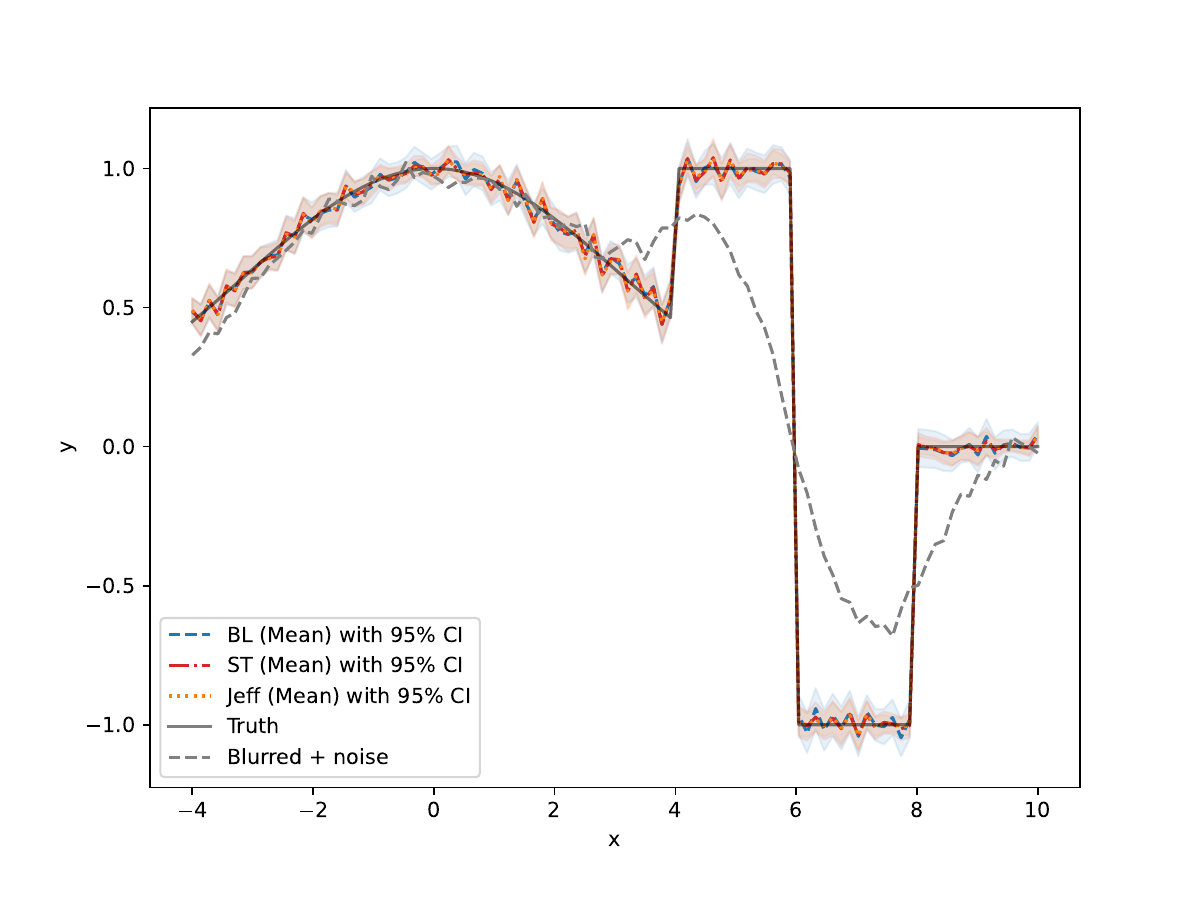}
	\includegraphics[width=0.49\textwidth]{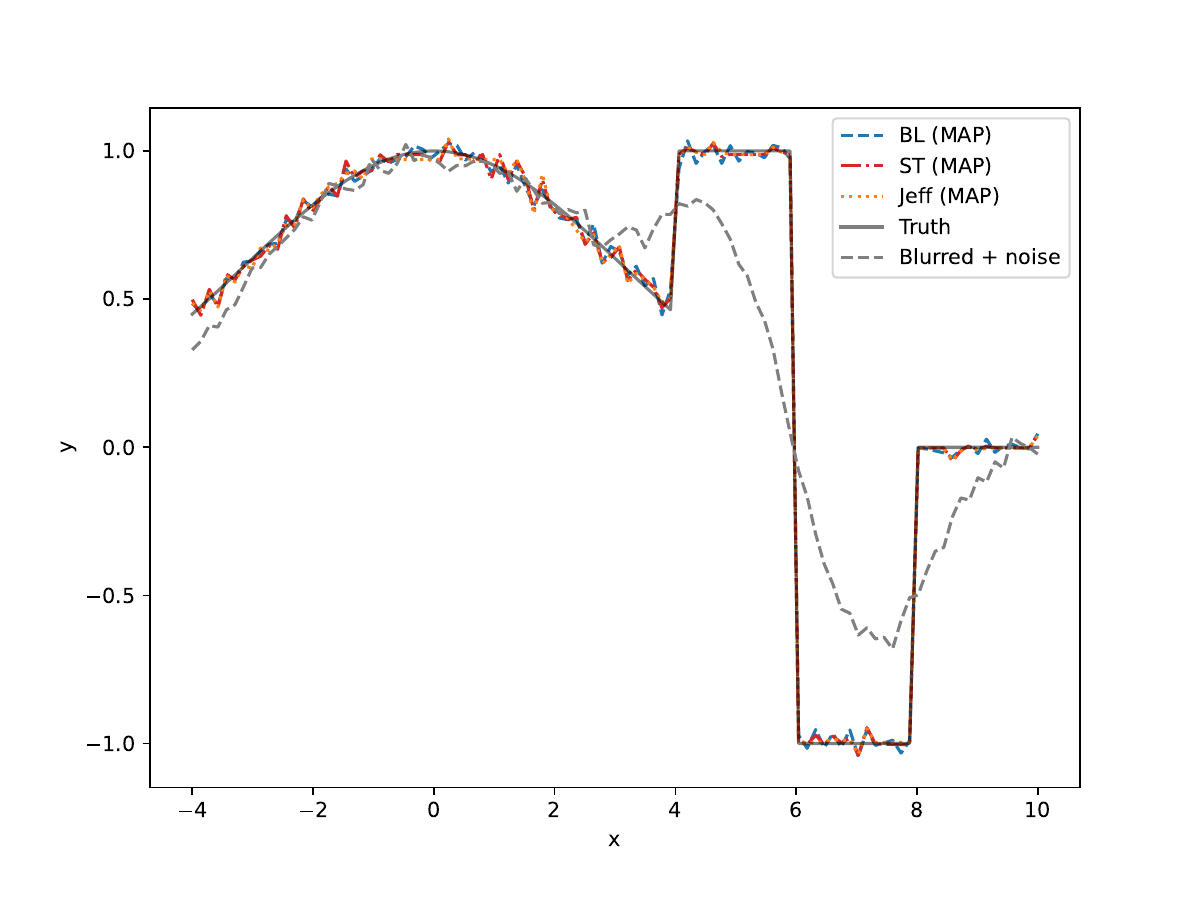}
	\caption{\note{
	The plot shows the original function, blurred and noisy one, and the reconstructions obtained applying BL, Jeffreys, and ST priors.}}
	\label{fig:1d_deblurring}
\end{figure}

\subsubsection{2D image deblurring}

\begin{figure}[!htbp]
	\centering\includegraphics[width=.24\textwidth]{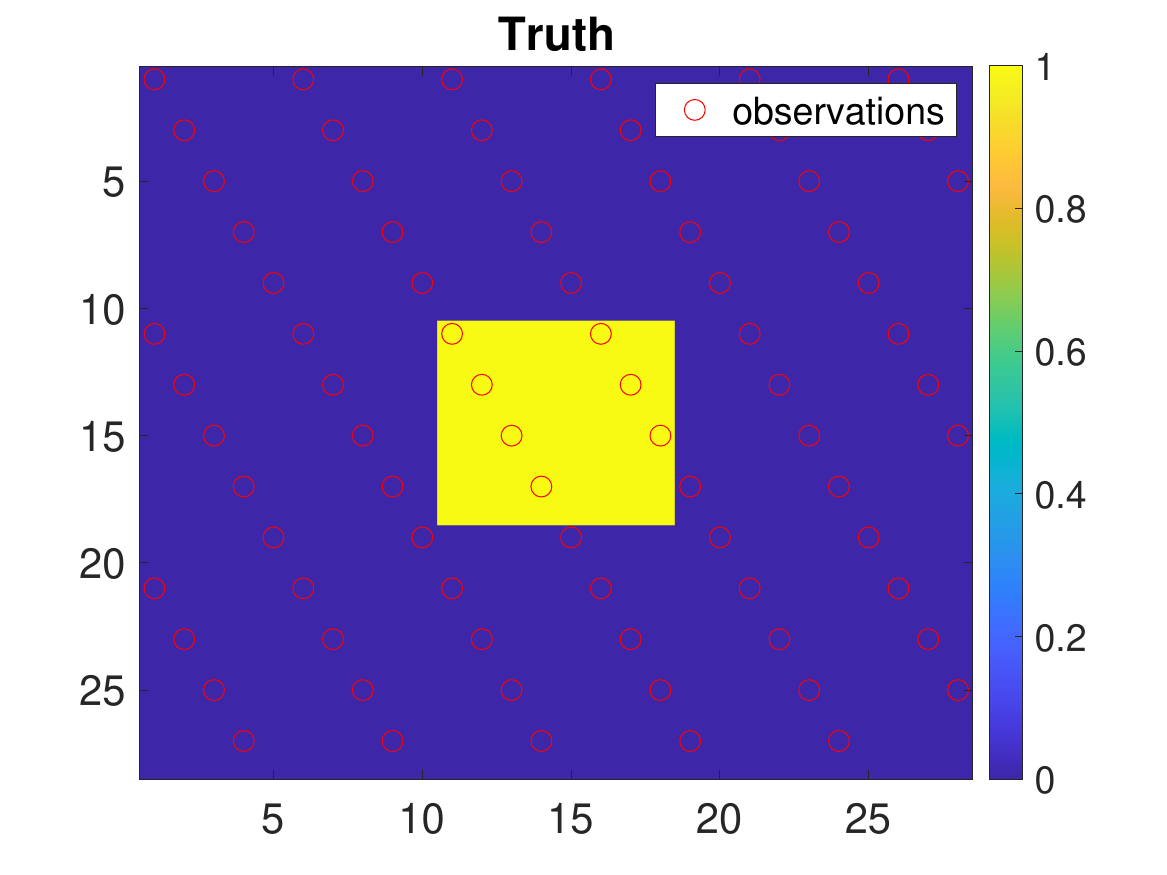}
	\includegraphics[width=0.24\textwidth]{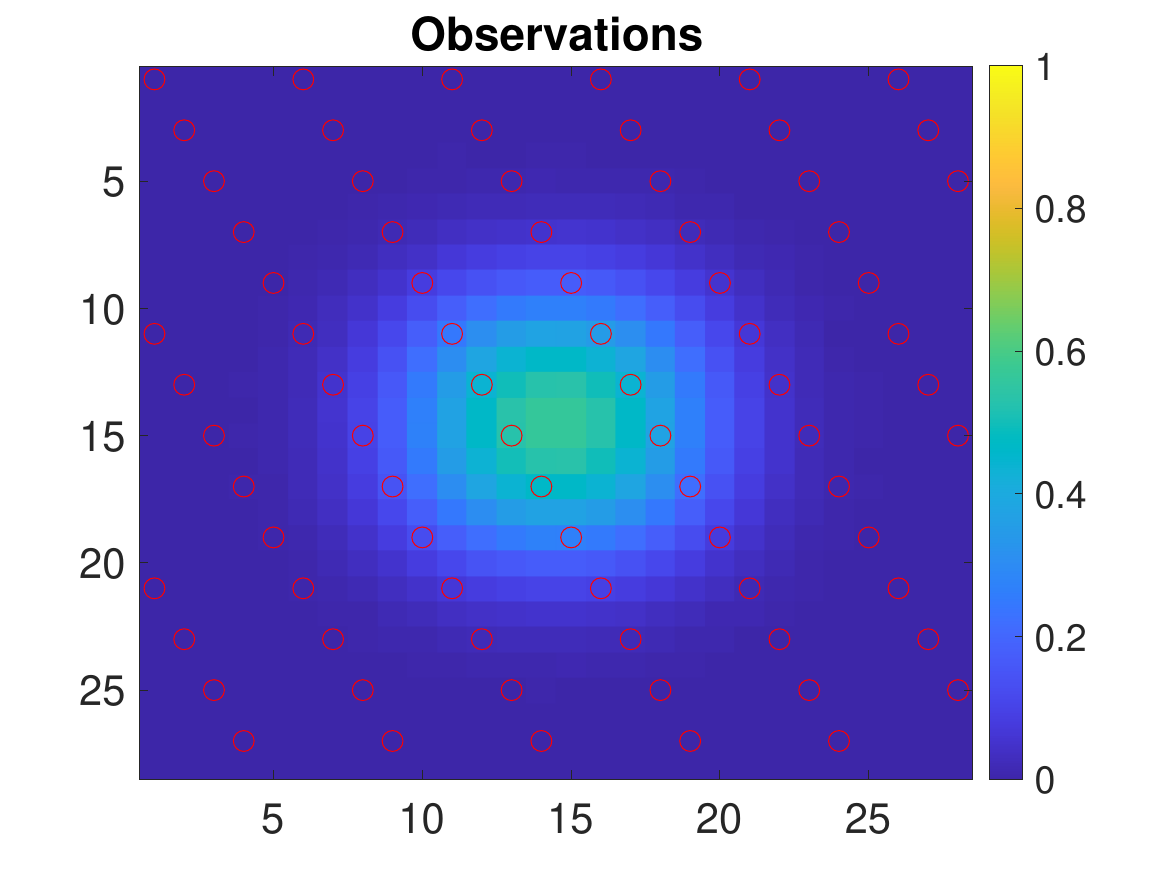}
	\includegraphics[width=0.24\textwidth]{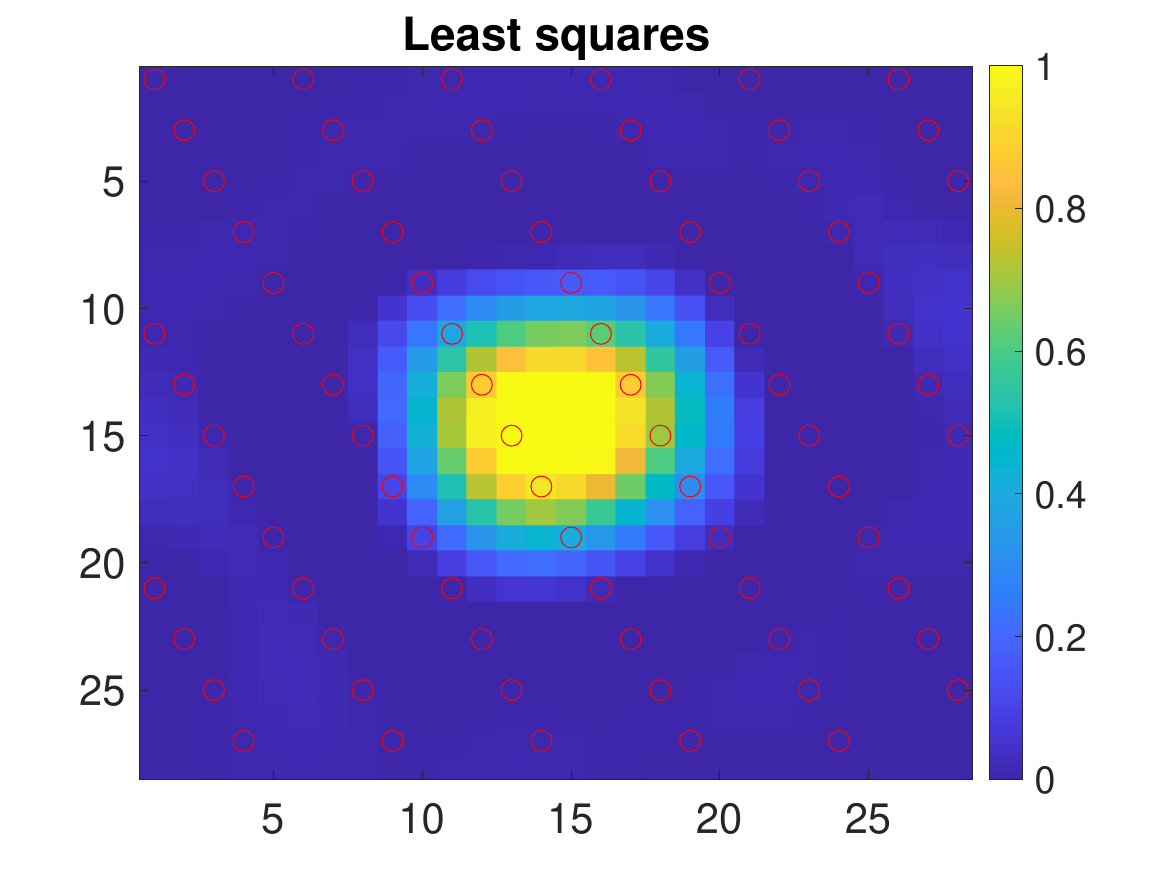} 
	\includegraphics[width=.24\textwidth]{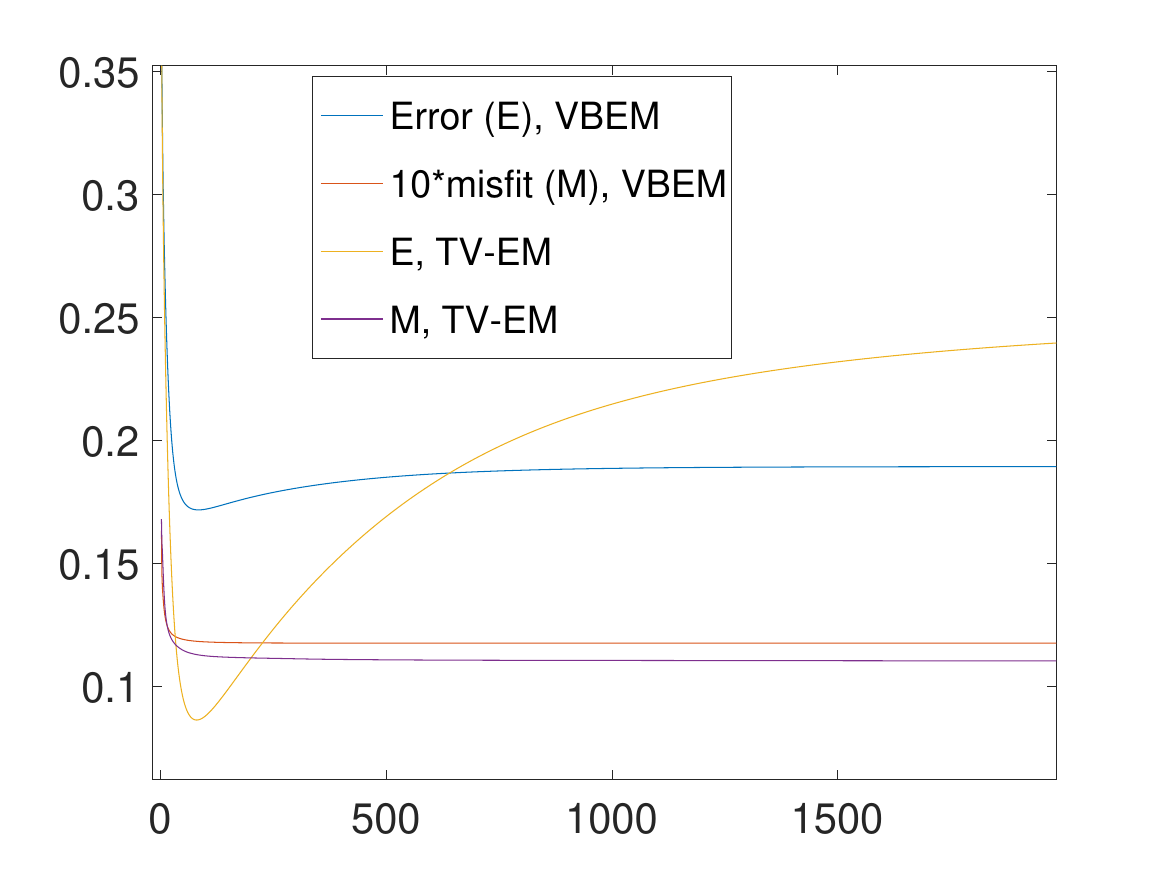} \\
	\includegraphics[width=.24\textwidth]{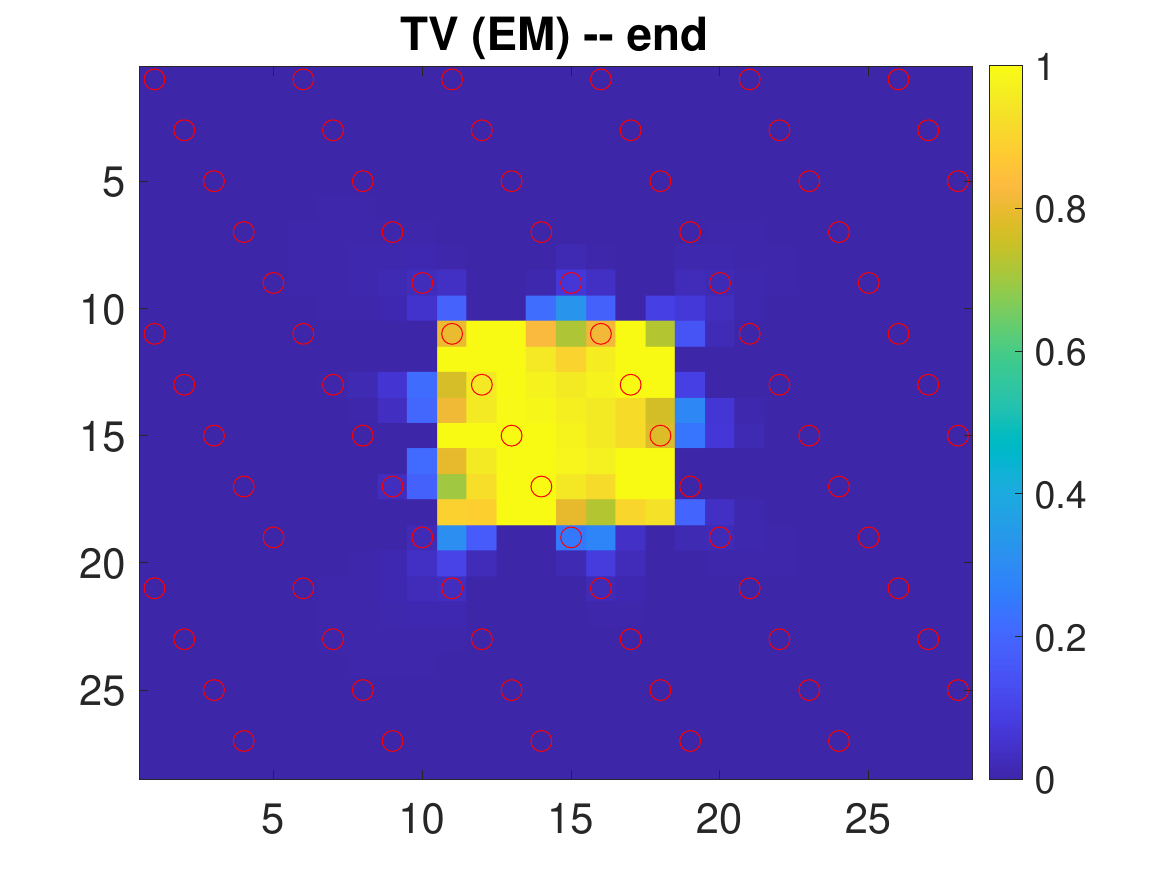}
	\includegraphics[width=0.24\textwidth]{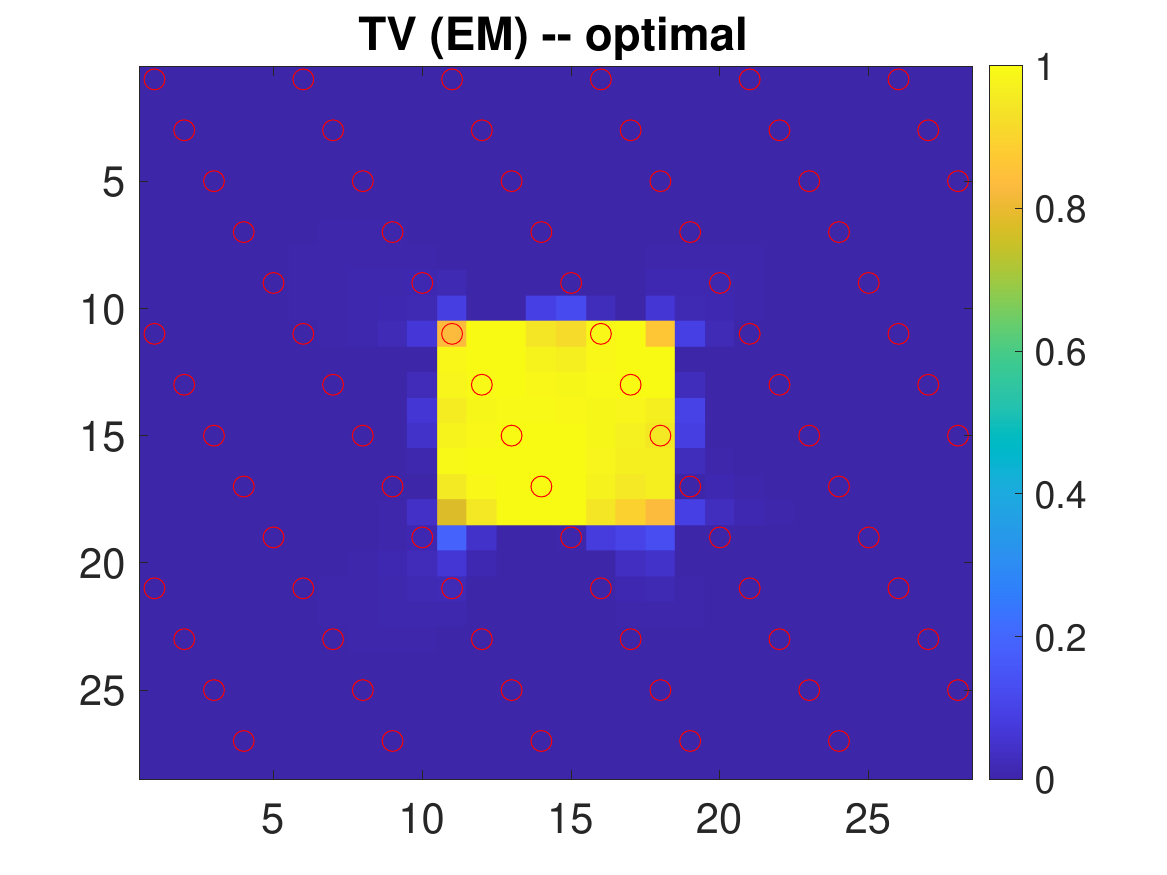}
	\includegraphics[width=0.24\textwidth]{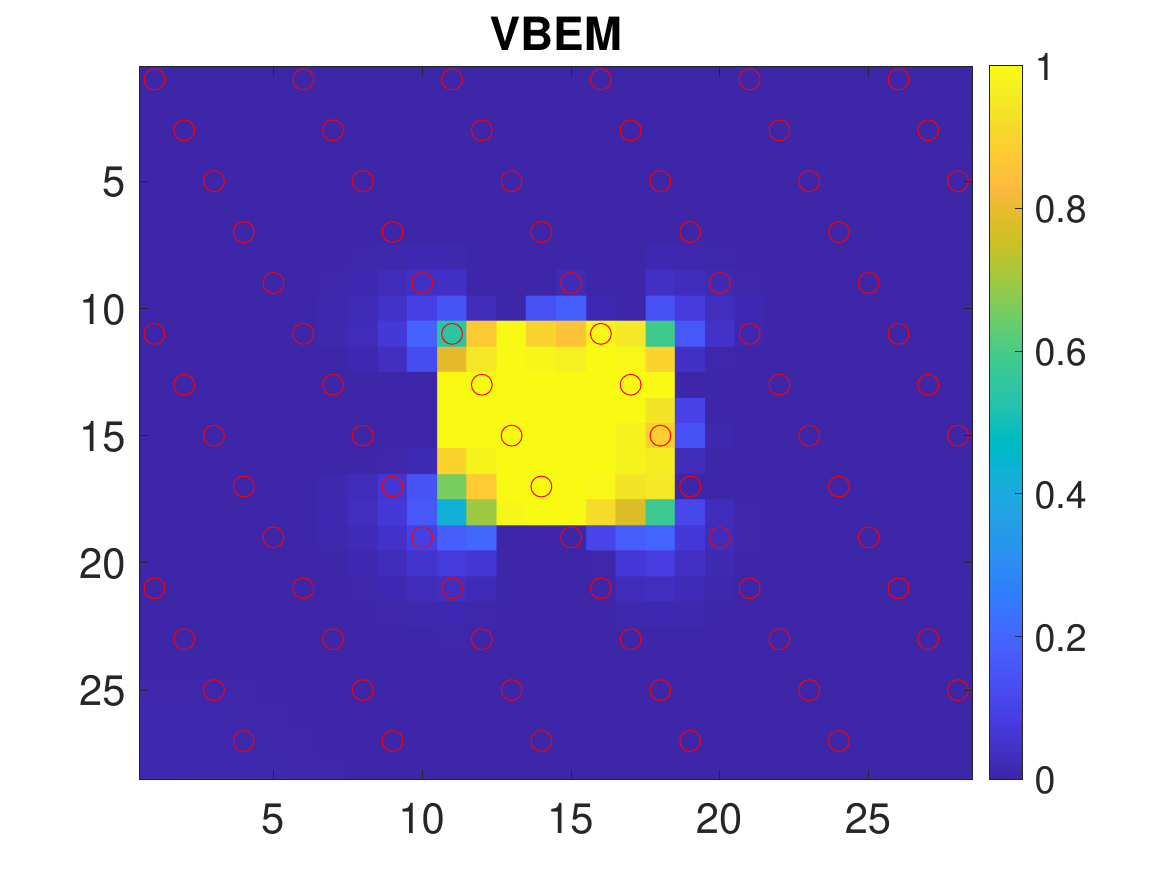} 
	\includegraphics[width=.24\textwidth]{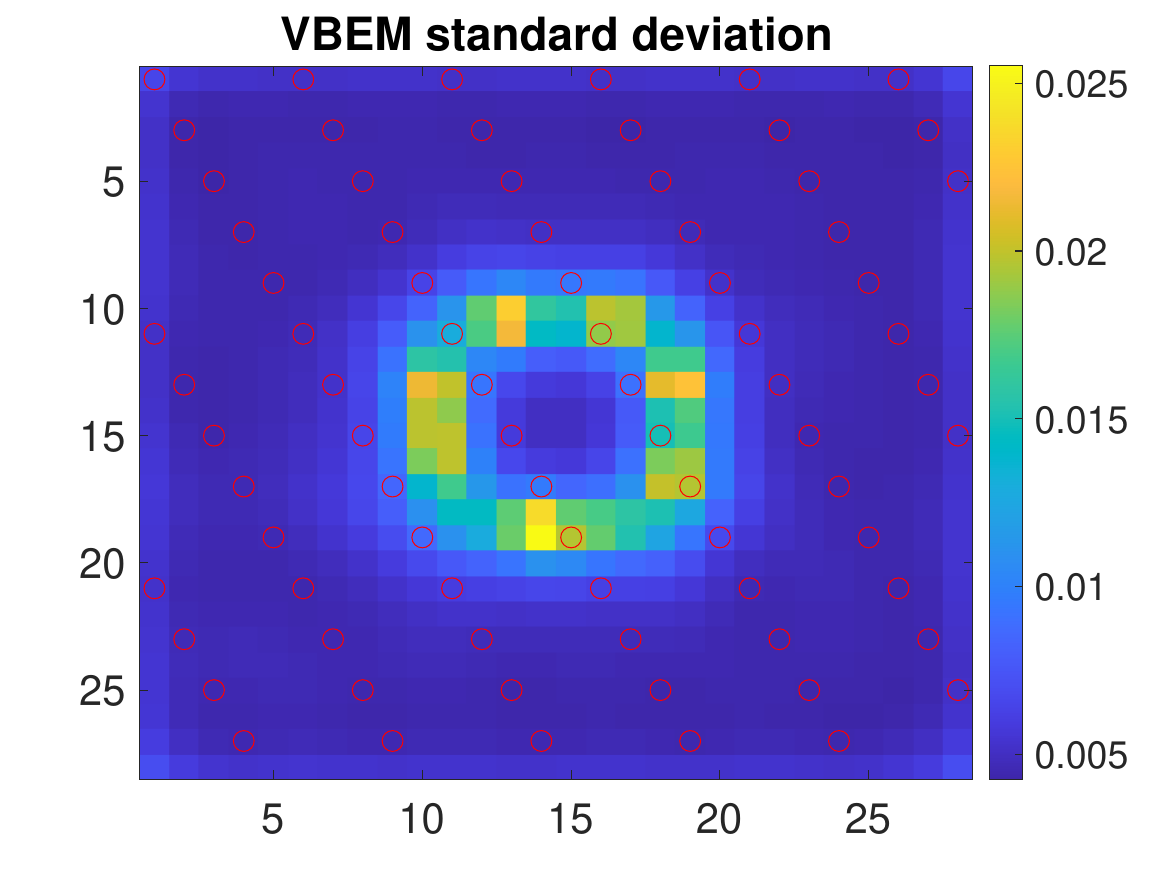}
	\caption{VBL TV-denoising illustrated on a simple toy image.
	The top row features (from left to right) the truth, blurry observations, 
	the Tikhonov-regularized (least squares) solution, and a plot of the data 
	misfit and reconstruction error over the EM/VBEM iterations for the various.
	The bottom row features (from left to right) the TV-regularized MAP estimator
	at convergence $\mu_n$, the TV-regularized MAP estimator at the minimum of 
	the error (around iteration 100 -- see top right), the VBL mean $m_n$, 
	and the VBL standard deviation.}
	\label{fig:VBL}
\end{figure}

We now conduct several experiments on images. 
First we consider a toy model with $n=\tilde{p}=p_0^2$ and $p_0=28$,
with strong blurring and small noise. With a large $\lambda$ we obtain very impressive
reconstructions (see \cref{fig:VBL}). It is notable that the uncertainty
is significantly underestimated, although relatively correct 
(in the sense that it is large and small where the error is). 
This is due to the fact that $\lambda$ has been chosen too large, 
which, on the other hand, provides the impressive reconstruction of the edges. 
Notice here the top right-hand plot which shows the (relative $L^2$) error as a function
of iteration -- its minimum may yield a much smaller error in comparison 
to the value at convergence, in particular for the MAP estimator (TV-EM).
The plot also shows the data misfit, where we can observe the classical
``L-curve'' and see that an appropriate stopping criterion can be derived 
from convergence of the misfit (in the more realistic scenario where we do 
not know the error).

\begin{figure}[!htbp]
	\centering\includegraphics[width=0.24\textwidth]{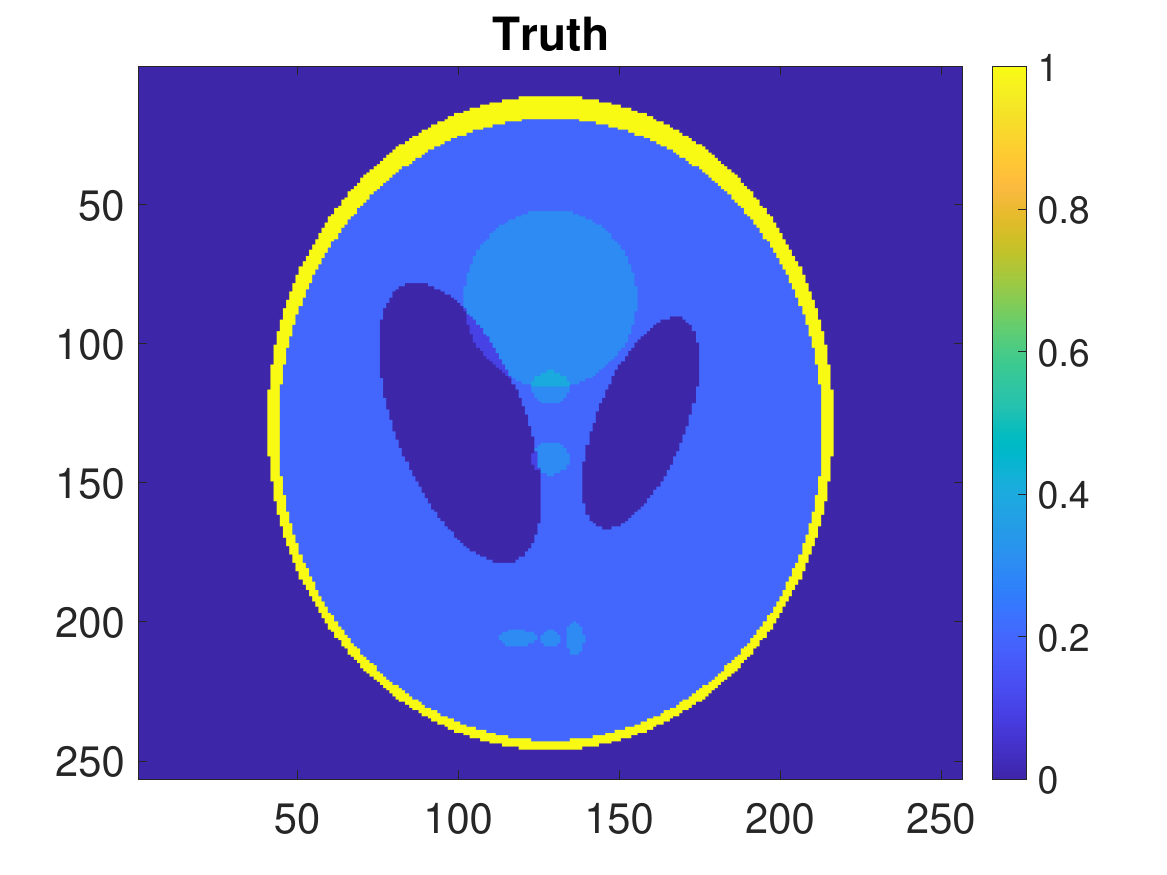}
	\includegraphics[width=0.24\textwidth]{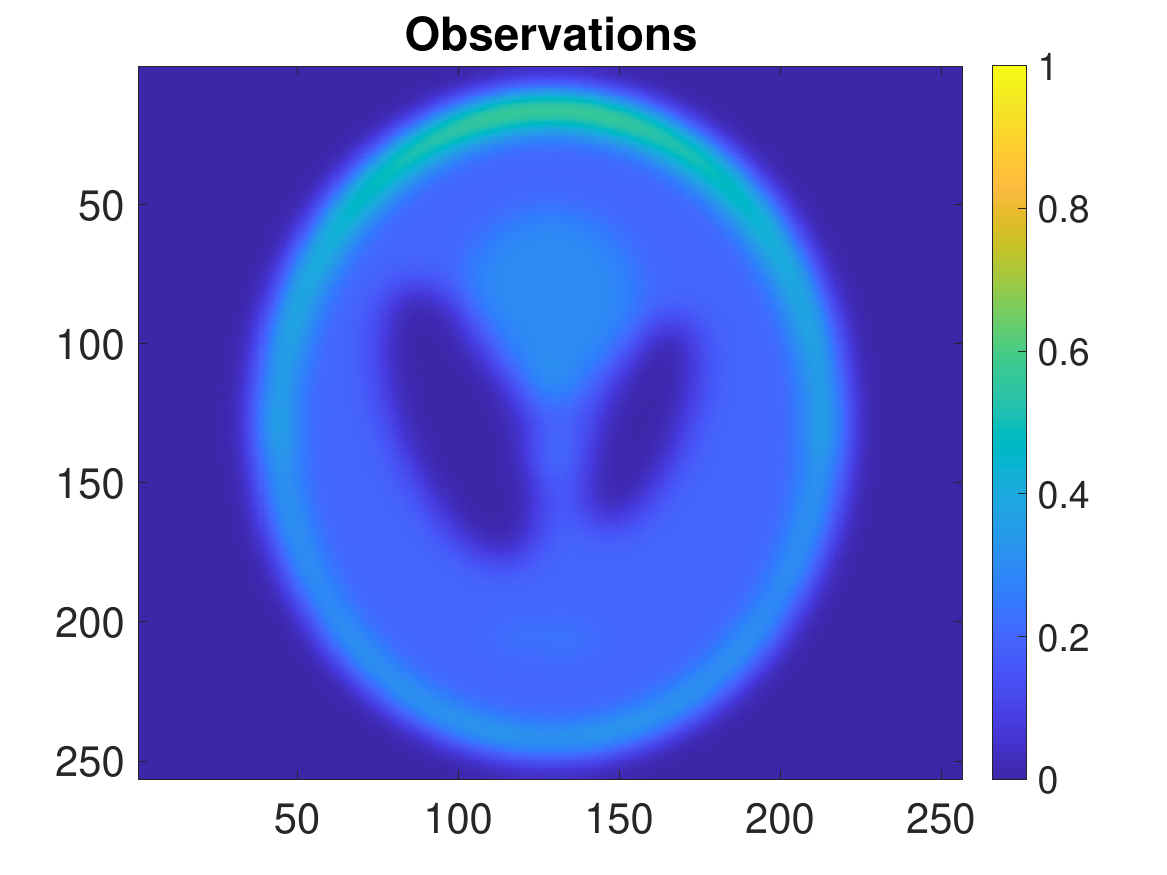}
	\includegraphics[width=0.24\textwidth]{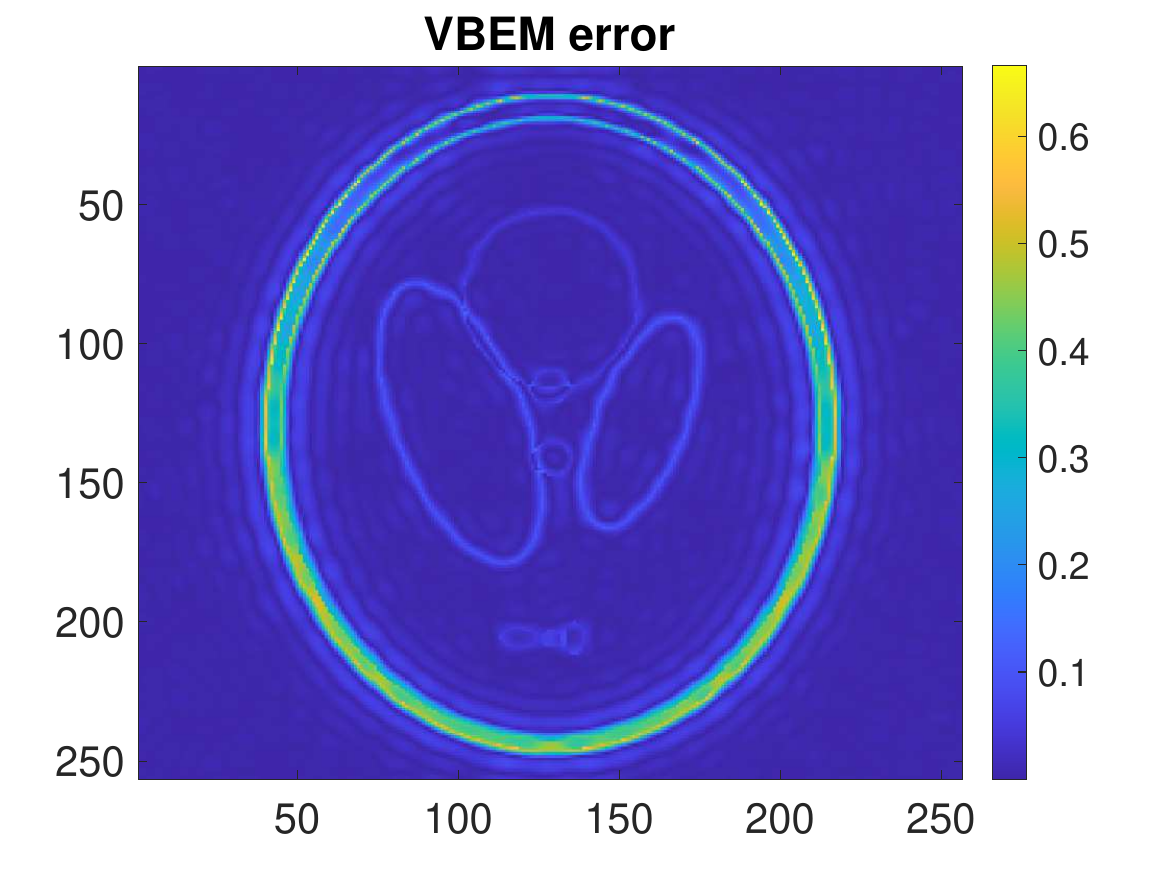} 
	\includegraphics[width=0.24\textwidth]{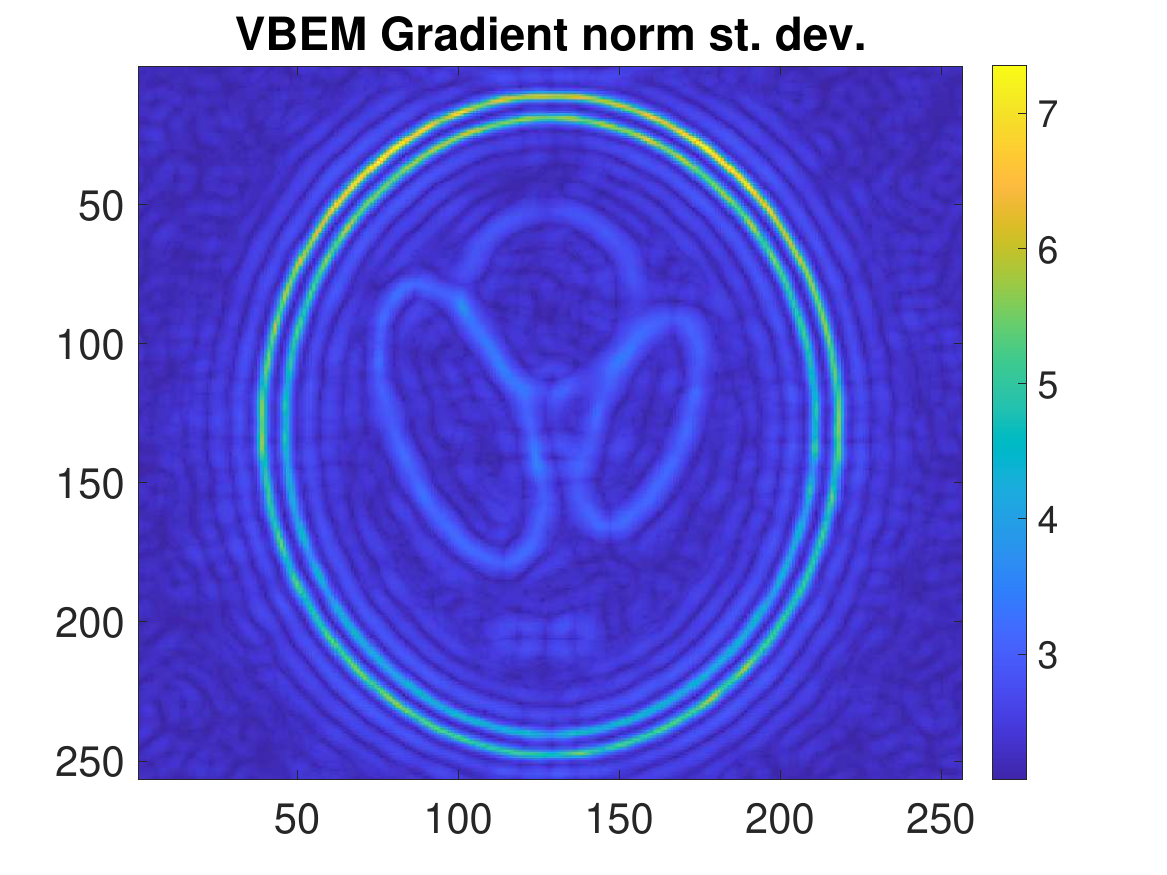} 
	\\
	\includegraphics[width=0.24\textwidth]{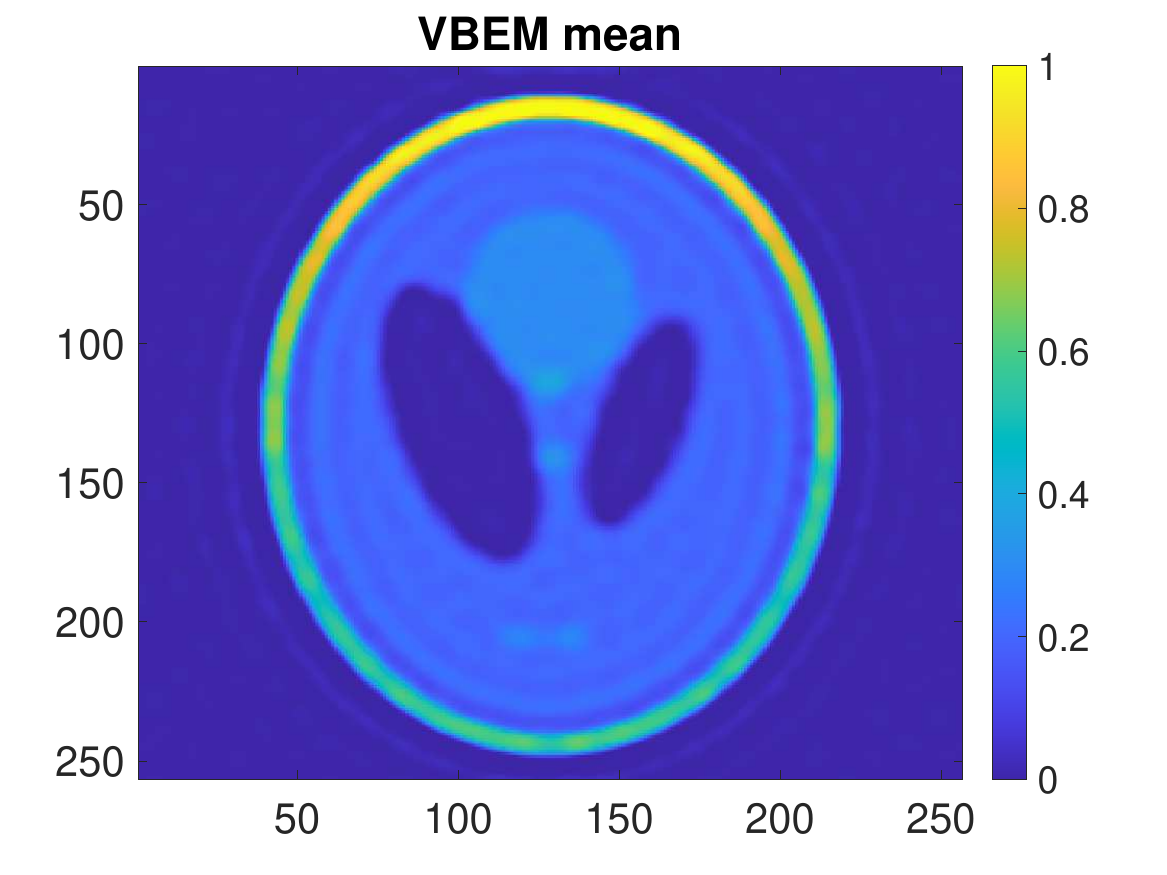}
	\includegraphics[width=0.24\textwidth]{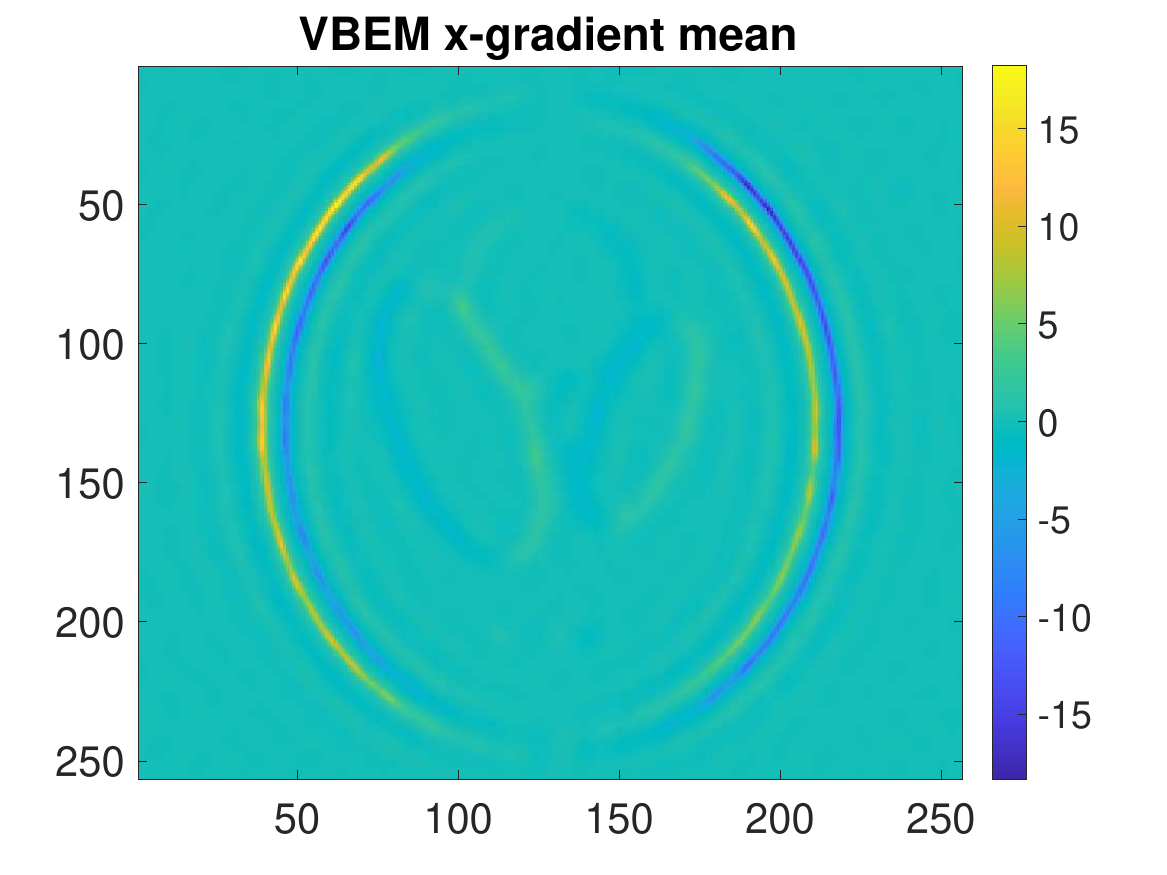}
	\includegraphics[width=0.24\textwidth]{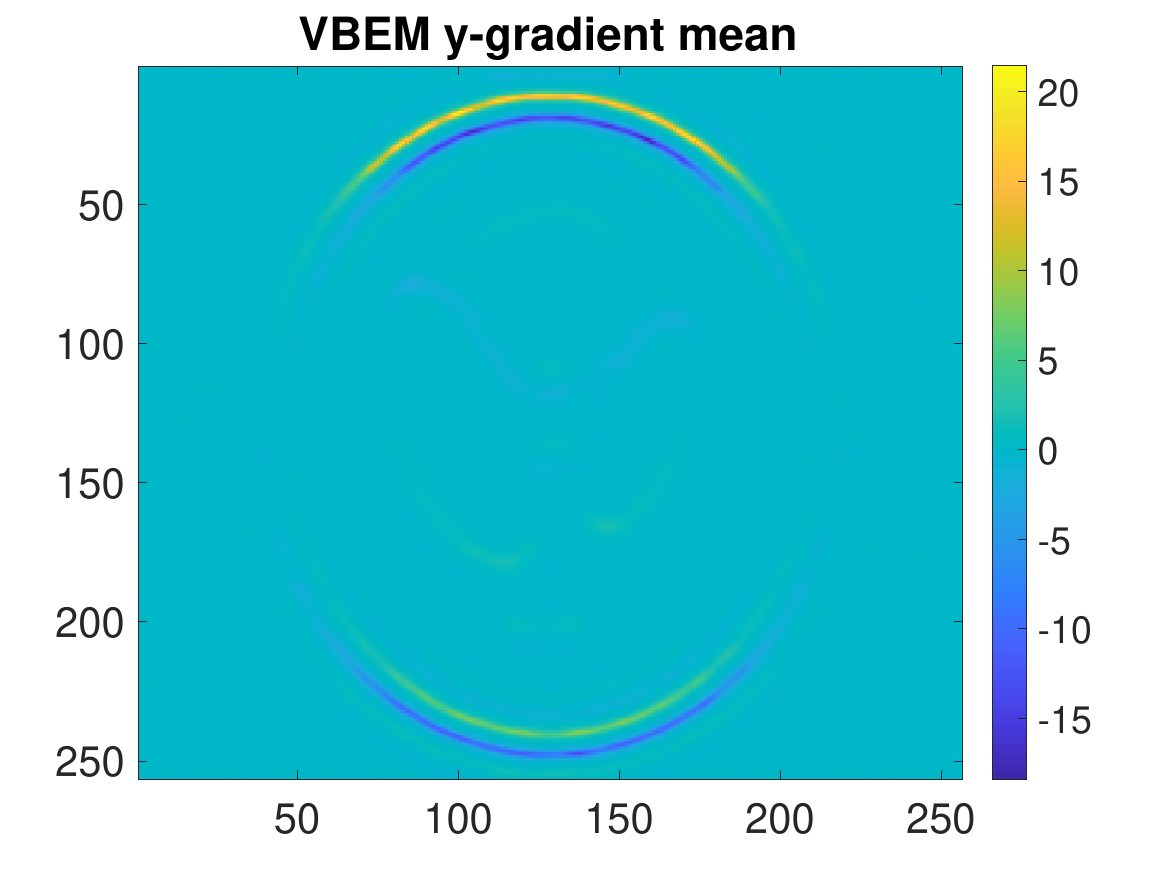} 
	\includegraphics[width=0.24\textwidth]{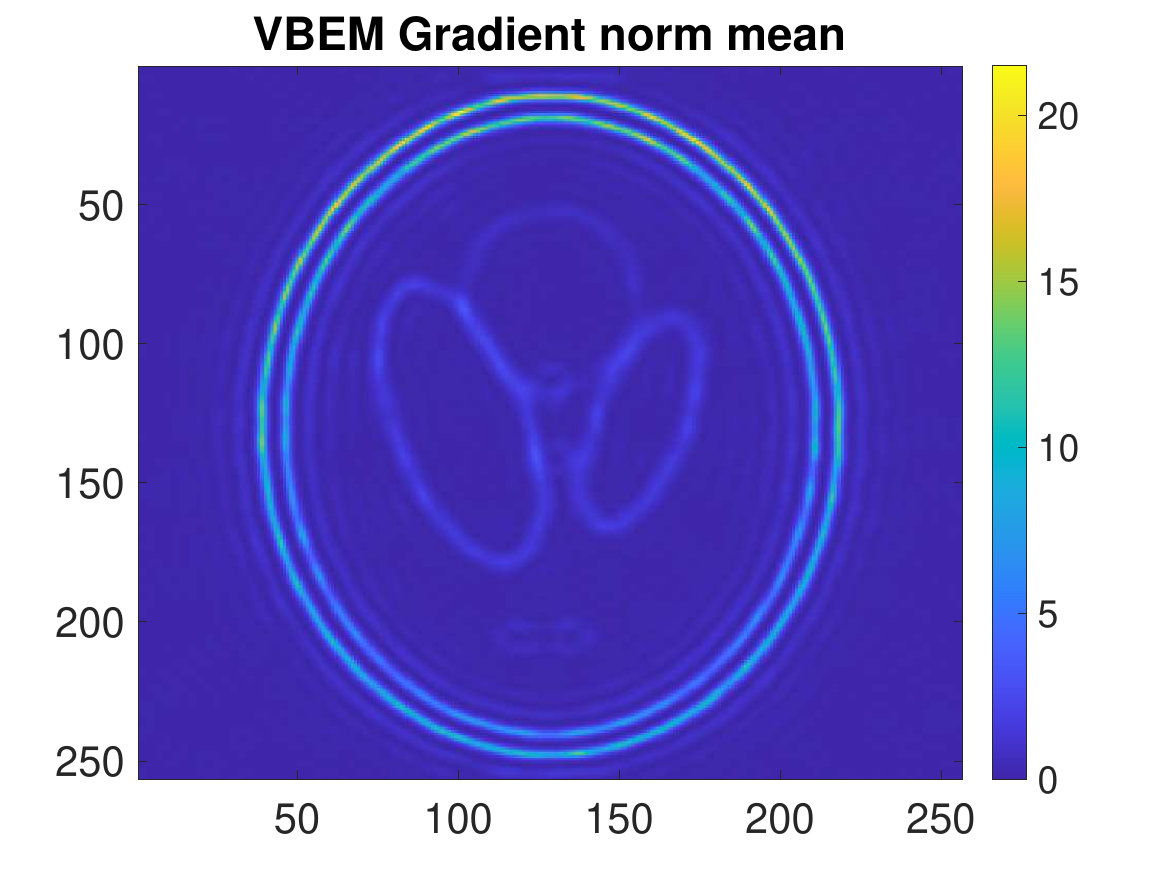}
	\\
	\includegraphics[width=0.24\textwidth]{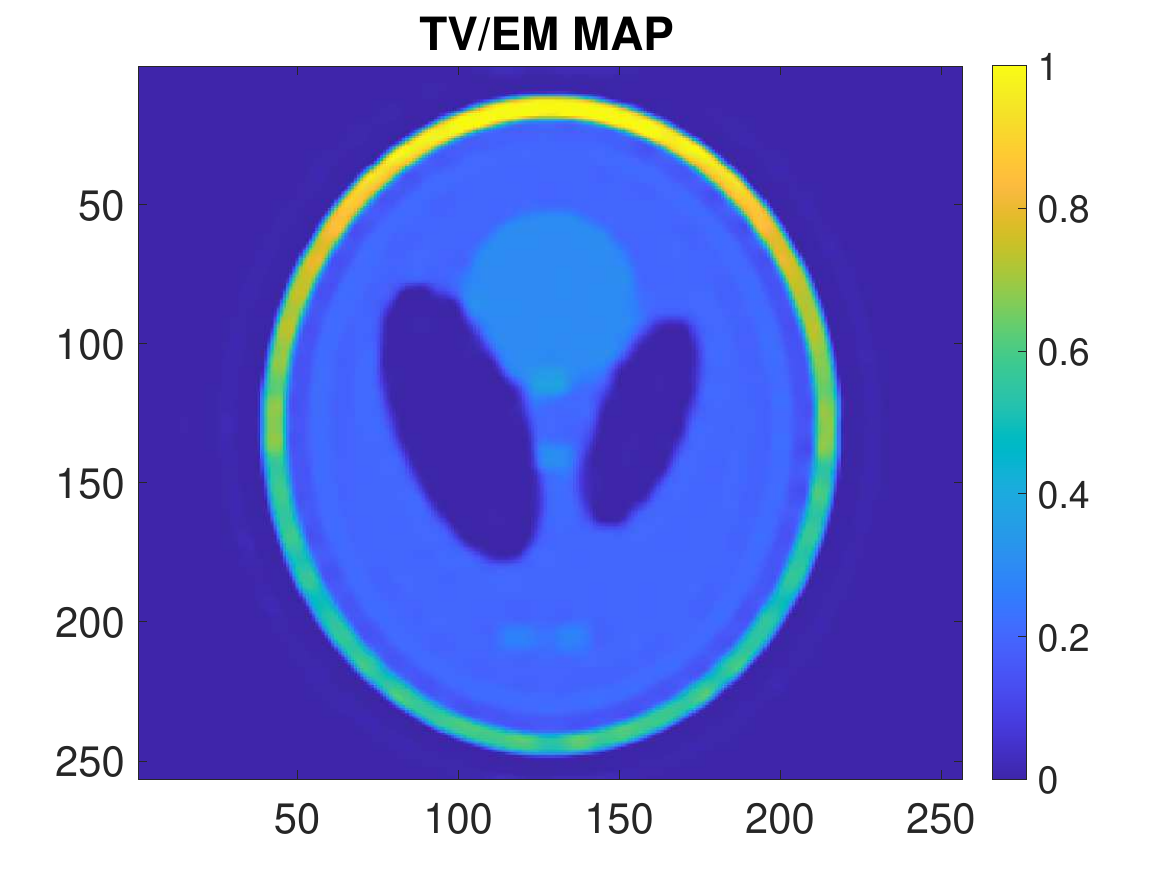}
	\includegraphics[width=0.24\textwidth]{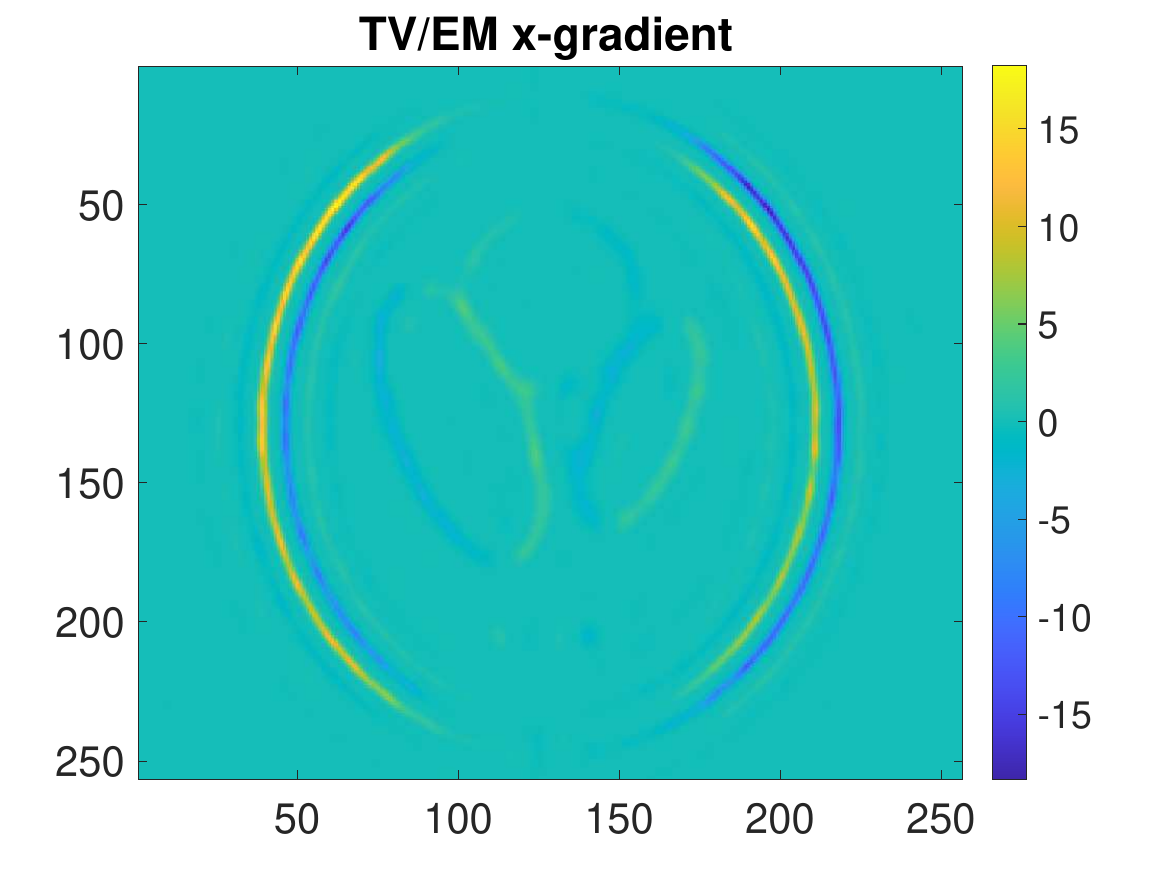}
	\includegraphics[width=0.24\textwidth]{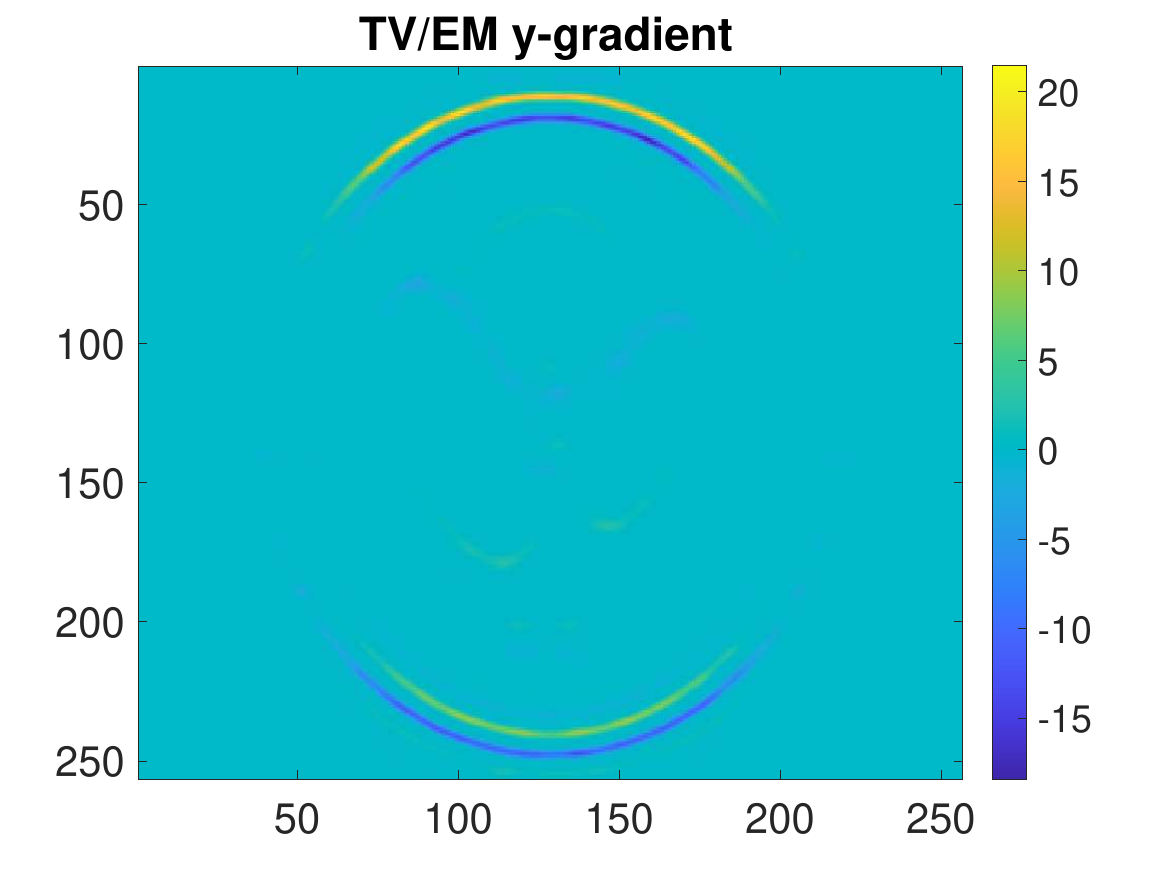} 
	\includegraphics[width=0.24\textwidth]{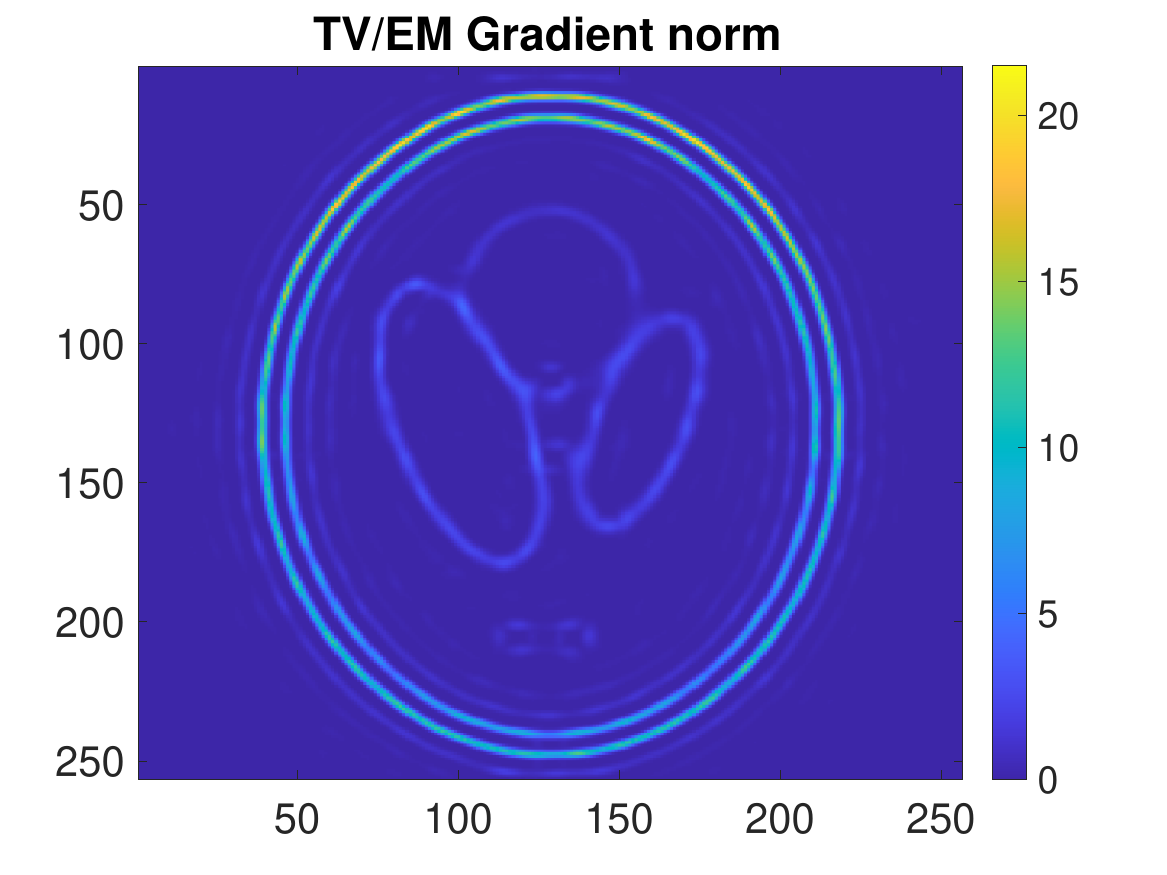}
	\\
	\includegraphics[width=0.24\textwidth]{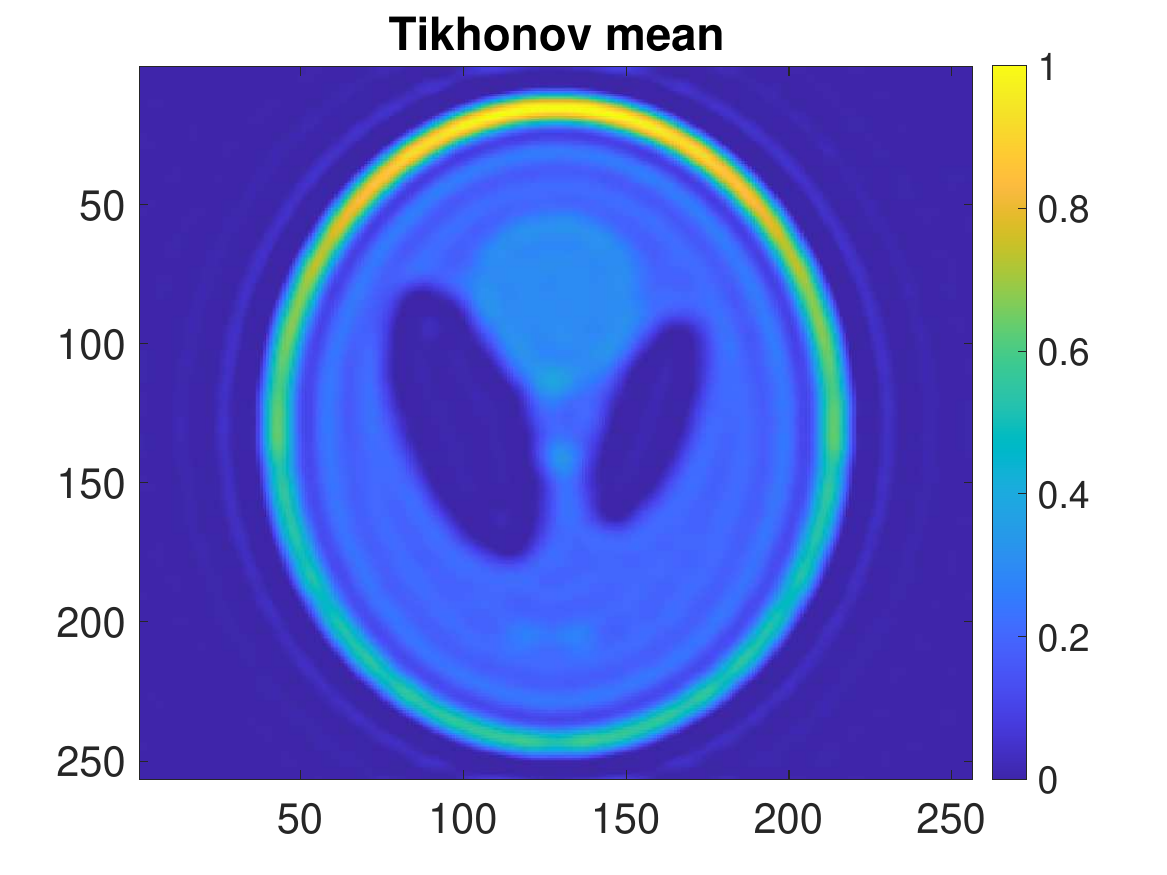}
	\includegraphics[width=0.24\textwidth]{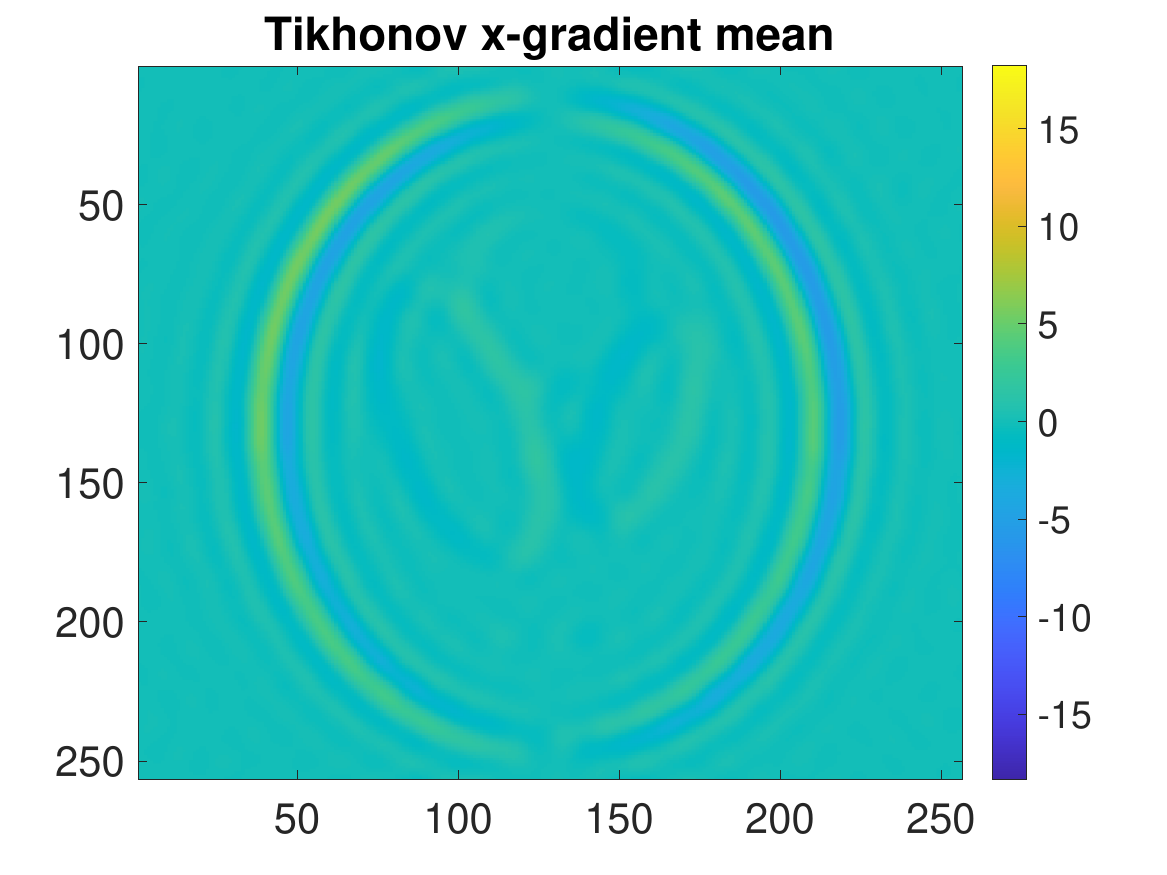}
	\includegraphics[width=0.24\textwidth]{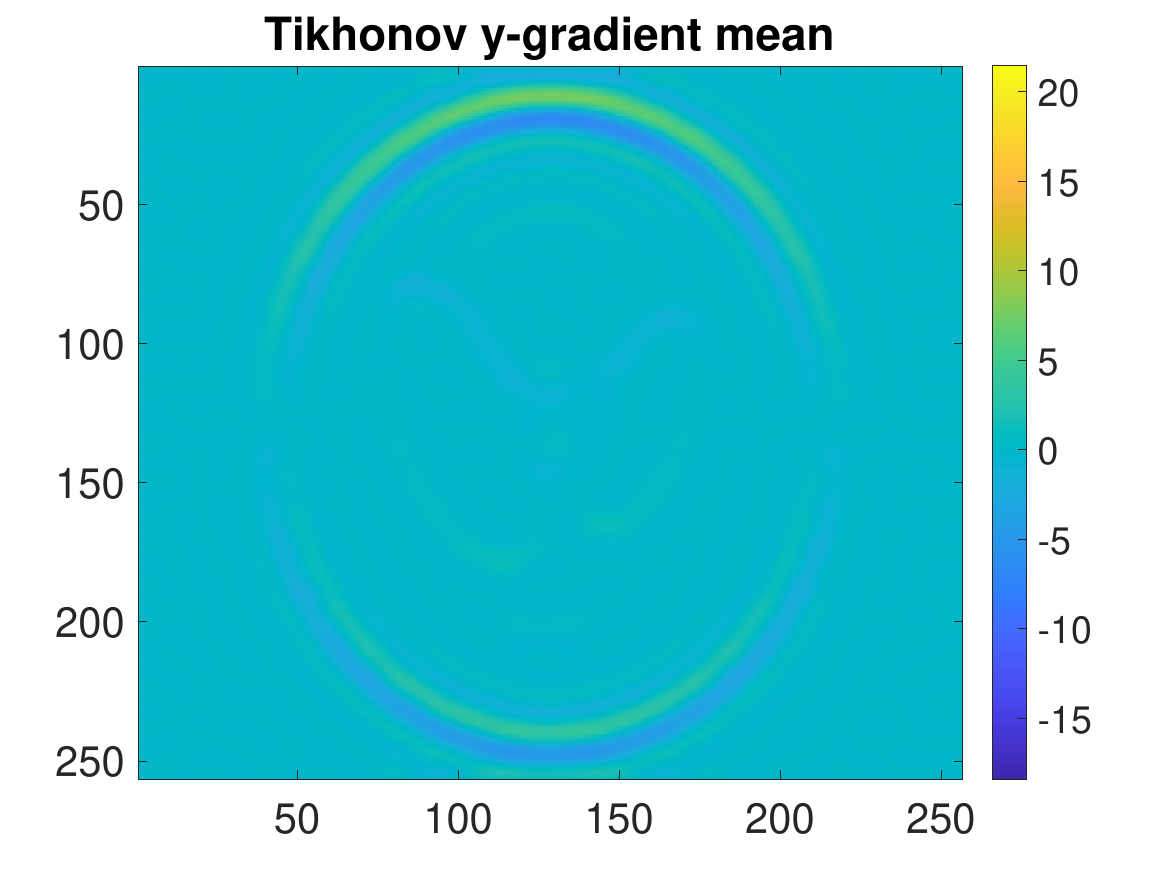} 
	\includegraphics[width=0.24\textwidth]{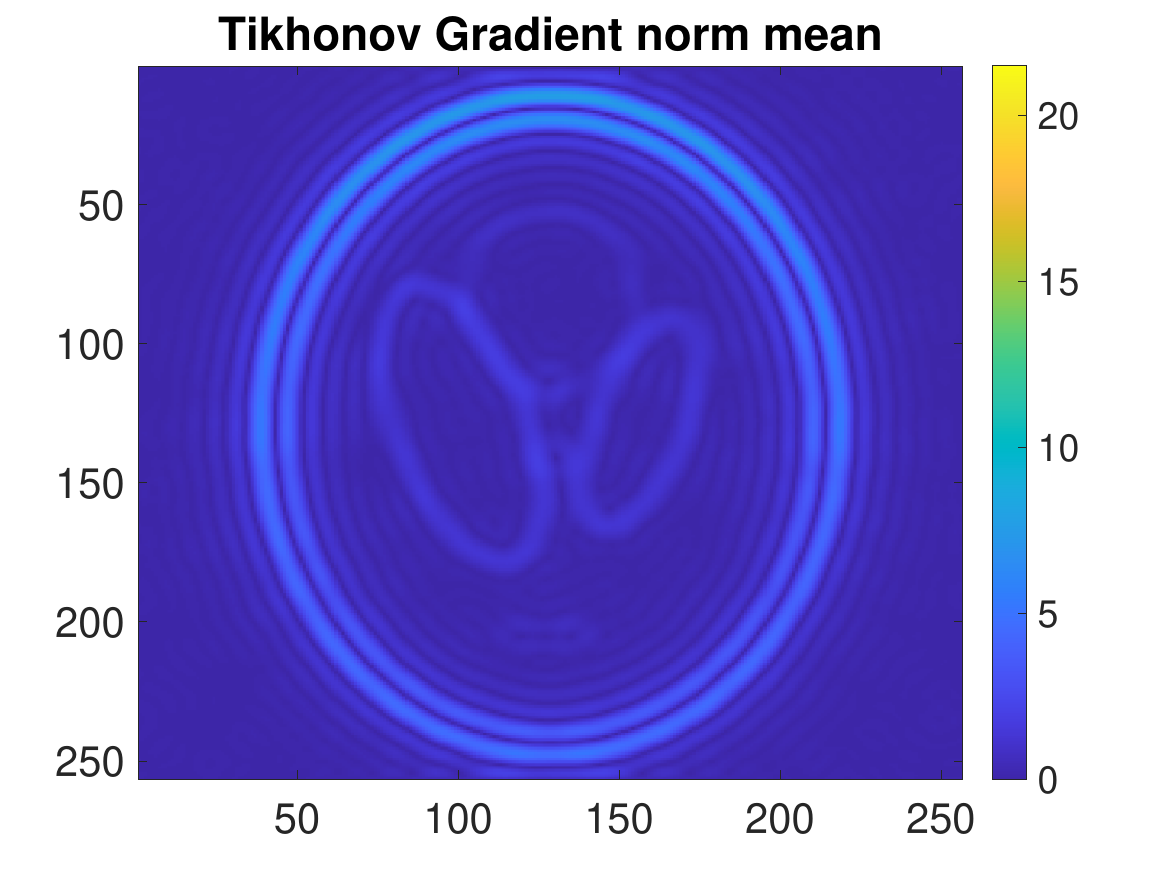}
	\caption{VBL TV-denoising illustrated on a high-resolution image of
	Shepp-Logan phantom.
	The top row features (from left to right) the truth, 
	blurry observations (every pixel), 
	the error w.r.t. VBL mean, and UQ in the form of 
	the standard deviation of the gradient norm of VBL (i.e. 
	$(\bbE (\beta_{1:\tilde{p}}-\bbE \beta_{1:\tilde{p}})^2 
	+ (\beta_{\tilde{p}+1:2\tilde{p}}-\bbE \beta_{\tilde{p}+1:2\tilde{p}})^2)^{1/2}$ 
	-- see the text).
	The second row features (from left to right) the VBL mean 
	($\bbE \tilde{\beta}$), 
	the x-gradient mean ($\bbE \beta_{1:\tilde{p}}$), 
	the y-gradient mean ($\bbE \beta_{\tilde{p}+1:2\tilde{p}}$),
	and the gradient mean norm
	($((\bbE \beta_{1:\tilde{p}})^2 + (\bbE \beta_{\tilde{p}+1:2\tilde{p}})^2)^{1/2}$).
	The next two rows correspond to the same quantities for the TV MAP 
	and the Tikhonov regularized problem.}
	\label{fig:VBL2}
\end{figure}

Now we will move to a higher dimensional example, with $p_0=256$.
Observe that the case $n=\tilde p$ cannot be handled
directly, which provides a testing ground for our sequential method.
However, it will be useful to have a ground truth, 
which is possible for appropriate choices of parameters. 
The calculation is provided in \cref{app:fouriertrunc}.
 The first set of experiments in \cref{fig:VBL2} shows the reconstruction
 corresponding to full osbervations, and illustrates that VBL is capable of 
 achieving edge sparsity as well as UQ.
 
\begin{figure}[!htbp]\centering
	\includegraphics[width=0.24\textwidth]{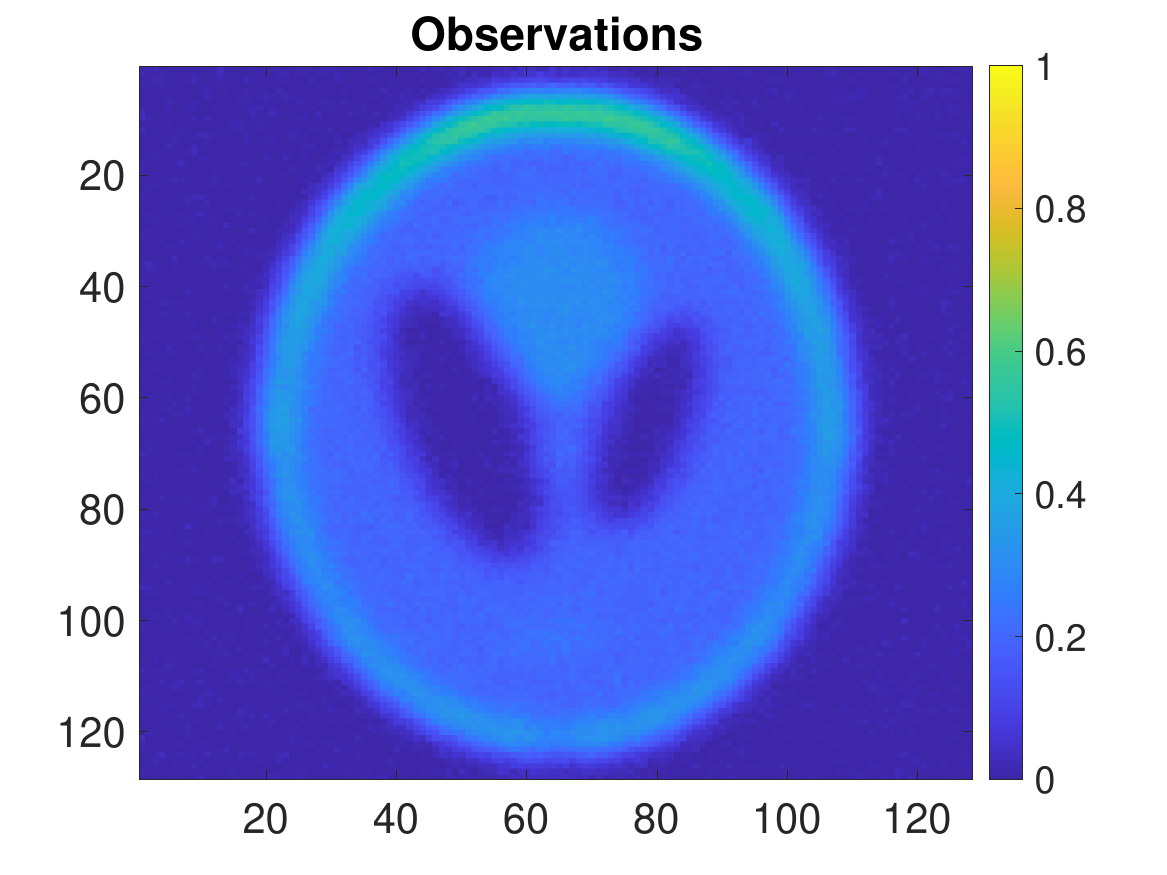}
	\includegraphics[width=0.24\textwidth]{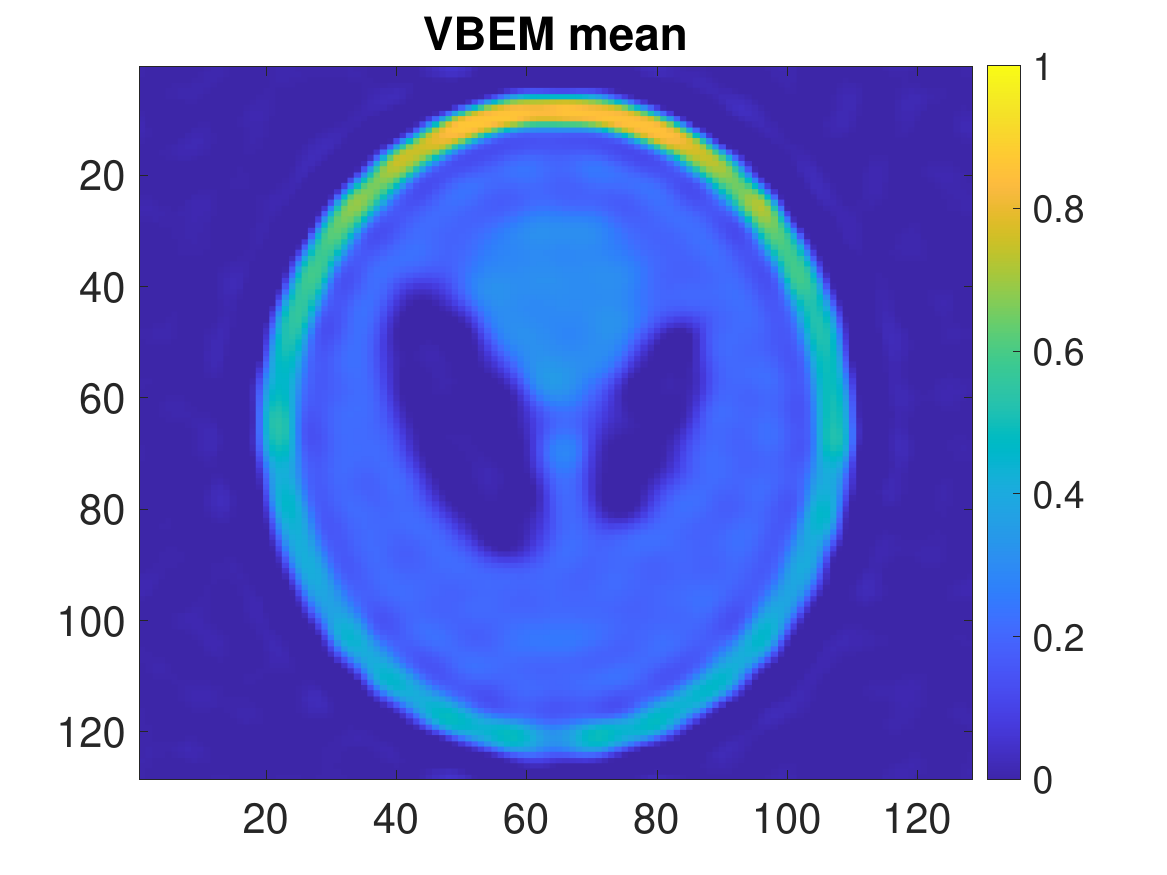} 
	\includegraphics[width=.24\textwidth]{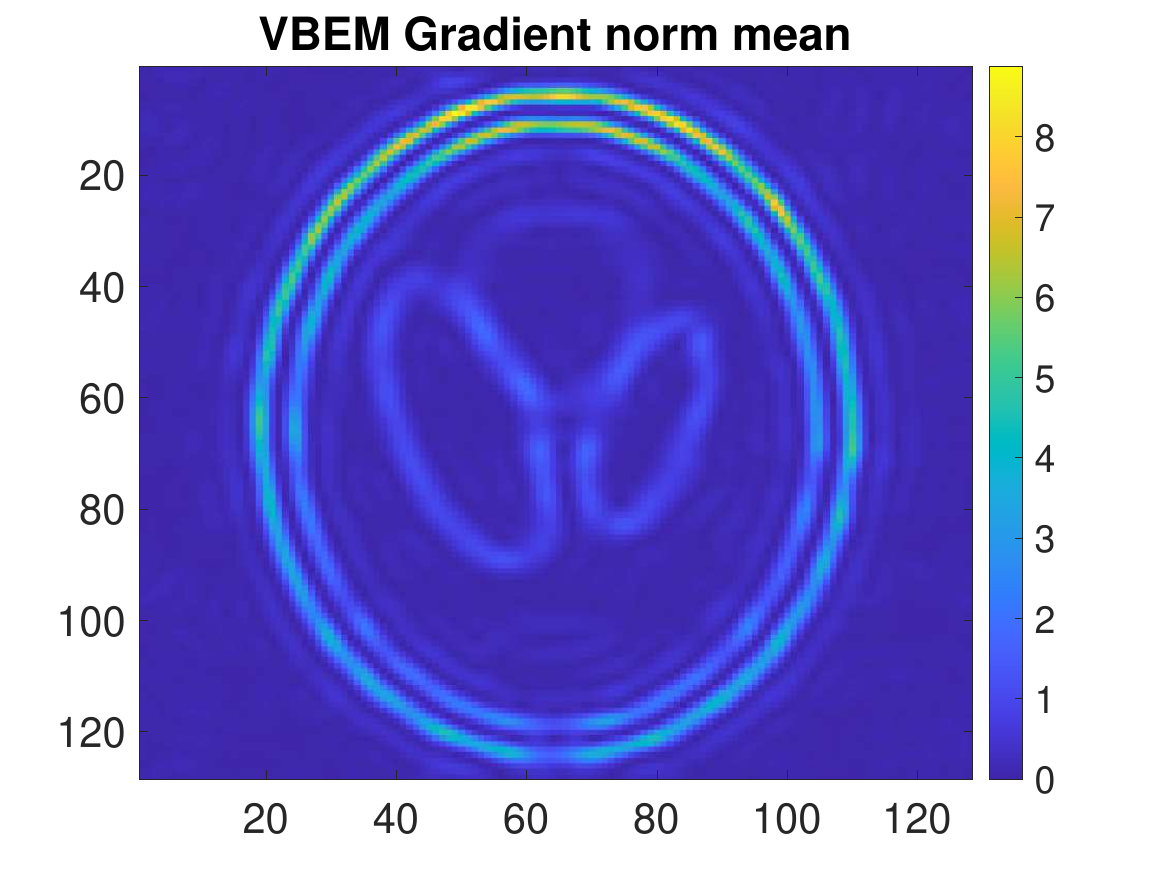} 
	\includegraphics[width=.24\textwidth]{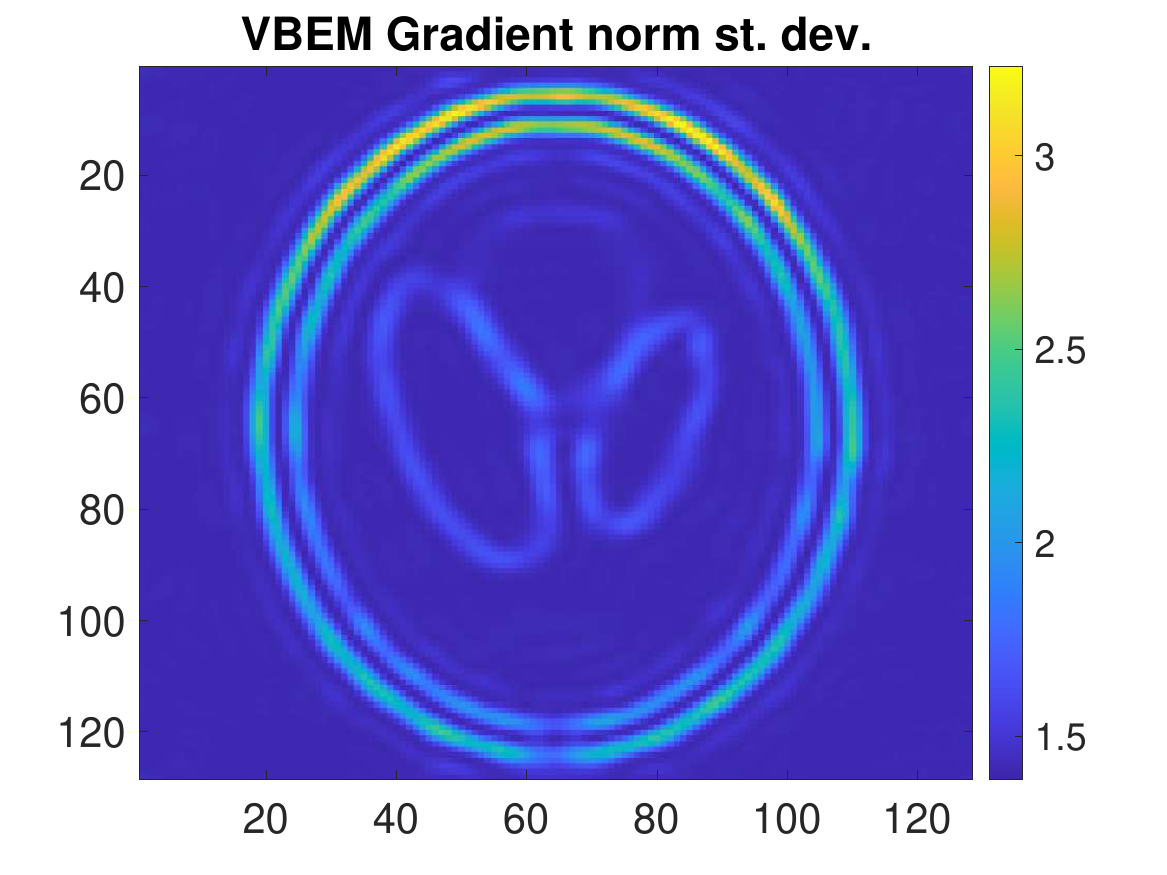}
	\\
	\includegraphics[width=.24\textwidth]{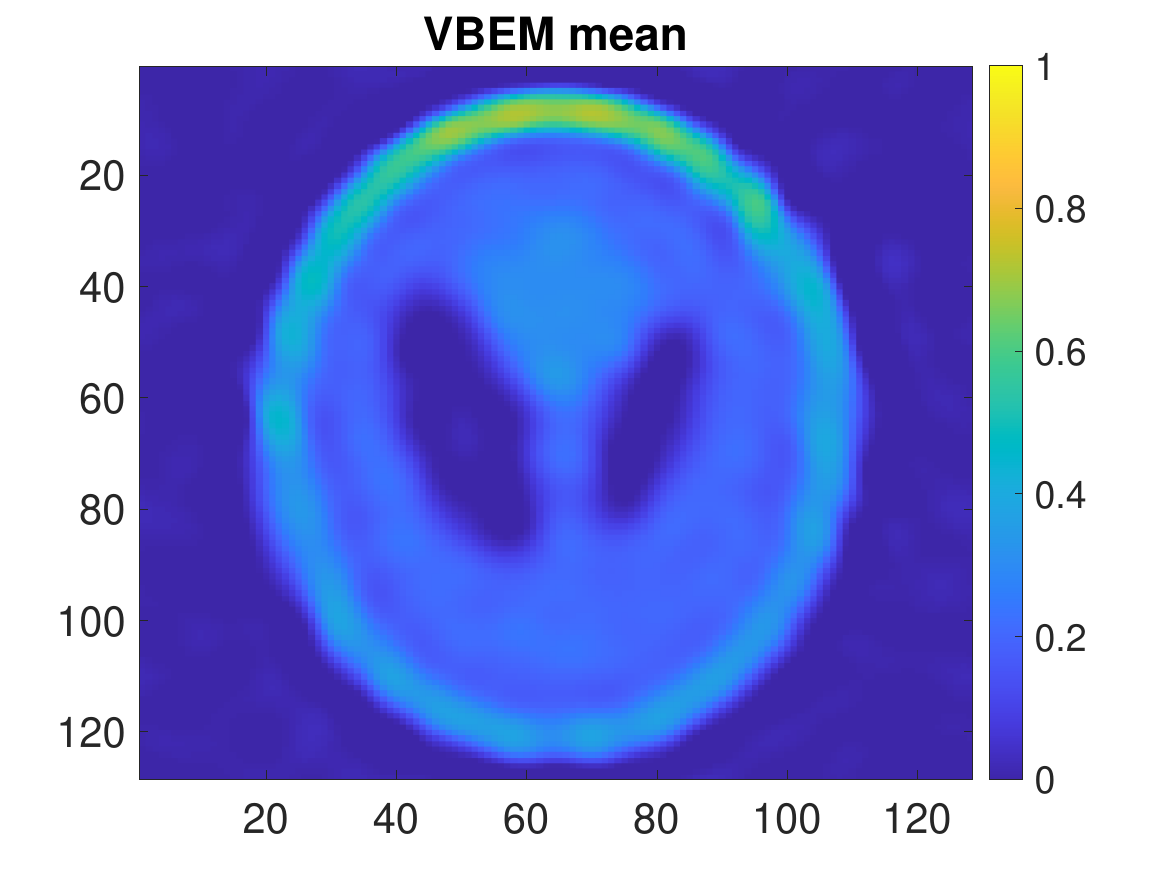}
	\includegraphics[width=0.24\textwidth]{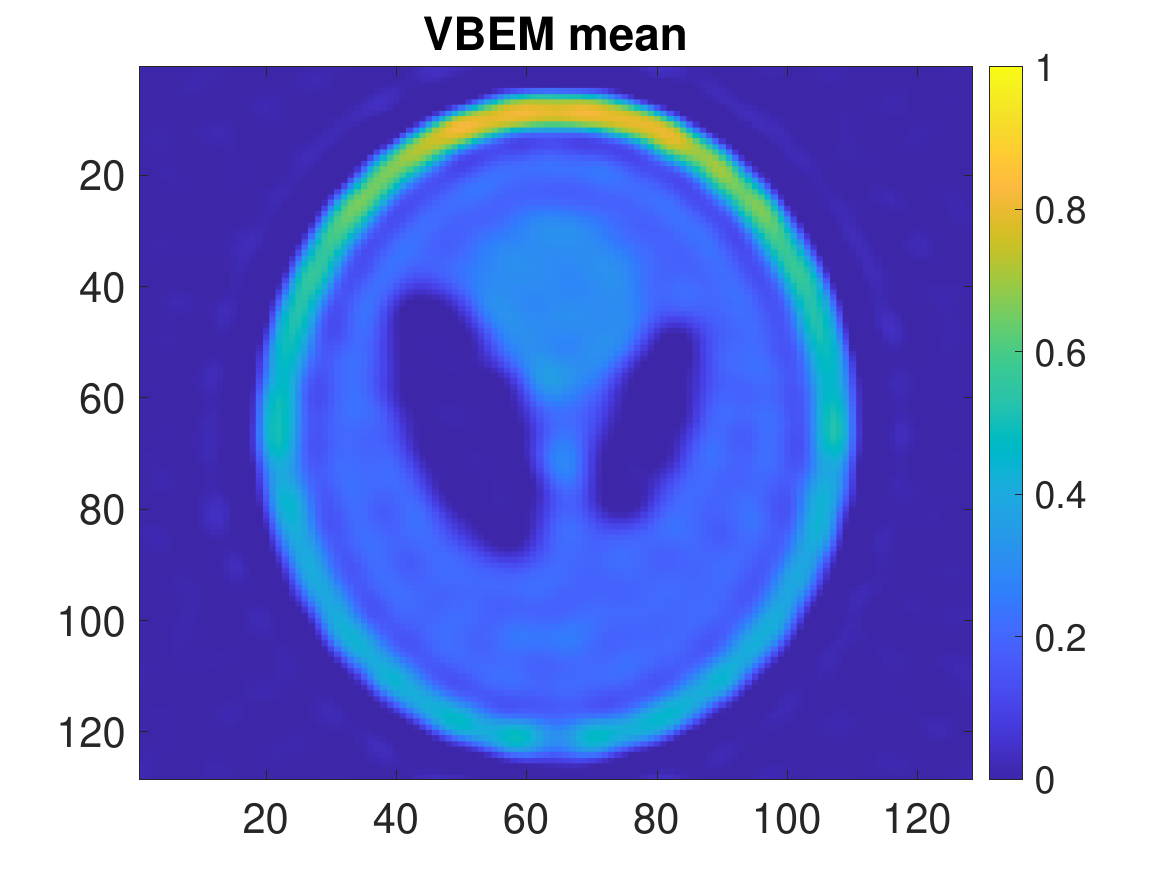}
	\includegraphics[width=0.24\textwidth]{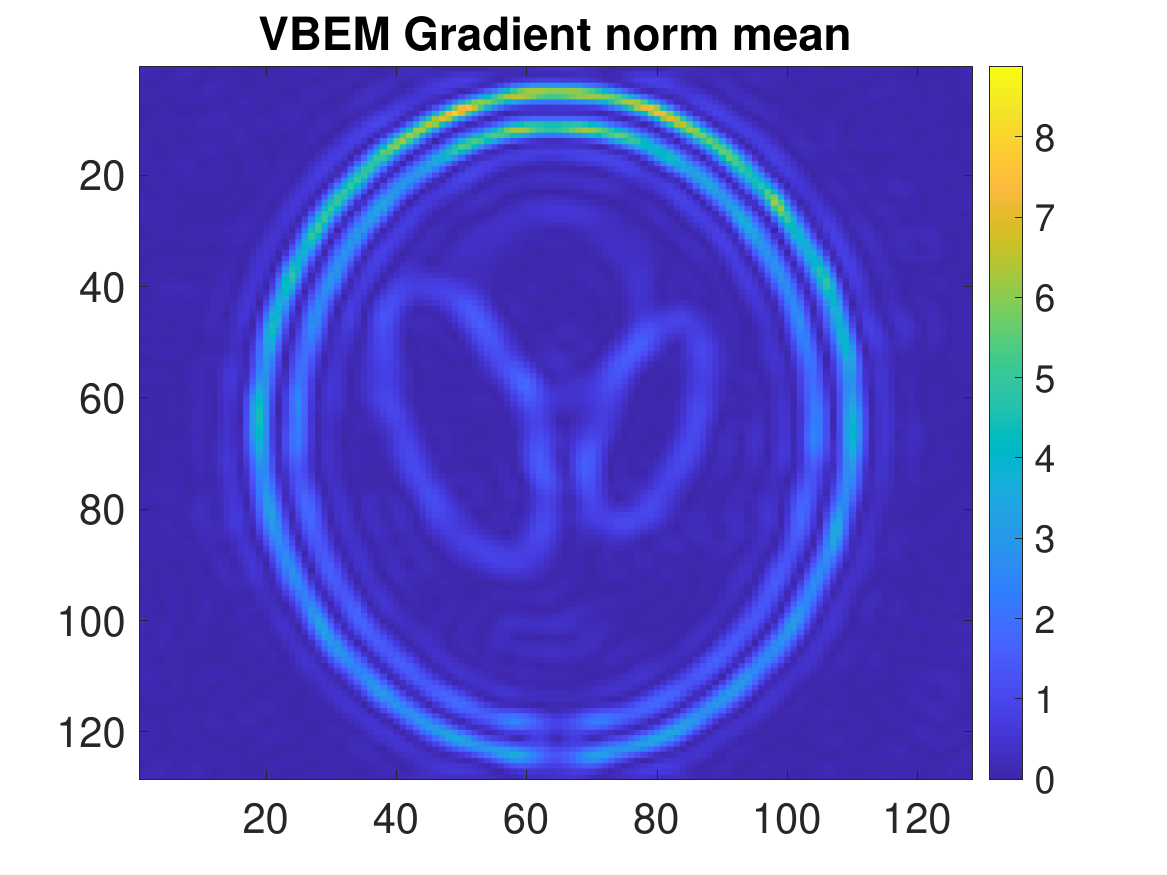} 
	\includegraphics[width=.24\textwidth]{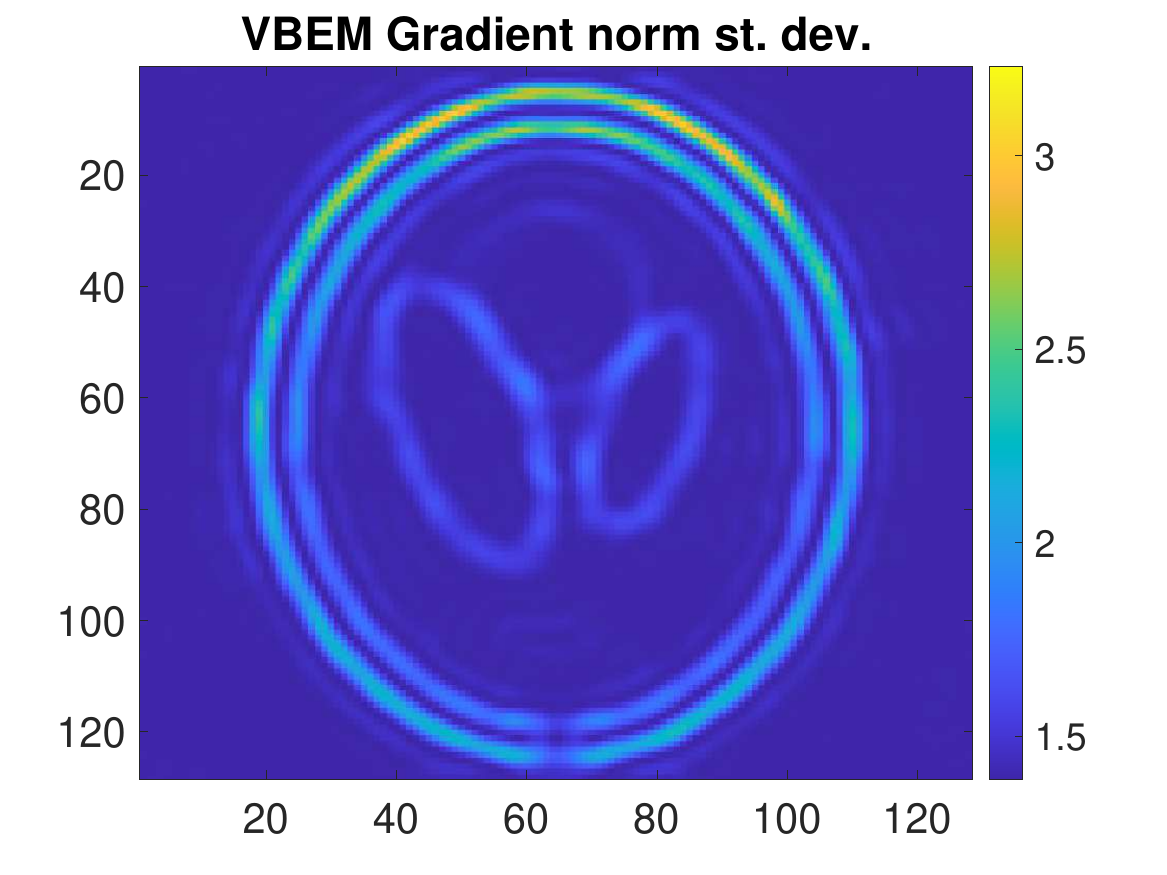}
	\caption{VBL TV-denoising (top row, from left to right -- observations, 
	mean, gradient mean, and gradient standard deviation) 
	in comparison to the recursive version as in \cref{alg:onlinevbl} (bottom row, from left to right -- mean after 1 iteration, 
	mean, gradient mean, and gradient standard deviation). $\omega=\gamma=0.01$.}
	\label{fig:online1}
\end{figure}

The next set of experiments is intended to illustrate two things.
First, in the case where the monolithic problem can be solved, as above,
the sequential version does a good job of getting close to the full monolithic
solution. For the choice of $\omega=\gamma=0.01$ and $\lambda=1$,
we recover a ground truth with 
$\tilde{n} = 1514$ for $\rho=0.8$ (see \cref{app:fouriertrunc}).
The relative $L^2$ error is $0.5$.
Letting $2M=1490$ for the recursive \cref{alg:onlinevbl}, 
a single iteration of $2M$ observations gives $0.58$, while
after completing the iterations, we get $0.52$. The results are shown in \cref{fig:online1}.

 \begin{figure}[!htbp]\centering
	\includegraphics[width=0.24\textwidth]{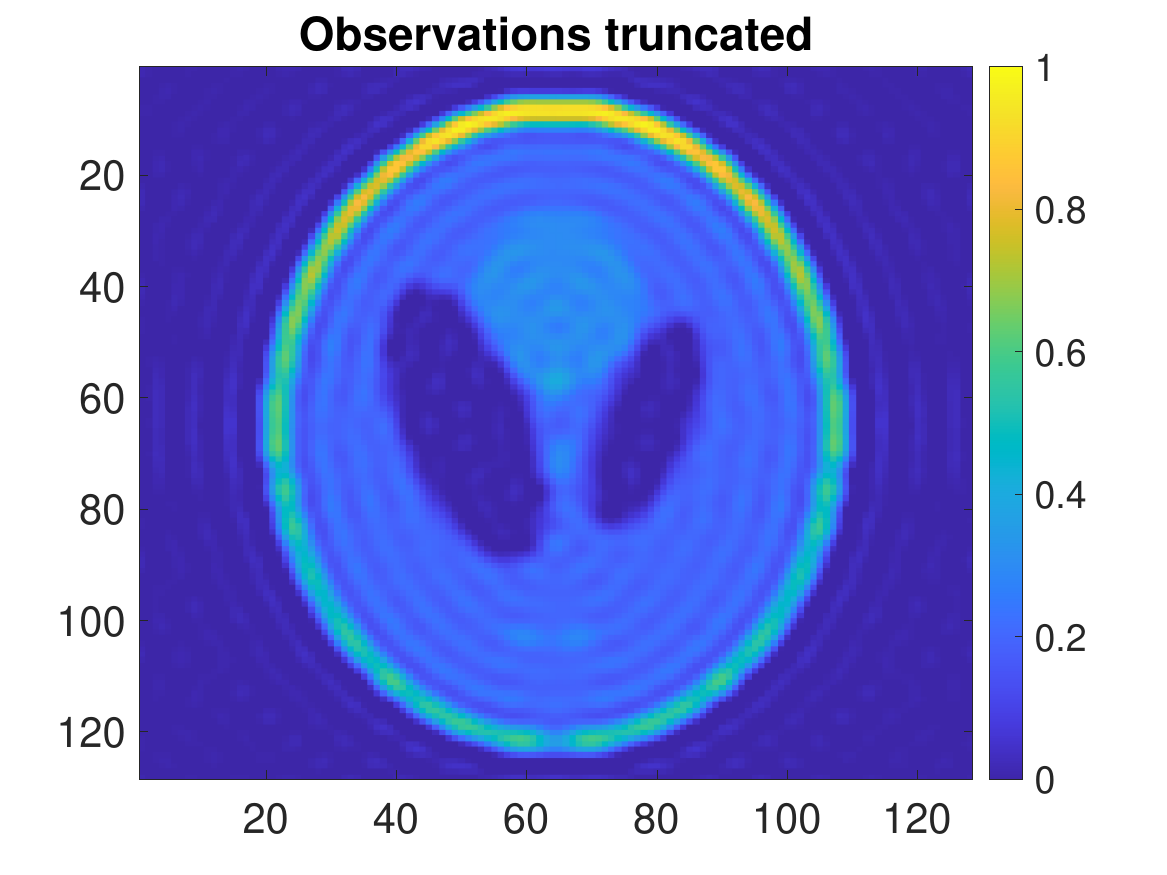}
	\includegraphics[width=0.24\textwidth]{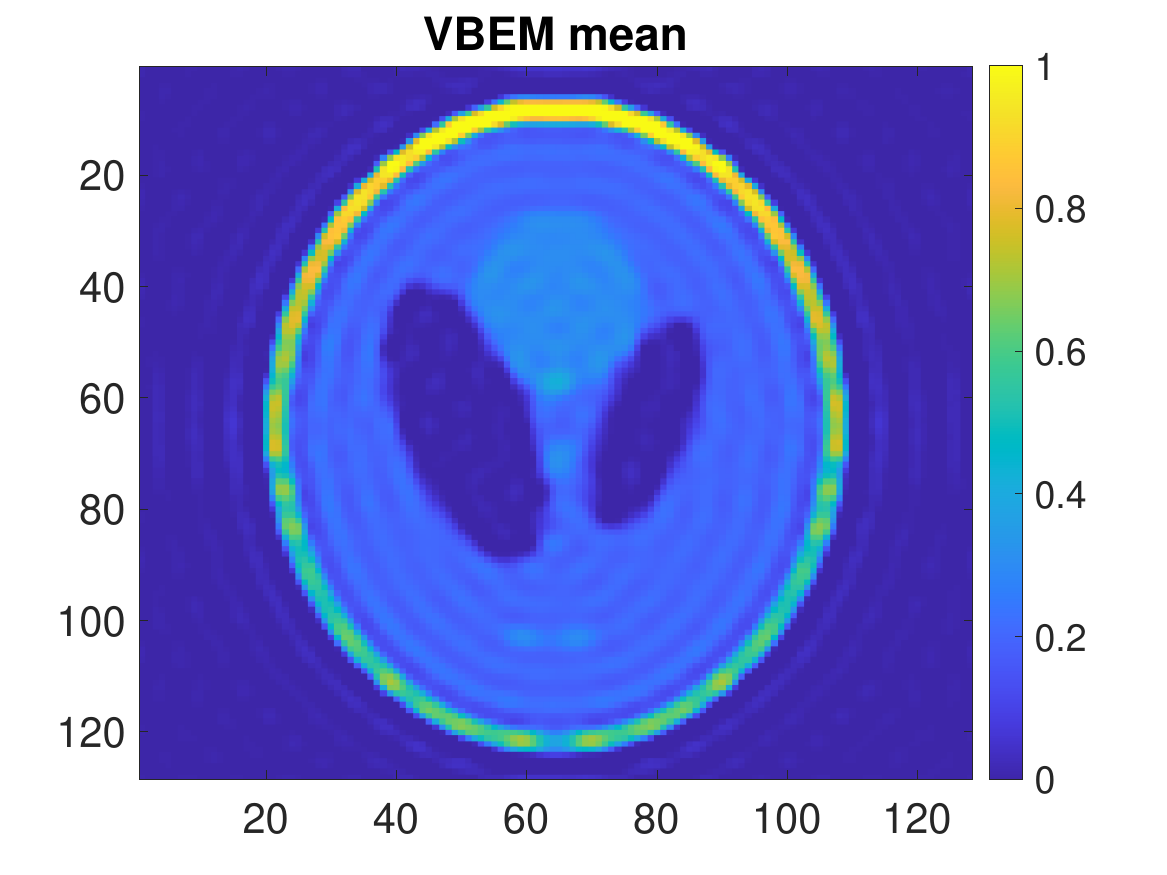} 
	\includegraphics[width=.24\textwidth]{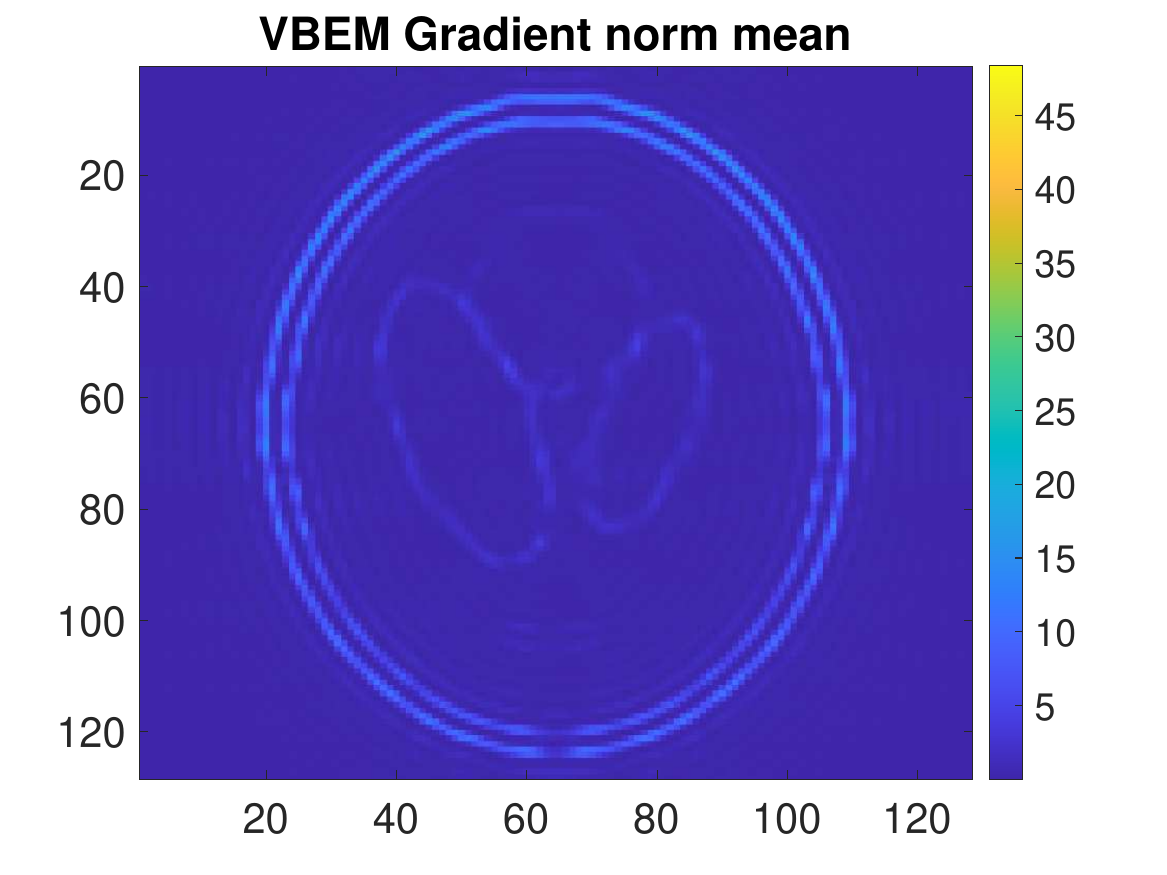} 
	\includegraphics[width=.24\textwidth]{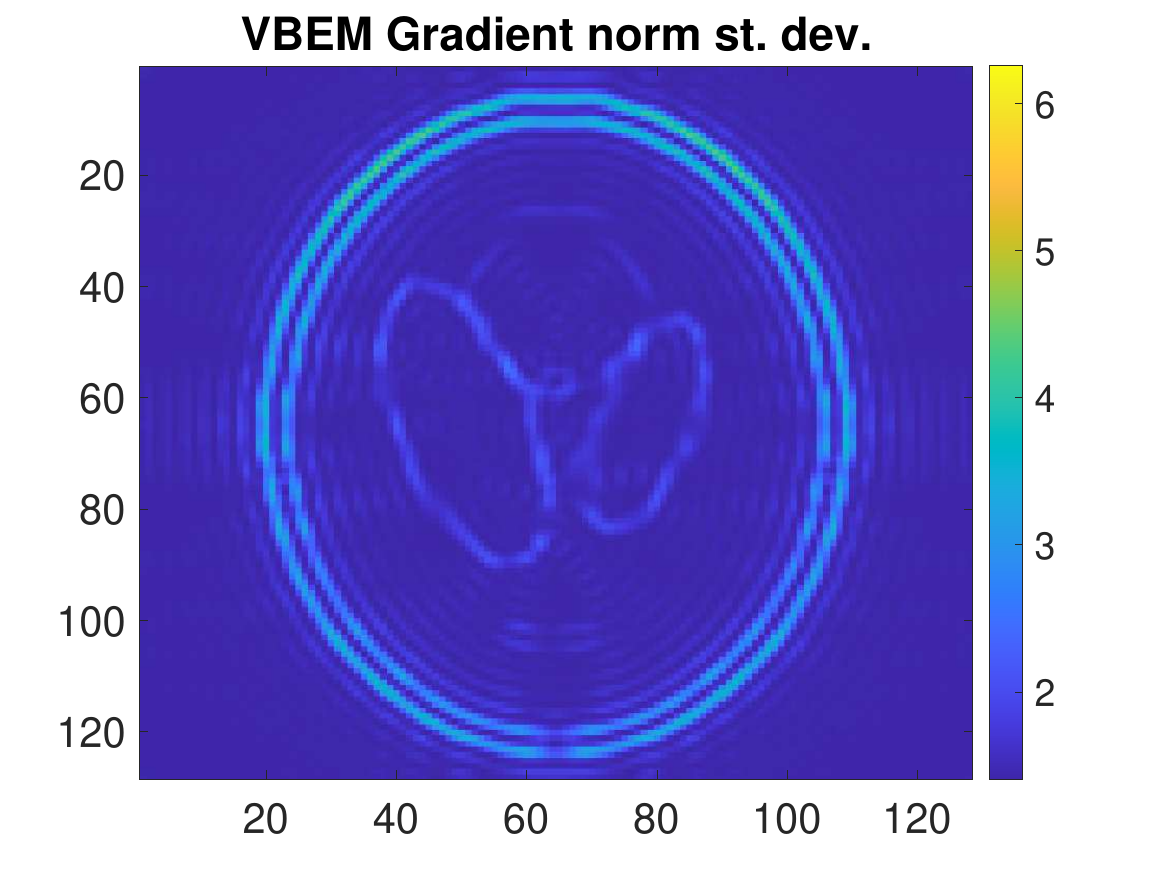}
	\\
	\includegraphics[width=.24\textwidth]{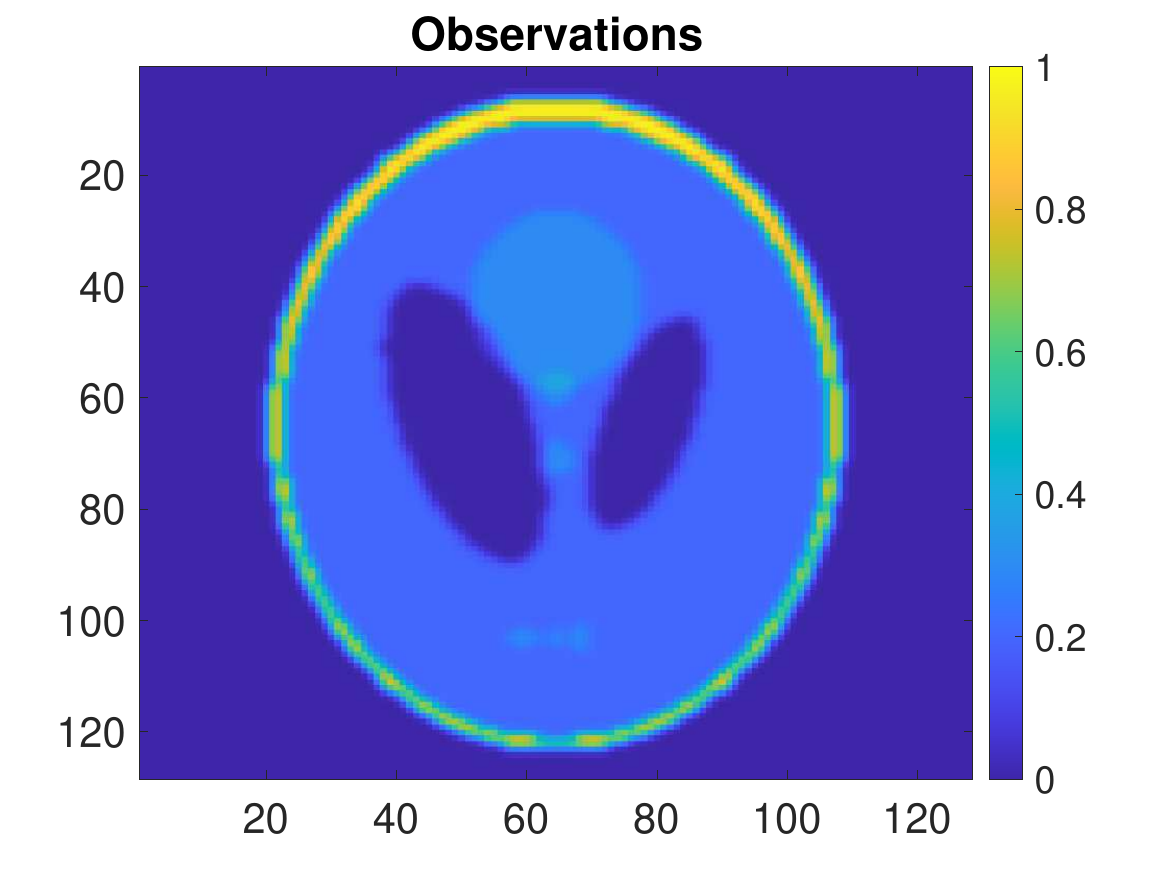}
	\includegraphics[width=0.24\textwidth]{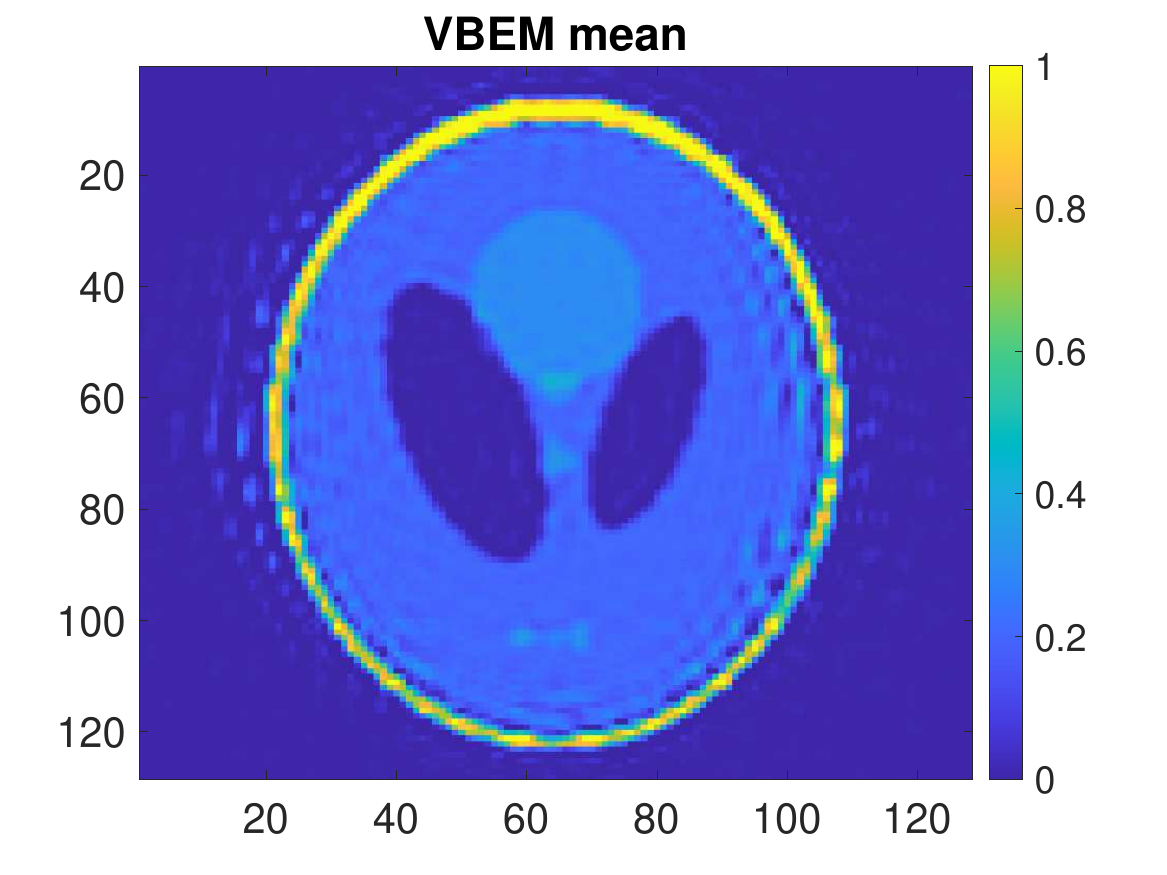}
	\includegraphics[width=0.24\textwidth]{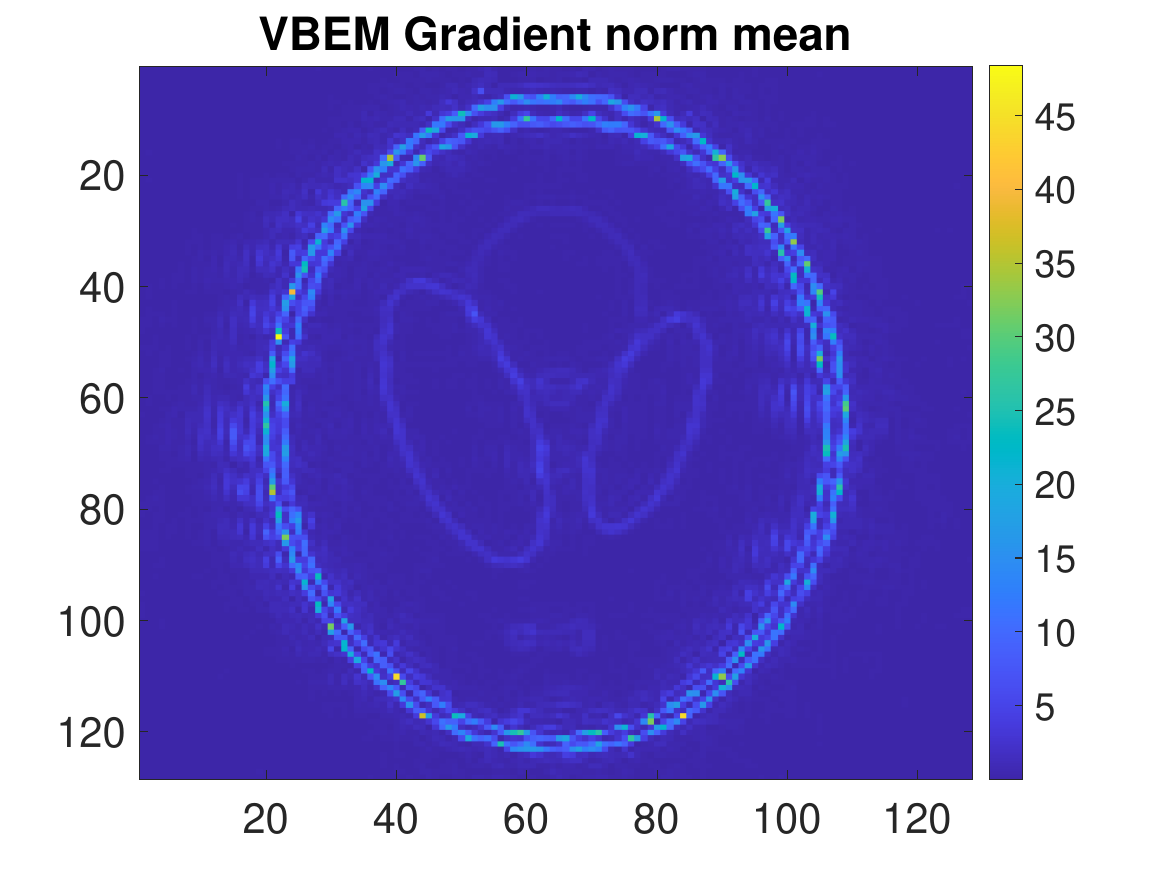} 
	\includegraphics[width=.24\textwidth]{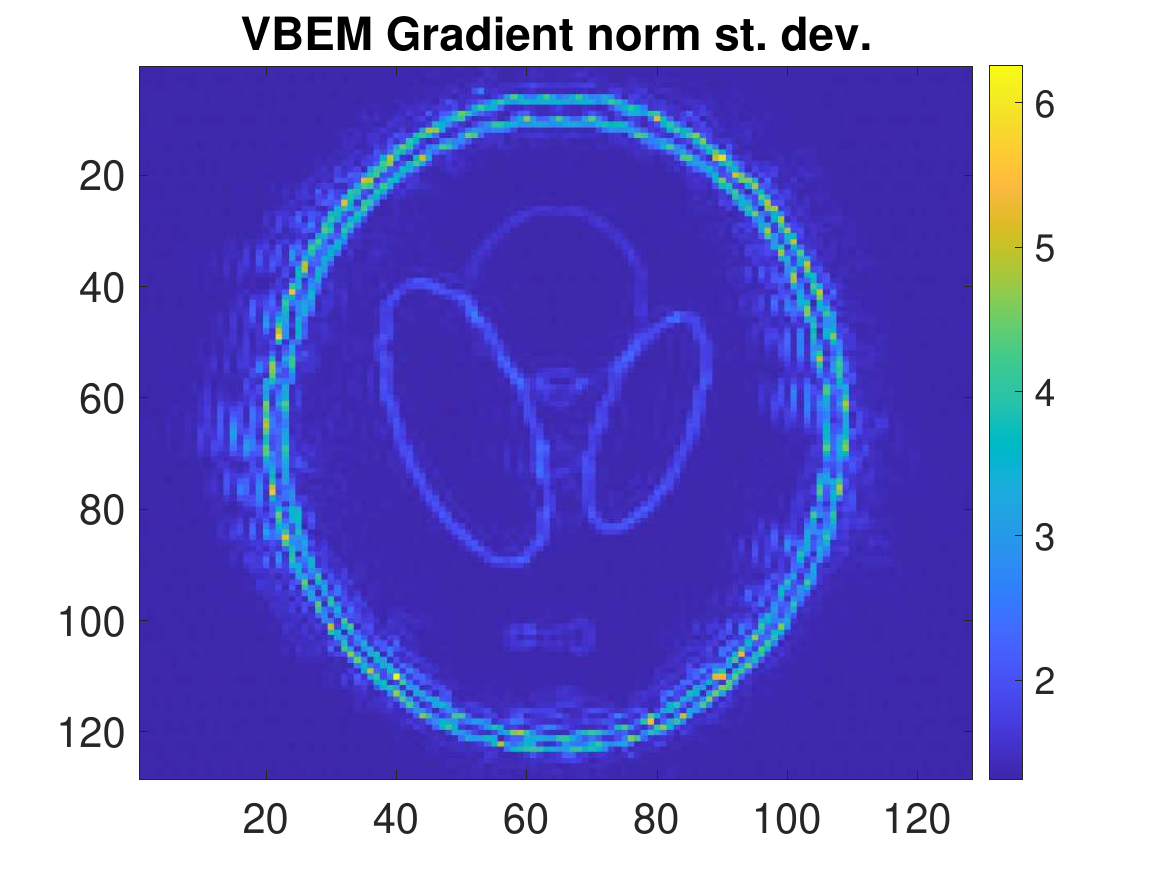}
	\caption{Truncated VBL TV-denoising (top row, from left to right -- 
	truncated observations, 
	mean, gradient mean, and gradient standard deviation) 
	in comparison to the recursive version as in \cref{alg:onlinevbl} (bottom row, from left to right -- observations, 
	mean, gradient mean, and gradient standard deviation). $\omega=\gamma=0.001$.}
	\label{fig:online2}
\end{figure}

Second, in the case of smaller values $\omega=\gamma=0.001$
where the monolithic problem cannot be solved and 
one must settle for either sparse observations or a truncation of $X$ 
above the desired threshold described in \cref{app:fouriertrunc},
the sequential version does significantly better than the 
monolithic approximation.
In this case, the desired threshold with $\rho=1$ would be $\tilde{n} \approx 16000$
which is not feasible. We use the coarse approximation with $\tilde{n} = 1640$
dominant modes, and achieve relative $L^2$ errors in the truncated observations of 
$0.1 \gg \gamma$ and in the solution $0.39$.
For the recursive implementation we let $2M=1640$, and achieve an error of $0.32$. 
The results are shown in \cref{fig:online2}. 
We notice in this case that, despite the fact that the reconstruction error is better and the edges
are more crisp, there is some strange radiation/noise in the recursive reconstruction.
It is a topic of further investigation to understand this better (and remove it).

The method is able to handle a larger problem with $p=10^6$ and we were able to 
assimilate $n=500$ batches of size $M=200$ overnight, reducing the relative
error from $0.76$ with a single pair of batches to $0.57$ and yielding a reasonable 
looking reconstruction. These results are not shown.
 
\begin{rem}
Note that we impose a sparsity constraint on $\beta_0$ as well, which is slightly
different from TV. This constraint could be easily removed but our aim is not 
to belabour the finer points of TV-denoising and rather to illustrate our method on this example.
\end{rem}

\section{Conclusion}

\note{
Here a variational Bayesian approach is adopted for solution
of Normal-Generalized-Inverse-Gaussian scale mixture models,
which includes some existing and some new models.
It is shown that the method delivers UQ at a cost much less than
fully Bayesian models, as well as comparable accuracy and variable 
selection capabilities. The method is presented in a condensed
and digestible form, and supplemented with an easy-to-implement 
code package, which will make this technology accessible to the wider
science and engineering community.
It is shown how it can be implemented online, which facilitates either batch
processing of data or streaming data, for example in the context of
sequential experimental design of computer simulations.
Furthermore, an approximation is presented which is able to 
recover comparable results for a linear cost in the number of 
parameters $p$. 
The method is implemented on several real and simulated
datasets, including a challenging high-dimensional image-deblurring 
example with $p=10^6$ and $n=10^5$.
It is compared with competing methods, where it is shown
to perform favourably. In particular, it provides a nice balance of
speed, accuracy, UQ, and ease-of-implementation. 
A parallel version will be presented in future work.}

\vspace{15pt}
\noindent{\bf Acknowledgements.} KJHL and VZ were supported by The Alan Turing Institute 
under the EPSRC grant EP/N510129/1.
KJHL and VZ were also supported in part by the U. S. Department of Energy, Office of Science, Office of Fusion Energy Sciences and Office of Advanced Scientific Computing Research through the Scientific Discovery through Advanced Computing (SciDAC) project on Advanced Tokamak Modeling under a contract with Oak Ridge National Laboratory.

\bibliographystyle{plain}
\bibliography{referencesRRLS,references}

\appendix

\section{Derivations related to EM}\label{ap:em}

Following from \cref{eq:qb}, 
note that (i) ${Q}(\beta|\beta^t) \leq \log \bbP(Y,\beta)$ and 
(ii) ${Q}(\beta^t|\beta^t) = \log \bbP(Y,\beta^t)$.
 Therefore 
\begin{eqnarray*}
\log \bbP(Y,\beta^{t+1}) & = & 
{Q}(\beta^{t+1}|\beta^t) + \log \bbP(Y,\beta^{t+1}) -  {Q}(\beta^{t+1}|\beta^t) \, \\
& \geq & {Q}(\beta^{t+1}|\beta^t) \, \\
& = & {Q}(\beta^{t+1}|\beta^t) - {Q}(\beta^t|\beta^t) + \log \bbP(Y,\beta^t) \, \\
& \geq & \log \bbP(Y,\beta^t) \, .
\end{eqnarray*}
The first inequality arises from property (i), the equality comes from property (ii)
and the final inequality is due to the optimality of $\beta^{t+1}$.
This shows that the EM algorithm provides a non-decreasing algorithm for the optimization of 
$\log \bbP(Y,\beta^{t+1})$. 
It is a particular case of what have come to be known as majorization minimization algorithms \cite{hunter2004tutorial}.

The full calculation of \cref{eq:e} is given by
\begin{eqnarray}\nonumber
Q(\beta | \beta^t) &:=& -\int \log(\bbP(\beta,Y_n | \theta, X_n)) \bbP(\theta | \beta^t, X_n,Y_n) d\theta +\kappa(\beta^t, X_n,Y_n) \\ \nonumber
&=& -\int \left(\log(\bbP(\beta| \theta))+\log(\bbP(Y_n | \beta, X_n))\right) \bbP(\theta | \beta^t, X_n,Y_n) d\theta +\kappa(\beta^t, X_n,Y_n) \\ \nonumber
&=& \frac12 \sum_j \beta_j^2 \bbE\left [ \frac{1}{\theta_j} \Big | \beta^t, X_n,Y_n \right ]  + 
\frac12\bbE\left [ \log (\theta_j) \Big | \beta^t, X_n,Y_n \right ] + \frac1{2\gamma^2} |Y_n - X_n \beta|_2^2
+\kappa(\beta^t, X_n,Y_n) \\
\nonumber
&=& \frac12 \beta^T D(1/\theta^t) \beta  + 
\frac1{2\gamma^2} |Y_n - X_n \beta|_2^2 + 
\kappa(\beta^t, X_n,Y_n) \, ,
\end{eqnarray}
\note{where $\kappa$ is a generic constant which changes from line to line 
and absorbs all irrelevant terms}.

\section{Discussion of other iterative methods for MAP estimation}
\label{app:othermap}

It is worth briefly discussing other standard iterative optimization algorithms for solving %
quadratic optimization problems.
In particular, gradient descent and quasi-Newton methods are very promising alternatives in the case where
the design matrix $X_n$ is sparse. 
This is not the primary context of interest in the present work, 
so the general case is discussed.
Gradient descent methods achieve {\em linear convergence}, 
which means that in terms of iterations the complexity is logarithmic in the desired accuracy
\cite{nocedal}, however the rate can get very close to one, particularly in high dimensions,
resulting in very slow convergence in practice.
Computation of the gradient %
incurs a cost of $\cO(np)$,
and if one uses a conjugate gradient approach \cite{nocedal, hestenes1952methods} 
(ensuring that each successive search direction is orthogonal to all previous ones) 
then the number of iterations required for convergence to {\em the exact solution} is bounded
above by $p$, i.e. the memory and computational complexity is no worse than the monolithic approach.
Stochastic gradient descent alleviates $n-$dependence per iteration 
by using an unbiased estimate of the gradient, 
i.e. a batch of $b=\cO(1)$ data is used at each iteration with a cost of $\cO(p)$. 
Under appropriate assumptions, this approach can converge \cite{kushner, gower},
but there are no tight theoretical complexity bounds. 
In the machine learning literature, one
often refers to {\em epochs}, or sweeps (plural) through the full data set, 
so one can expect a complexity of at least $\cO(np)$. 
An alternative in similar spirit is the (randomized) Kaczmarz algorithm \cite{kaczmarz, strohmer},
which also enjoys a per iteration cost of $\cO(p)$, 
but would also typically require $\cO(n)$ 
iterations until convergence. 
The latter may be improved with a $\cO(np)$ pre-processing step
which replaces a uniform distribution on the data with one scaled by the row norms of $X_n$.

Of course none of these methods provides an uncertainty estimate. 
Quasi-Newton methods, such as BFGS, provide an approximation of the covariance
$C_n$ from equations \cref{eq:monolithC}, \eqref{eq:KFC} as well as super-linear convergence.
In this case, one has a per iteration complexity cost of $\cO(p(n+p))$, 
and a memory requirement of $\cO((n+2k)p)$, for $k$ iterations 
(a rank 2 update to the approximation of the Hessian and its inverse is performed at each iteration). 
The limited memory alternative limits $k \leq k_{\rm max}$.
One expects the method to converge very rapidly, for $k =\cO(1)$, so it can still be competitive.
A one-off cost of $\cO(np^2)$ to compute $X_n^TX_n$ and $X_n^TY_n$
can reduce the $n-$dependence of either method to a $p-$dependence.
Furthermore, the computation of $X_n^TX_n$ can be split into $n/m$ batches of size $m$
to be computed in parallel, yielding $\cO(mp)$ memory and $\cO(p^2m)$ computation cost for each,
for the price of an additional $\cO(p^2n/m)$ cost to combine at the end.

\section{Ensemble Kalman filter formulation}
\label{app:enkf}

In an online context, the Kalman filter provides recursive equations below,
analogous to \cref{eq:monolith} and \cref{eq:monolithC},
for either the covariance or the precision
\begin{eqnarray}\nonumber
m_n &=&  \left(\frac1{\gamma^2}x_nx_n^T + C_{n-1}^{-1}\right)^{-1}
\left(\frac1{\gamma^2}x_n y_n + C_{n-1}^{-1}m_{n-1}\right) \\ \label{eq:KF}
&=& m_{n-1} + C_{n-1} x_n \left(\gamma^2 + x_n^TC_{n-1}x_n\right)^{-1}\left(y_n - x_n^T m_{n-1}\right) \,  , \\
C_n &=& \left(\frac1{\gamma^2}x_nx_n^T + C_{n-1}^{-1}\right)^{-1} = %
C_{n-1} - C_{n-1} x_n \left(\gamma^2 + x_n^TC_{n-1}x_n\right)^{-1} x_n^T C_{n-1} \, .
\label{eq:KFC}
\end{eqnarray}
We have the following incremental update formula for \cref{eq:ctheta}, 
which incurs a cost of $\cO(p)$
\begin{equation}\label{eq:prectheta}
(C_{n}^{\theta})^{-1} = (C_{n}^{\theta'})^{-1} - D(1/\theta') + D(1/\theta) \, .
\end{equation}
Unfortunately, the solution of \cref{eq:mtheta} requires inversion of a 
$p\times p$ to compute \eqref{eq:KF}, at a premium cost of $\cO(p^3)$ 
(for exact solution and in the absence of sparsity).

In this context it is natural to consider the ensemble Kalman filter
as a low-rank and cost-efficient alternative.
The EnKF
was introduced in \cite{evensen1994sequential, burgers1998analysis} and has since exploded in popularity, 
largely due to its remarkable success in providing an efficient approximation to the Kalman 
filter in very high dimensional geophysical applications. 
Many versions of EnKF exist, but in this case the version which makes the most sense
is the deterministic, or square root, EnKF \cite{law2015data}.
The method is initialized with an ensemble
$\beta_0^{(1)},\dots, \beta_0^{(K)} \sim N(m_0,C_0)$, 
and then the Kalman filter equations \eqref{eq:KF} are replaced with 
the following, for $n \geq 1$
\begin{align}
m_{n-1} &= \frac1K \sum_{i=1}^K \beta_{n-1}^{(i)} \, , \quad 
C_{n-1} = \frac1K \sum_{i=1}^K ( \beta_{n-1}^{(i)} - m_{n-1} )( \beta_{n-1}^{(i)} - m_{n-1} )^T \, , \label{eq:enave} \\
\widehat m_{n} &= m_{n-1} + 
C_{n-1} x_{n} (\gamma^2 + x_{n}^TC_{n-1}x_{n})^{-1}(y_{n} - x_{n}^T m_{n-1}) \,  , \notag \\
\widehat C_{n} &= (I_p - C_{n-1} x_{n} (\gamma^2 + x_{n}^TC_{n-1}x_{n})^{-1})C_{n-1} \, , \notag \\
\beta_{n}^{(i)} &\sim \cN(\beta \, ; \, \widehat m_{n}, \widehat C_{n}) \, , \quad i=1,\dots, K \, , \label{eq:ensamp}
\end{align}

The most common regime of application is $K \ll p$, 
which admittedly looks dubious from a statistical perspective. 
However, the cost of this method is now $\cO(Kp)$ in both computation and memory,
so the impetus is clear from a purely computational perspective. 
The remarkable thing is that it actually often works quite well, although
we note that the more common regime of application is dynamical systems
in which some particle-wise (often nonlinear) forward propagation occurs in 
between \cref{eq:ensamp} and \cref{eq:enave}. 
The stochastic version can be used directly in the absence of the sparsity considerations of \cref{sec:sparse}. 
However, in order to use the identify \cref{eq:prectheta} 
we need the precision. 
One potential, and common, solution is to modify/inflate \cref{eq:enave} 
for some small $\epsilon>0$ with
\begin{equation}\label{eq:pertcov}
C_n = \frac1K \sum_{i=1}^K ( \beta_n^{(i)} - m_n )( \beta_n^{(i)} - m_n )^T + \epsilon I_p \, .
\end{equation}
In our case, however, there is by design a more sensible choice of approximation
by a diagonal matrix plus low-rank correction. 
The whole program can be carried out, but due to this fact, we will not consider EnKF further here.
Note that such adjustments, known generally as covariance inflation 
in the data assimilation literature \cite{law2015data}, 
prevent convergence of the model to the Kalman filter
in the limit of an infinite sample size, so exactness is lost.

\section{Full observations Fourier truncation for TV denoising}
\label{app:fouriertrunc}

Thanks to the diagonalization of $\tilde{X}$ we can identify an approximation as follows.
Let $\mathcal{I} := \{ k ; \exp(-\omega|k|^2) > \rho \gamma\}$, for $\rho<1$, 
such that for $z=\tilde X \tilde\beta$
and $\widehat{z} = \tilde X_\mathcal{I} \tilde \beta$, we have $|z - \widehat{z}| < \gamma$, i.e. 
the observed signal is less than the observational noise.
Here $\tilde X_\mathcal{I}$ is shorthand notation for the rank $\tilde n = | \mathcal{I}|$
approximation of $\tilde X$ obtained by truncating wavenumbers $k \notin \mathcal{I}$.
For appropriate choices of $\omega, \gamma >0$, this provides a tractable scenario for full 
observations (in the sense that the solution is close to the actual full observation case).
The situation is slightly complicated however, since $X \in \bbR^{\tilde p \times p}$ 
despite being rank $\tilde{n}$.
We therefore approximate $X \approx X_\ell X_r$, 
where $X_\ell \in \bbC^{\tilde p \times \tilde n}$ and $X_r \in \bbC^{\tilde n \times p}$ 
are defined as follows
$$
X_\ell := F^{-1} (\exp(-\omega |k|^2))_{k\in \mathcal I} \, , \qquad
X_r^\dagger := \begin{pmatrix} 
F^{-1} (i k_1 |k|^{-2})_{k\in \mathcal I} \\
F^{-1} (i k_2 |k|^{-2})_{k\in \mathcal I} \\
\tilde{p}\delta_{k=0}
\end{pmatrix} \, .
$$
Now $X^\dagger X = X_r^\dagger (X_\ell^\dagger X_\ell) X_r$
and we simply redefine \cref{eq:monolithC} with an alternative application
of the Sherman Morrison Woodbury matrix identity
$$
C_n = (I_p - \tilde K X_r)C_0 \, , \qquad 
\tilde{K} = C_0 X_r^\dagger (X_r C_0 X_r^\dagger + \gamma^2 (X_\ell^\dagger X_\ell)^{-1})^{-1} \, , 
$$
and \cref{eq:monolith} becomes
$$
m_n = \tilde K (X_\ell^\dagger X_\ell)^{-1} X_\ell^\dagger Y \, .
$$
We note that $X_\ell^\dagger X_\ell \in \bbR^{\tilde n \times \tilde n}$ can be easily computed 
and inverted for a cost $\cO(p\tilde{n}^2)$.

\end{document}